%% file: main.tex
\documentclass[fleqn,usenatbib]{mnras}

\usepackage{newtxtext,newtxmath}

\usepackage[T1]{fontenc}

\DeclareRobustCommand{\VAN}[3]{#2}
\let\VANthebibliography\thebibliography
\def\thebibliography{\DeclareRobustCommand{\VAN}[3]{##3}\VANthebibliography}

\usepackage{graphicx}	
\usepackage{amsmath}	
\usepackage{hyperref}
\usepackage{booktabs}

\newcommand{\cm}{\,$\text{cm}$}	
\newcommand{\ms}{\,$\text{m}\,\text{s}^{-1}$}	
\newcommand{\kms}{\,$\text{km}\,\text{s}^{-1}$}	

\newcommand{\mearth}{M$_\oplus$}
\newcommand{\rearth}{R$_\oplus$}

\defcitealias{rescigno_hot_2023}{R24}

\title[A study of the TOI-2134 system]{Understanding eccentric temperate giants: an in-depth study of the architecture and stellar obliquity of the TOI-2134 system}

\author[F. Rescigno et al.]{
Federica Rescigno$^{1}$\thanks{E-mail: f.rescigno@bham.ac.uk},
Manu Stalport$^{2,3}$,
Ancy Anna John$^{1}$,
Tiger Lu$^{4}$,
Daisy A. Turner$^{1}$,
\newauthor
Lorena Acu\~{n}a-Aguirre$^{5}$,
Anand Bhongade$^{1}$,
Anjali A.A. Piette$^{1}$,
Vedad Kunovac$^{6,7}$,
Michael Cretignier$^{8}$,
\newauthor
Andrew Vanderburg$^{9}$,
Ken Rice$^{10,11}$,
Annelies Mortier$^{1}$,
Rishikesh Sharma$^{12}$,
Guillaume H\'ebrard$^{13}$,
\newauthor
Abhijit Chakraborty$^{12}$,
Alessandro Sozzetti$^{14}$,
Andrew Collier Cameron$^{15}$,
P\'ia Cort\'es-Zuleta$^{15}$,
\newauthor
Rosario Cosentino$^{16,17}$,
Florian Destriez$^{18,13}$,
Mercedes L\'opez-Morales$^{19}$,
Luca Malavolta$^{20,21}$,
\newauthor
Jes\'us Maldonado$^{22}$,
Giacomo Mantovan$^{23,21}$,
Francesco Pepe$^{24}$,
Matteo Pinamonti$^{14}$,
Andr\'e M. Silva$^{25,26}$,
\newauthor
Stephane Udry$^{27}$,
Shreyas Vissapragada$^{28}$,
Thomas G. Wilson$^{7,15}$
\\
$^{1}$School of Physics \& Astronomy, University of Birmingham, Edgbaston, Birmingham B15 2TT, UK\\
$^{2}$Space Sciences, Technologies and Astrophysics Research (STAR) Institute, Universit\'e de Li\`ege, All\'ee du 6 Ao\^ut 19C, 4000 Li\`ege, Belgium\\
$^{3}$Astrobiology Research Unit, Universit\'e de Li\`ege, All\'ee du 6 Ao\^ut 19C, B-4000 Li\`ege, Belgium\\
$^{4}$Center for Computational Astrophysics, Flatiron Institute, 162 5th Avenue, New York, NY 10010, USA\\
$^{5}$Max Planck Institut f\"ur Astronomie, K\"onigstuhl 17, 69117, Heidelberg, Germany\\
$^{6}$Centre for Exoplanets and Habitability, University of Warwick, Coventry, CV4 7AL, UK\\
$^{7}$Department of Physics, University of Warwick, Gibbet Hill Road, Coventry CV4 7AL, United Kingdom\\
$^{8}$Department of Physics, University of Oxford, OX13RH Oxford, UK\\
$^{9}$Center for Astrophysics | Harvard \& Smithsonian, Cambridge, MA 02138, USA\\
$^{10}$Institute for Astronomy, University of Edinburgh, The Royal Observatory, Blackford Hill, Edinburgh, EH93HJ, UK\\
$^{11}$Edinburgh Centre for Planetary Sciences, University of Edinburgh, Edinburgh, EH93HJ, UK\\
$^{12}$Astronomy \& Astrophysics Division, Physical Research Laboratory, Ahmedabad 380009, India\\
$^{13}$Institut d'astrophysique de Paris, UMR 7095 CNRS universit\'e Pierre et Marie Curie, 98 bis, boulevard Arago, 75014, Paris, France\\
$^{14}$INAF  -- Osservatorio Astronomico di Torino, Strada Osservatorio 20, Pino Torinese, Italy\\
$^{15}$Centre for Exoplanet Science, SUPA School of Physics and Astronomy, University of St Andrews, North Haugh, St Andrews KY16 9SS, UK\\
$^{16}$Fundaci\'on Galileo Galilei - INAF\\
$^{17}$INAF -- Osservatorio Astrofisico di Catania\\
$^{18}$LIRA, Universit\'e Paris Cit\'e, Observatoire de Paris, Universit\'e PSL, Sorbonne Universit\'e, CY Cergy Paris Universit\'e, CNRS, F-92190 Meudon, France\\
$^{19}$Space Telescope Science Institute, 3700 San Martin Drive, Baltimore, MD 21218, USA\\
$^{20}$Dipartimento di Fisica e Astronomia “Galileo Galilei”, Universit\` a degli Studi di Padova, Vicolo dell’Osservatorio 3, 35122, Padova, Italy\\
$^{21}$INAF, Osservatorio Astronomico di Padova, Vicolo dell’Osservatorio 5, 35122,Padova, Italy\\
$^{22}$INAF -- Osservatorio astronomico di Palermo\\
$^{23}$Centro di Ateneo di Studi e Attivit\`a Spaziali ``G. Colombo'' -- Universit\`a degli Studi di Padova, Via Venezia 15, IT-35131, Padova, Italy\\
$^{24}$D\'epartement d'astronomie, Universit\'e de Gen\`eve, Chemin Pegasi 51, 1290 Versoix, Switzerland\\
$^{25}$Instituto de Astrof\'isica e Ci\^encias do Espa\c{c}o, Universidade do Porto, CAUP, Rua das Estrelas, 4150-762 Porto, Portugal\\
$^{26}$Departamento de F\'isica e Astronomia, Faculdade de Ci\^encias, Universidade do Porto, Rua do Campo Alegre, 4169-007 Porto, Portugal\\
$^{27}$Geneva Observatory, Geneva University, Chemin Pegasi 51, 1290 Versoix, Switzerland\\
$^{28}$Carnegie Science Observatories, 813 Santa Barbara Street, Pasadena, CA 91101, USA
}
\date{Accepted XXX. Received YYY; in original form ZZZ}
\pubyear{2026}

\begin{document}
\label{firstpage}
\pagerange{\pageref{firstpage}--\pageref{lastpage}}
\maketitle

\begin{abstract}
We revisit the TOI-2134 planetary system with three new high-cadence TESS sectors and 98 more spectra. This new analysis confirms the two orbiting planets by simultaneously modelling a total of eight sectors of corrected TESS photometry and 280 HARPS-N and SOPHIE radial velocities: an inner mini-Neptune in a near-circular \mbox{$9.229198\pm0.000003$} days orbit, and an outer temperate sub-Saturn orbiting with a \mbox{$95.852840\pm0.000042$ days} period and eccentricity of $0.31\pm0.01$. The masses and radii of the planets were computed to be \mbox{$9.37\pm0.54$ \mearth} and \mbox{$2.735\pm0.068$ \rearth} for planet b, and \mbox{$58.3\pm1.9$ \mearth} and \mbox{$7.35\pm0.18$ \rearth} for planet c. The new data not only improves the detection significance and precisions on the planetary orbits, but also breaks the original multimodality in the eccentricity solution for the outer planet. We also detect a long-term trend in the radial velocity data, which we attribute to a stellar magnetic cycle. We investigate the spin-orbit alignment of the system via observations of the Rossiter-McLaughlin effect for TOI-2134~b with EXPRES and TOI-2134~c with PARAS-2. No RM effect was detected for planet b, but we find a 4.7$\sigma$ detection of a $59\pm31^{\circ}$ obliquity for planet c. Finally, we examine the architecture of the system, assess its completeness, investigate the planetary interior, and their suitability for follow-up atmospheric analysis.

\end{abstract}




\section{Introduction}

Of the over 6,000 exoplanets discovered to date\footnote{As shown in \url{https://exoplanet.eu}, accessed on 30/03/2026}, gas giants with inner low-mass companions are among the most informative, yet poorly understood populations.
In the Solar System, our gas giants Jupiter and Saturn reside well beyond the ice line. This is consistent with core-accretion theories which require them to form in a cold, solid-rich environment \citep[e.g.][]{Pollack1996I, Hubickyj2005}. Exoplanet surveys have, however, detected gas giants spanning a wide range of orbital periods, with the majority of them residing inside their system's ice line \citep{Winn2015}, putting in-situ formation into question, and suggesting that their orbits evolved either during or after formation. Overall, detected gas giants appear divided into three sub-categories: hot Jupiters ($P<10$ days), warm or temperate giants \mbox{(10 days $<P<$ 100-400 days)}, and cold giants (orbiting outside the ice line). Each group presents distinct characteristics: hot Jupiters are strongly irradiated, are often tidally locked to their stars, and are subject to intense atmospheric escape \citep[e.g.,][]{Murray2009}. Temperate gas giants orbit further away from their star, and are spared the majority of these extreme phenomena. Irradiation-driven processes do not significantly affect their atmosphere and composition, allowing them to preserve their primordial characteristics for longer. The presence of either these classes of planets challenges current formation theories. The current consensus is that they must have undergone some type of migration within their system via, for example, disk-driven processes or high-eccentricity interactions \citep[e.g.,][]{Lin1996, Winn2015, Nelson2018, Schulte2024}. Yet, the contributions of different types of migrations remain topic of debate.

Transiting gas giants are particularly valuable laboratories, especially when their masses can be measured with radial-velocity (RV) follow-ups, enabling the study of their bulk density and composition. Gas giants are well described by two-layer models, with heavy metal cores and H/He-dominated envelopes. Population-level analysis of their metal mass fractions can be important to distinguish between run-away gas accretion and alternative pathways such as envelope accretion in gas-depleted disks \citep[e.g.,][]{Pollack1996I, Hubickyj2005, Lee2015, Thorngren2016, Chachan2025, Thomas2025}. Their large atmospheric scale heights also facilitate atmospheric analysis via transmission or emission spectroscopy, providing insights into their chemical compositions, cloud properties, and escape mechanisms \citep{Madhusudhan2019}. Analysis of the compositions and atmospheres of transiting giants over a wide range of temperatures and orbital periods can thus shed some much-needed light on possible formation or migration theories.
However, the majority of well-characterised transiting giants fall into the hot Jupiter category. Their short orbital periods and large RV amplitudes make them much easier to detect. In fact, 49.8\% of all confirmed exoplanets have $P<10$ days, and only 20\% have $P>90$ days\footnote{According to data at \url{https://exoplanet.eu}, accessed on 30/03/2026}.
Precious few temperate gas giants have been detected and we have high-significance mass determinations for even less. Any new detection is thus incredibly valuable.

Within the temperate gas giants population, Saturn- and sub-Saturn-like planets occupy a critical transitional regime between the very common sub-Neptunian exoplanets \citep[thought to follow different formation mechanisms, e.g.,][]{Helled2014,Lee2019} and Jovian worlds. They are typically defined to have radii between 4 and 8 \rearth\, and masses within $\sim$15 and 60 \mearth\, \citep{Petigura2017}. For this sub-population, a correlation between planet mass and stellar metallicity has been found \citep{Petigura2017, Nowak2020}, consistent with findings suggesting that more massive planets preferentially orbit higher-metallicity stars \citep[e.g.][]{Santos2001, Fischer2005, Buchhave2012,Rodriguez2025}. These trends imply that more metal-rich disks are more likely to be able to support the formation of massive cores able to accrete large gaseous envelopes \citep[also seen for massive sub-Neptunes in][]{Wilson2022}. The relationship between the mass of the substellar companions and their stars has also been investigated, suggesting that formation by core accretion is most efficient for planets with masses $0.2-2$ $M_{\rm J}$ \citep{Maldonado2019}.
Like all temperate gas giants, sub-Saturns were also found to span a wide range of eccentricities, with an overdensity in their occurrence rates around $e\sim0.3$ \citep{Fairnington2025}, supporting multiple eccentricity excitation mechanisms. However, the sample of temperate Saturns remains comparatively small, limiting the relevance of any population-level studies.

Measurements of orbital obliquity offer an additional and complementary formation and evolution diagnostic. The angle between the planet's plane of orbit and the rotational axis of the star can be measured by observing the Rossiter-McLaughlin effect \citep[RM effect:][]{Rossiter1924, McLaughlin1924} during transit. A large fraction of hot Jupiters have been found to be spin-orbit misaligned \citep{Albrecht2022}, supporting migration scenarios involving dynamical excitation and tidal evolution. In contrast, recent studies found temperate gas planets to be preferably aligned with the stellar spin axis \citep{Rice2022, Wang2024, Espinoza2026}, despite orbiting too far from their star to have re-aligned via tidal dissipation, raising the possibility of the population being primordially aligned and having instead possibly migrated via disk-interactions \citep[e.g.][]{Mantovan2024}. Extending obliquity analysis to a wider mass range as well as longer orbital periods is thus vital to assess the validity of different migration pathways.

Therefore, the characterisation of transiting temperate gas giants is of particular relevance. With well-defined radii, masses, and orbital characteristics, they enable a wealth of follow-up analysis, from ensemble studies to atmospheric observations. When such giants are also found to have inner lower-mass companion planets, they provide a unique opportunity to study dependence on multi-planetarity, as well as a chance to probe atmospheric evolution across a wide irradiation range within the same system. In this context, well-characterized multi-planet systems containing both a close-in planet and a temperate giant companion are particularly powerful laboratories for testing coupled theories of core growth, gas accretion, dynamical sculpting, and atmospheric mass loss.

\subsection{The TOI-2134 System}
The TOI-2134 system was first analysed in \cite{rescigno_hot_2023} (from here on referred to as \citetalias{rescigno_hot_2023}). The star, also known as G204-45, is a bright, high-proper motion, K5V dwarf \citep{giclas_lowell_1979,stassun_revised_2019}. Through this work, we employed the stellar parameters retrieved by \citetalias{rescigno_hot_2023}, and reported in Table \ref{tab:star}.

\input{Tables/stellar_characteristics}

With five TESS sectors, three years of WASP photometry monitoring, and 219 spectra observed with HARPS-N and SOPHIE, \citetalias{rescigno_hot_2023} detected and confirmed two transiting exoplanets. The inner TOI-2134~b was determined to be a mini-Neptune in a \mbox{9.2292005$\pm$0.0000063 day} circular orbit, with radius and mass equal to $2.69\pm0.16$ \rearth\, and $9.13\pm 0.77$ \mearth. The planet was also subsequently observed during transit with Keck/NIRSPEC and XMM-Newton. With these data, \cite{Zhang2023} detected escaping helium for the fifth-ever time from a mini-Neptune, and for the first time from a mature one. TOI-2134~b was found to be particularly important as it anchors the relationship between observed and energy-limited mass-loss rates, as the one with the lowest value for both.

For the outer planet, TOI-2134~c, TESS photometry was able to capture only a mono-transit, and although an extensive follow-up campaign was undertaken both on the ground and in space, no second transit was detected due to bad luck and worse weather. From a joint analysis of the available mono-transit and the RVs, the exoplanet was concluded to be a Saturn-like temperate gas giant with orbital period of 95$^{+0.36}_{-0.25}$ days. Its radius and mass were computed to be \mbox{7.27$\pm0.42$ \rearth} and \mbox{41.89$^{+7.69}_{-7.83}$ \mearth}. The orbital configuration of TOI-2134~c was hard to pinpoint, as two equally probable solutions were found: one with eccentricity of 0.45$\pm$0.05 and a second with more extreme one of 0.67$\pm$0.05. After analysing the shape of the mono-transit and validating the stability of the orbit, the larger eccentricity case was deemed more likely and was thus presented with the note that the planetary mass and radius agreed to 1$\sigma$ for both models.

In the years following that publication, TOI-2134 was observed in three additional TESS sectors with high cadence, and a new transit of the outer planet was detected. Together with the long-term RV follow-up campaign undergone by HARPS-N and SOPHIE, this new amount of data kick-started a re-analysis of the system. Particular attention was paid to the eccentricity of the outer planet, given the two originally competing models (see Section 6.2 of \citetalias{rescigno_hot_2023}).
In this work, we thus revisit the TOI-2134 system. We confirm and refine the orbital parameters of both planets, reaching a more precise and better defined unique eccentricity solution for TOI-2134~c. We also investigate the orbital obliquity of the planets via an RM analysis.\\

This publication is structured as follows. The new data is presented in Section \ref{Sec:data}. In Section \ref{Sec:st_act} the stellar activity signal is re-analysed. We first fit the photometric data in Section \ref{Sec:transit}, then we analyse the RVs in Section \ref{Sec:rv}. Section \ref{Sec:joint} covers the final joint photometry and RV analysis. Section \ref{Sec:rm} focuses on the study of the RM signals. We present the new planetary results and discuss the characteristics of the system in Section \ref{Sec:disc}. 
In particular, Section \ref{sec:ecc} discusses the new eccentricity solution of planet c and places it in its wider context.
Sections \ref{sec:detection_limits} and \ref{sec:migration} focus on the possible formation and migration pathways of the system, also considering possible undetected outer companions.
The interior compositions of both planets are assessed in Section \ref{sec:interiors}.
We also highlight the potential of the system for atmospheric analysis follow-ups in Section \ref{sec:atmo}. Finally, we conclude in Section \ref{Sec:concl}.

\section{Data}
\label{Sec:data}
This work makes use of both archival and unpublished data.

\begin{figure*}
    \centering
    \includegraphics[width=16cm]{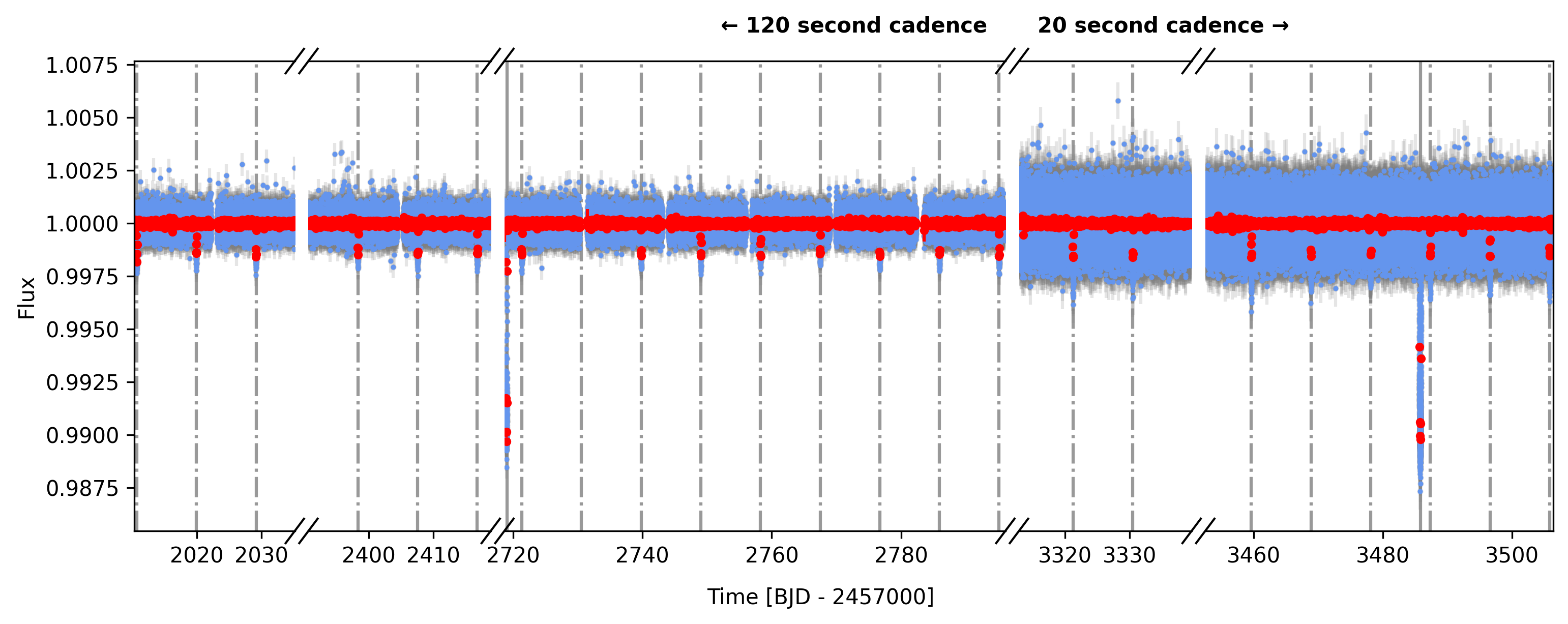}
    \caption{TESS systematics corrected light curve over eight sectors, with corrected Barycentric Julian Date on the x-axis and normalised flux on the y-axis. All data is plotted in blue, with errorbars in light gray. Overplotted in red is the 30-minute mean of the observations. The last three sectors were observed at higher cadence. The 23 visible transits of TOI-2134~b are highlighted by vertical dash-dotted lines. The two transits of TOI-2134~c are shown with solid gray lines.}
    \label{fig:tess}
\end{figure*}

\subsection{TESS Photometry}
\label{Sec:data_phot}
NASA's Transiting Exoplanet Survey Satellite \citep[TESS:][]{ricker_transiting_2014} is a space-based photometric telescope designed for all-sky survey of transiting exoplanets. It consists of four wide-field refracting cameras with 10\cm\, apertures. TESS has observed TOI-2134 sporadically over the past 5 years. Five sectors (Sectors 26, 40, 52, 53, and 54) were observed with 2-minute cadence only, between BJD 2,459,010 and 2,459,035, BJD 2,459,390 and 2,459,418, and BJD 2,459,718 and 2,459,797 (9 Jun to 4 Jul 2020, 24 Jun to 22 Jul 2021, and 18 May to 5 Aug 2022). Subsequently, three sectors (Sectors 74, 79, and 80) were also observed at 20-second cadence between BJD 2,460,312 and 2,460,340, and BJD 2,460,452 and 2,460,506 (2 to 30 Jan 2024, and 21 May to 14 Jul 2024). In total there are 330,359 data points. 

The data were originally processed and calibrated using the TESS Science Processing Operation Centre (SPOC) pipeline based at NASA Ames Research Center \citep{jenkins_overview_2010}. Two sets of planetary transits were detected: 23 transits for an inner planet, and two transits for the outer one (contained in the original Sector 52, and the new Sector 80). However, strong systematics in Sectors 40, 79 and 80 remained. To correct for these, we performed a second systematics correction of the SPOC Simple Aperture Photometry (SAP) light curves \citep{twicken_photometric_2010, morris_kepler_2020}. We modelled the systematics as the sum of moments of the spacecraft quaternion time series, while long-term variations were described by a basis spline \citep{vanderburg_tess_2019}. After producing the systematics-corrected light curves, we flattened them by simultaneously modelling the transit and the slow variations, as described in \cite{Vanderburg2016}. These corrected data are shown in Fig. \ref{fig:tess}, with all the planetary transits highlighted by vertical lines.

\subsection{Radial Velocities}
\label{Sec:data_rv}
We obtained a total of 315 RVs with four different instruments: HARPS-N, SOPHIE, EXPRES, and PARAS-2.

\subsubsection{HARPS-N Spectroscopy}
\label{Sec:data_hn}

\begin{figure*}
    \centering
    \includegraphics[width=17.5cm]{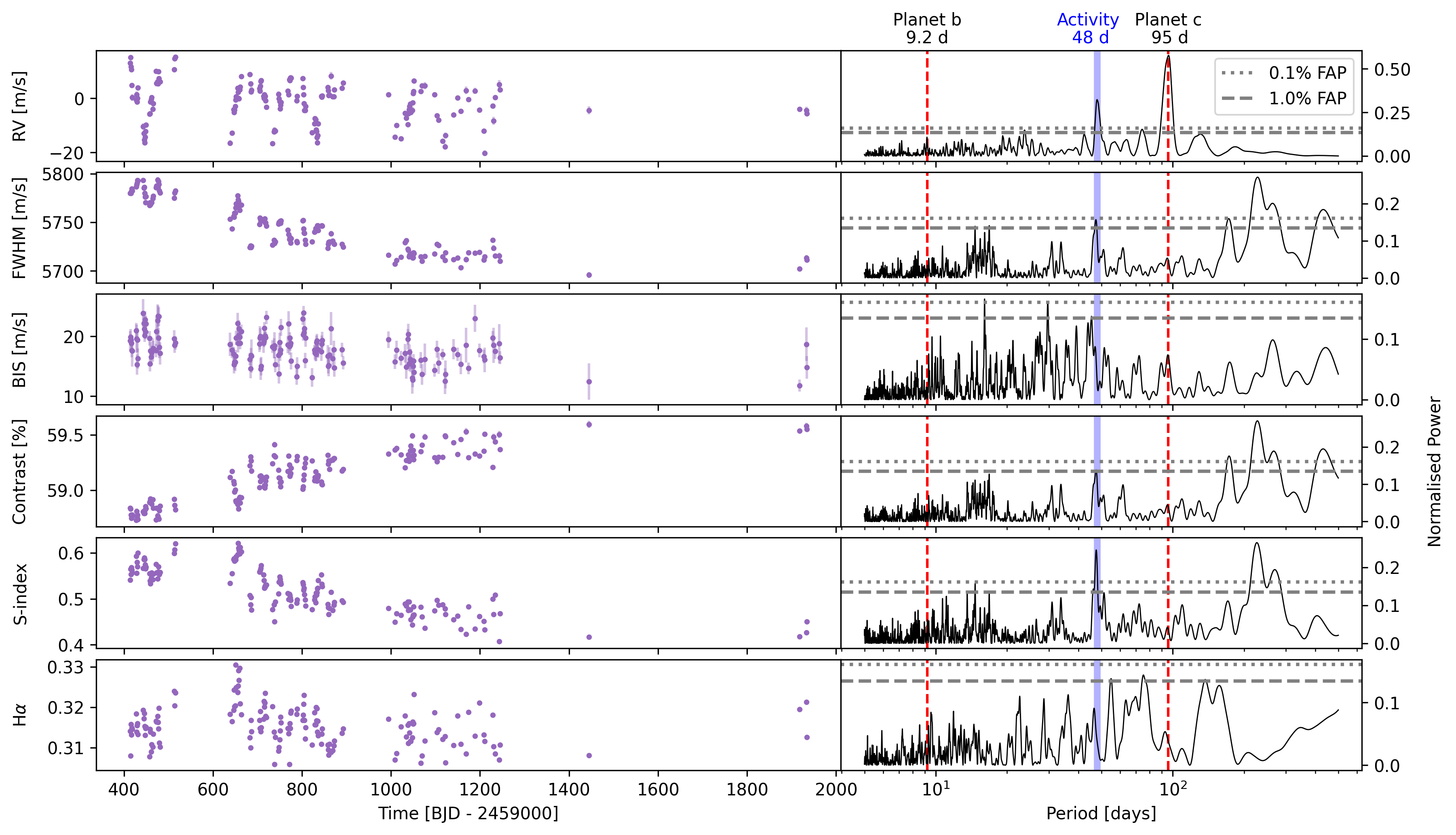}
    \caption{\textit{Left}: Data products of the 3.0.1 HARPS-N DRS plotted against corrected Barycentric Julian Date. In order from top to bottom: radial velocities, full-width at half-maximum, bisector span (all in \ms), contrast, S-index, and H$\alpha$ index. Uncertainties are plotted as errorbars, but may be too small to be visible. The first two seasons were published in \citetalias{rescigno_hot_2023}, and the latter two are presented for the first time in this work. \textit{Right}: Generalised Lomb-Scargle periodograms of the corresponding time series, plotted as period of the signal in days against normalised power. False alarm probabilities levels of 1 and 0.1$\%$ for each periodogram are depicted as dashed and dotted horizontal gray lines, respectively. Vertical red dashed lines highlight the orbital periods of the two planets in the system as computed from \citetalias{rescigno_hot_2023}. The suspected activity signal is shown by a blue solid vertical band.}
    \label{fig:hn_rv}
\end{figure*}

On top of the previously published 111 RVs, we collected 49 additional spectra of TOI-2134 (for a total of 160 spectra) with the pressure- and temperature-stabilised \'echelle High-Accuracy Radial-velocity Planet Searcher for the Northern hemisphere spectrograph \citep[HARPS-N:][]{cosentino_haprsn_2012,cosentino_haprsn_2014}, installed on the 3.6-m Telescopio Nazionale Galileo (TNG) at the Observatorio de Los Muchachos on La Palma, Spain. HARPS-N's spectral range spans between 383 and 691 nm, and has average resolution $R=115,000$. The spectra were collected as part of the HARPS-N Collaboration Time for the RV follow-up campaign for TESS planet candidates. The first 32 spectra were collected between BJD 2,459,417 and 2,459,515 (27 Oct to 21 Jul 2021), 79 were collected between BJD 2,459,638 and 2,459,892 (27 Feb to 11 Nov 2022), 45 of the new spectra were observed between BJD 2,459,995 and 2,460,245 (19 Feb to 27 Oct 2023), while the last 4 were collected as low-cadence follow-ups between BJD 2,460,445 and 2,460,934 (14 May 2024 to 15 Sep 2025). The average exposure time was $\sim$15 minutes, with an average signal-to-noise ratio (S/R) at 550 nm of $\sim$110. The spectra were reduced using the 3.0.1 version of the standard Data Reduction Software (DRS) adapted from the ESPRESSO pipeline \citep{dumusque_three_2021}, and the radial velocities were extracted using the cross-correlation function (CCF) method computed with a K6-type numerical mask. The DRS also outputs a series of activity indicators derived from the stellar spectra (the S-index and the H$\alpha$ index) and their CCFs (the full-width at half-maximum FWHM, the bisector span BIS, and the contrast) as well as their uncertainties. All data are plotted in Fig. \ref{fig:hn_rv}.
The derived RVs show a peak-to-peak variation of 35.65 \ms, and a standard deviation of 7.46 \ms. We subtract the median RV of -20.622 \kms\, for illustrative purposes. The mean RV uncertainty is computed to be 0.76 \ms.

\subsubsection{SOPHIE Spectroscopy}
\label{Sec:data_sp}

\begin{figure*}
    \centering
    \includegraphics[width=17.5cm]{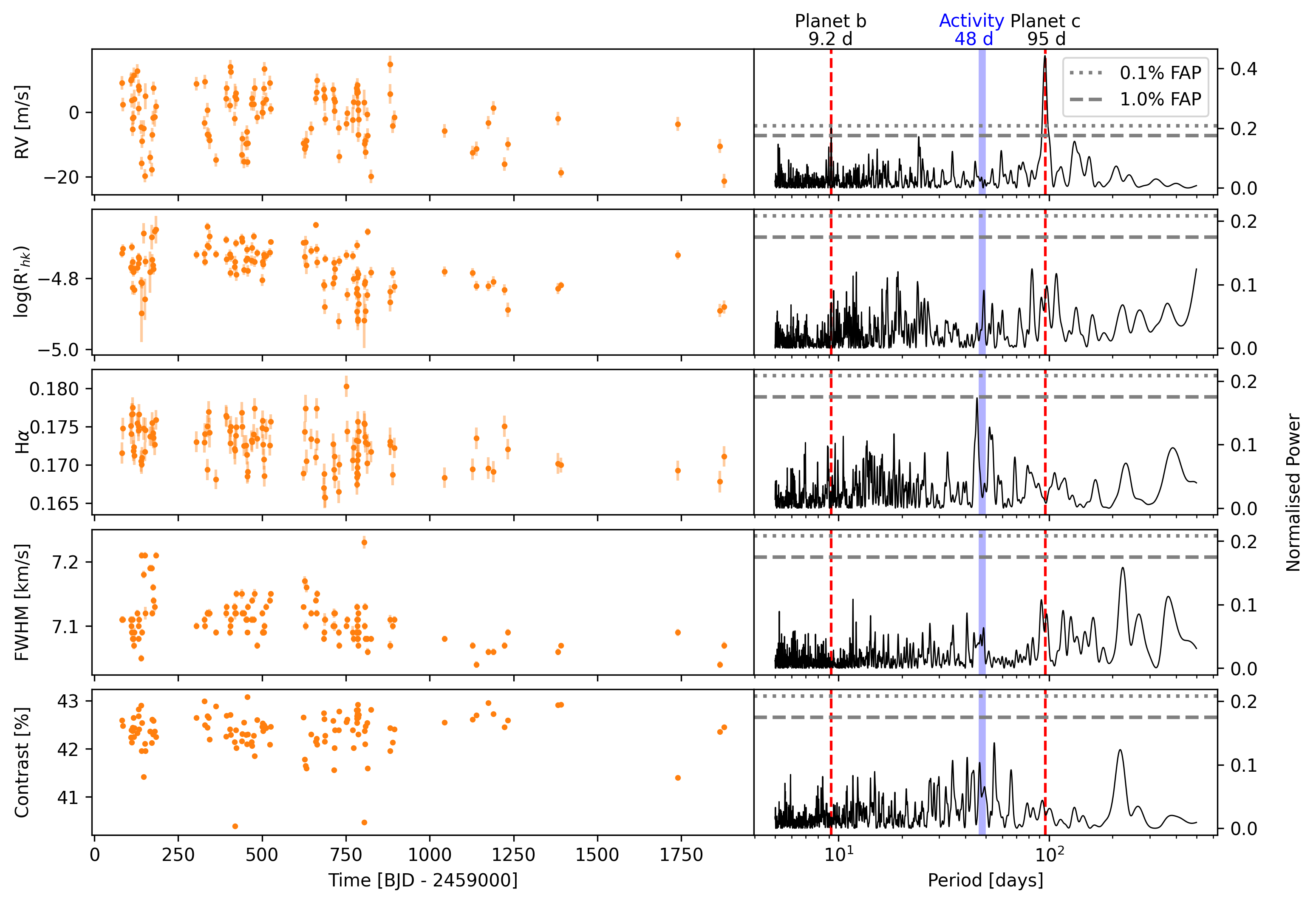}
    \caption{Same as Fig. \ref{fig:hn_rv}, but for the SOPHIE data products. In order from the top: radial velocities, $\log R'_{\rm hk}$, H$\alpha$ index, FWHM and contrast of the CCF, with associated periodigrams on the right. \citetalias{rescigno_hot_2023} included data from the first three seasons, while the last 12 data-points have been obtained since its publishing. Uncertainties for all timeseries are included, but errorbars may be too small to be visible.}
    \label{fig:sp_rv}
\end{figure*}

Further follow-up spectra were also taken by the Spectrographe pour l’Observation des
Ph\'enom\`enes des Int\'erieurs stellaires et des Exoplan\`etes \citep[SOPHIE:][]{perruchot_sophie_2008}. SOPHIE is a stabilised \'echelle spectrograph mounted on the 193 \cm\, telescope at the Observatoire de Haute-Provence, France \citep{bouchy_sophie_2013}. Similarly to HARPS-N, SOPHIE observes the entire optical range (387-694 nm). \citetalias{rescigno_hot_2023} presented 113 measurements of TOI-2134, of which 108 passed quality assessments. For this work, we obtained 13 additional SOPHIE spectra between BJD 2,460,044 and 2,460,391 (8 Apr 2023 to 20 Mar 2024). The data were acquired under the program ID 19A.PNP.HEBR. None of them were found to be significantly polluted by sky background. Observations were taken with SOPHIE's high-resolution mode (R=75,000), and the fast mode of the CCD reading. The data were reduced as described in \citetalias{rescigno_hot_2023}, in particular following the procedures presented by \cite{Heidari2024, Heidari2025}. One radial velocity observation was removed after quality cuts, yielding 12 new spectra. The final SOPHIE dataset for TOI-2134 thus includes a total of 120 measurements, and spans a longer baseline than the HARPS-N data.
Together with the corrected RVs, a series of activity indicators derived from the spectra ($\log R'_{\rm hk}$ and H$\alpha$ index) and the CCFs (FWHM and contrast) are produced. Uncertainties for the first two were derived directly from the fitting of the spectra, while uncertainties for the CCF indicators are defined starting from uncertainties on the RV measurements following the scaling convention described in \cite{santerne_pastis_2015}. All data are plotted in Fig \ref{fig:sp_rv}.
The derived RVs show peak-to-peak variations of 36.20 \ms, and a standard deviation of 8.46 \ms, both comparable to those derived from the HARPS-N radial velocities. The average uncertainty is however three times larger, equal to 2.28 \ms.

\subsubsection{EXPRES Spectroscopy}
We observed TOI-2134 during a transit of planet b on UT night 28 April 2023 (circa BJD 2,460,062.5) using the EXtreme PREcision Spectrometer \citep[EXPRES:][]{jurgenson2016} mounted on the 4.3m Lowell Discovery Telescope \citep{levine2012,levine2016}, located near Flagstaff, AZ, USA. EXPRES is a stabilized high-resolution ($R\sim 137,000$) optical (390-780 nm) spectrograph with a single measurement precision of <32 cm$\,\mathrm{s}^{-1}$\, in ideal circumstances \citep{Blackman2020}. We took 21 exposures over a duration of \mbox{${\sim}5.8$ hours} covering the full $3$-hour transit with exposure times of 900 seconds. The data were acquired as part of the Lowell Observatory Staff program.
Due to target visibility we only acquired one pre-transit exposure. The observations were taken in moderate conditions, with S/N ranging from 39 to 80 at 550 nm, and target S/N of 100. The start of the observing sequence was particularly affected by poor seeing and high airmass. A laser frequency comb and Th-Ar lamp were used for wavelength calibration throughout the observations. We computed CCFs using a K2 line mask. Radial velocities and their uncertainties were derived from the centre of a Gaussian fit to the CCF. The peak-to-peak variation of the data was computed to be 15.8 \ms, and the standard deviation was found to be \mbox{3.5 \ms}. The median value of the RVs (-20.6\ms) was subtracted from the data. We computed the mean uncertainty value to be 1.5 \ms. These uncertainties reflect the combination of bad seeing and the target being observed at high airmass.

\subsubsection{PARAS-2 Spectroscopy}
\label{Sec:data_p2}
The PRL Advanced Radial-velocity Abu-sky Search 2 \citep[PARAS-2:][]{chakraborty_paras-2_2018,chakraborty_prl_2024} is a high-resolution fibre-fed spectrograph mounted on the 2.5 m telescope at the PRL Mount Abu Observatory, Gurushikhar, India, as the successor of the \mbox{PARAS-1} spectrograph \citep[]{chakraborty_prl_2008, chakraborty2014_PARAS}. The spectrograph has been functioning since late 2022 and is able to achieve on-sky RV dispersion of $\sim$2 \ms\, \citep{toi6651}. It operates in the wavelength range of 380-690 nm at a median resolution of $\sim$107,000. It uses Uranium-Argon hollow-cathode lamps for simultaneous referencing and wavelength calibration \citep{toi6651, Sharma_2021_uar}. 
Given its unique latitudinal position, PARAS-2 proved to be the only high-resolution spectrograph able to observe TOI-2134 over the past few years during any of the night-time transits of planet c to measure its RM. The selected transit fell on the UT night 7 April 2025 ($\sim$ BJD 2,460,773.5) and was expected to last $\sim$5 hours. We obtained 13 spectra during the transit, as well as 3 further spectra the nights before and after to build a baseline. Unfortunately, due to the limited available dark hours, no baseline data could be collected on the night of the transit. The data were obtained as part of the PRL Exoplanet Legacy Program.
The RVs are derived by cross-correlating the wavelength-calibrated spectra with a K5 line mask. A detailed description of the data reduction procedures and analysis pipeline is presented in \cite{toi6651}. The exposure time for all the spectra acquired was 900 seconds, except for one spectrum, acquired the night before the transit, which was 1200 seconds. Although the observations span the majority of the planetary transit, the weather at the site was sub-optimal, and the achieved S/N ranges between 20 and 34 (per pixel), which resulted in a mean uncertainty of 5.5 \ms. The peak-to-peak variation of the RV was computed to be 54.1 \ms, with a standard deviation of \mbox{3.6 \ms.} The mean seeing during transit was approximately 1.25", for airmasses between 1.72 and 1.04. Two datapoints were found to be outliers by several tens of \ms and were excluded from all analyses. The final dataset used for RM analysis comprised of 14 spectra.

\section{Stellar Activity Signal}
\label{Sec:st_act}
We started by measuring the projected rotational velocity $v\sin(i_\star)$ of TOI-2134 from the obtained spectra. In agreement with the results of \citetalias{rescigno_hot_2023}, we were only able to define the higher limit of \mbox{$v\sin(i_\star)<2$\kms} through spectral synthesis \citep{Tsantaki2014}, which yielded a minimum stellar rotation period of $\sim$18 days (assuming $i_\star=90^\circ$). In order to further refine this measurement we also used a method similar to that described in \cite{Rainer2023}, which extracts the projected rotational velocity using the FWHM of the spectral CCFs. We computed the $v\sin(i_\star)$ to be \mbox{$1.32\pm0.32$ \kms.}

The mean S-index of the observations was computed to be 0.520$\pm$0.001, equivalent to a mean $\log R'_{\rm hk}$ of -4.87$\pm$0.41 using the \cite{Noyes1984} relations \citep[compared to the Sun's mean of circa -5.02, e.g.][]{lubin_frequency_2012}.

We computed the Spearman correlation coefficient between both RV datasets and their corresponding activity indicators\footnote{A significant caveat to this analysis is that the RVs are dominated by the signal of the outer planet, which will inherently weaken the correlation with activity indicators.}. The correlation plots are shown in the Appendix in Fig. \ref{fig:corr}. Weak but non-zero correlations were only found between a subsection of the \mbox{HARPS-N} activity indicators with their RVs. In order of significance: the contrast showed a correlation coefficient of -0.26, followed by the FWHM with 0.25, the S-index with 0.24, the BIS with 0.14, and finally the H$\alpha$ index with 0.05. The first three proxies also presented a clear temporal dependency, hinting at the presence of a long-period stellar signal. SOPHIE RVs showed close to no correlation with any of their indicators, with Spearman correlation coefficients of 0.11, 0.07, 0.04, and 0.02 for the H$\alpha$ index, FWHM, $\log (R'_{\rm hk})$, and contrast respectively. Overall, no strong correlation were thus found. Part of this behaviour is expected, as \citetalias{rescigno_hot_2023} showed that the amplitude of the stellar variability signal is comparable to the semi-amplitude of TOI-2134~c.

To begin assessing the impact of stellar variability as well as to isolate the rotation period of TOI-2134, we computed Generalised Lomb-Scargle periodograms \citep[GLS: ][]{zechmeister_generalised_2009} of the radial velocities and their proxies, plotted next to their respective time series in Figs. \ref{fig:hn_rv} and \ref{fig:sp_rv} for HARPS-N and SOPHIE, respectively. In all cases, we included false alarm probabilities (FAPs) of 1$\%$ and 0.1$\%$. Starting from the HARPS-N radial velocities, the strongest signal is found at $\sim$95 days, the orbital period of TOI-2134~c extracted in \citetalias{rescigno_hot_2023}. No strong peak is found at the period of the other known planet at 9.2 days. The second most significant peak, and the only other peak to surpass the 0.1$\%$ FAP level is instead found at $\sim$48 days, the postulated stellar rotational period. However, it is important to note that 48 days is almost exactly half of the period of the strongest peak at 95 days. This result made distinguishing between planetary signals and stellar activity contributions difficult. In fact, finding peaks at harmonics for planets and activity will significantly bias simple periodogram whitening analysis, and especially in cases where eccentricity is non-negligible. We proved this by simulating a Keplerian signal with period 95 days with significant eccentricity, subtracting a sinusoidal of the same period, and taking the periodogram of this remaining signal, which showed a statistically significant peak at around 48 days. This coupled behaviour created the original difficulties in determining the eccentricity of planet c, as the signal of an eccentric Keplerian can also be modelled by two sinusoidals with periods of $P$ and $P/2$. Nonetheless, we could ascertain the presence of a stellar periodic signal from the analysis of the periodograms of the activity indicators.
This $\sim$48 days period in fact also corresponds to the most significant peak under 100 days for the FWHM, the contrast, and especially in S-index, where the peak exceeds FAP of 0.01$\%$.
Most proxies also show power in the periodogram at long periods, indicative of a long-term trend (as a visual inspection of the time series also highlights). This trend is very consistent between S-index and FWHM, and strongly anti-correlated with the contrast. This signal is most likely an unresolved stellar magnetic cycle with a period longer than 3.3 years.

SOPHIE radial velocities also present their most significant peak at the orbital period of the outer planet, with a second relevant signal found at the period of the inner planet. No other significant peaks are found. The activity indicators also do not show strong signs of activity, with no significant periodic signal detectable.

We also investigated the presence of activity-induced signals in the available photometry. We found no signs of rotational modulation or significant contamination by flare events.

\subsection{Modelling the Activity Indicators}
\label{Sec:st_act_gp}
In order to better constrain the stellar rotation period, we also modelled the activity indicators with a Gaussian process (GP) described by a quasi-periodic (QP) kernel \citep{haywood_corot_2014} with a white noise "jitter" term added in quadrature. 
We define the QP kernel $k$ following the formulation of \cite{rajpaul_gaussian_2015} with an added white noise term as:
\begin{equation}
    \label{eq:GP}
    k(x_{\rm i}, x_{\rm j}) = A^2 \exp \left( -\frac{\sin^2(\pi (x_{\rm i} - x_{\rm j})/P_{\rm rot})}{2\lambda^2} - \frac{(x_{\rm i} - x_{\rm j})^2}{2P_{\rm dec}^2} \right) + \delta_{\rm i,j}\beta,
\end{equation}
in which $x_{\rm i}$ and $x_{\rm j}$ are two data points of our time series, $A$ represents the amplitude, $P_{\rm rot}$ is the period of the "periodicity" (in our case the rotational period of the star), $\lambda$ is the inverse harmonic complexity, $P_{\rm dec}$ is the timescale of the evolution (often tied to the decay timescale of active regions), $\delta_{\rm i,j}$ is the Kronecker Delta, and $\beta$ is the amplitude of the white noise.
We fitted for the best covariance hyperparameters with an affine invariant Markov Chain Monte Carlo \citep[MCMC: e.g.,][]{foreman-mackey_emcee_2013} parameter space searching algorithm. We used the {\sc MAGPy-RV}\footnote{Available at: \url{https://github.com/frescigno/magpy rv}} pipeline for one-dimensional GP regression \citep{rescigno_magpy-rv_2023}. We chose to only model the two activity proxy time series whose periodograms showed the strongest signal at a similar period as the RVs: the S-index and the FWHM of the HARPS-N spectra.

We defined similar priors for both datasets. The period was bound by a Gaussian prior centred on 48 days, with a width of 10 days, based on the results of the periodogram analysis. The timescale of the quasiperiodicity was required to be larger than the period, to ensure the interpretability of the results. All other variables were bound by wide uninformative positive uniform priors.

We simultaneously evolved 200 chains for 50,000 steps each, discarding a burn in of 10,000 steps. The models for both time series converged for all hyperparameters defined by a Gelman-Rubin statistic being less than 1.01. The S-index was best fit by a period of $50.5^{+2.2}_{-2.6}$ days, while the FWHM was best described with a period of $48.3\pm1.6$ days. In both cases, the derived stellar rotations agree within 1$\sigma$ with the rotation period extracted from the RVs in \citetalias{rescigno_hot_2023}, and with each other.

\section{Transit Photometry}
\label{Sec:transit}
To refine the orbital characteristics of the system, we underwent a full re-analysis of all the available data. We started with the photometry. We modelled the TESS data using \cite{mandel_analytic_2002} transit models, implemented in the package {\sc batman}\footnote{Available at: \url{https://github.com/lkreidberg/batman}} \citep{kreidberg_batman_2015}. In this formulation, we fit three parameters for the host star: its density, $\rho_\star$, and its linear and quadratic limb-darkening terms, $q_1$ and $q_2$, parametrised following \cite{kipping_efficient_2013}. The two transiting planets are described by six parameters each: their orbital period $P$, time of inferior conjunction $t_0$, orbital inclination $i$, the logarithm of the ratio of the radii of planet and star $\log(R_{\star}/R_{\rm pl}$), and eccentricity and argument of periastron $e$ and $\omega$, parametrised to $S_k = \sqrt{e}\sin(\omega)$ and $C_k = \sqrt{e}\cos(\omega)$ \citep{eastman_exofast_2013}.

We imposed a series of informative priors: we firstly bound the stellar density with a Gaussian prior centred on the known stellar density value and with a width defined by its computed uncertainties. The period of planet b was bound by a uniform prior between 9 and 9.4 days, and its time of inferior conjunction was allowed to vary only between BJD 2,459,010 and 2,459,019, based on visual inspection of the time series and on the results of the periodogram analysis. Similarly, the period and time of inferior conjunction of planet c were bound by uniform priors between 80 and 110 days and BJD 2,459,700 and 2,459,790.
We restricted the impact parameter of both planets to be greater than 0 to allow for a visible transit, and their $S_k$ and $C_k$ values to be between -1 and +1, as per definition. The limb-darkening parameters were both only allowed to vary between 0 and 1, following \cite{kipping_efficient_2013}. All other parameters were allowed to explore the entire space.

\subsection{Photometry Results}
\label{Sec:transit_res}
This analysis was undertaken using Version 10.11 of the {\sc PyOrbit} framework\footnote{Available at: \url{https://github.com/LucaMalavolta/PyORBIT/}} \citep{malavolta_gaps_2016,malavolta_ultra-short_2018}, which allows for parameter fitting using
dynamic nested sampling techniques \citep[{\sc dynesty}\footnote{Available at: \url{https://github.com/joshspeagle/dynesty}}: ][]{speagle_dynesty_2020,koposov_dynesty_2024}.
We evolved 5,000 live points until convergence ($\Delta \log \hat{Z}_i < 10^{-2}$). All model comparison was undergone via the logarithmic Bayes factor $\Delta \log B$ (calculated as the difference in logarithmic Bayesian evidence of the two considered models). A $\Delta \log B>5$ defines a  significant statistical preference.

In this first analysis, the eccentricity of the outer planet was confirmed to be significant, and TOI-2134~b was found to have a near-circular orbit. We thus chose to test whether a simpler forced-circular model would be preferred. For this run we kept all the same priors, fixed eccentricity and argument of periastron of the inner planet to 0 and 90$^\circ$ respectively, and evolved the same amount of live points for a nested sampling analysis. This simpler model was actually strongly disfavoured ($\Delta \log B$ = 36), and the more complex model was reconfirmed as the preferred solution to the photometry-only analysis.

The radius ratios between the planets and the star were derived to be $0.03580\pm0.00043$ for TOI-2134~b and $0.09374_{-0.00079}^{+0.00097}$ for \mbox{TOI-2134~c} (83$\sigma$ and 107$\sigma$ detections respectively). We thus computed their radii as \mbox{$2.770\pm0.075$ \rearth} for planet b, and \mbox{$7.26\pm0.19$ \rearth} for planet c. The stellar radii uncertainties were the lead contribution to the error associated with the derived planetary radii, capping their precision at $35 \sigma$, highlighting the need of very precise stellar characterisation. We not only precisely re-confirmed the orbital period for TOI-2134~b as $9.2291966_{-0.0000029}^{+0.0000036}$ days, but were also able to measure the orbital period of TOI-2134~c to be $95.852851\pm0.000039$ days from the two transits available. The times of inferior conjunction were also very precisely determined to be BJD $ 2,459,010.68941\pm0.00048$ and $2,459,718.96923\pm0.00031$ days for the inner and outer planets respectively. While the eccentricity of the outer planet was measured to be statistically significant, it could not be strongly constrained ($e_{\rm c}=0.41_{-0.10}^{+0.13}$).

\subsection{Transit Timing Variations}
We performed a second photometric analysis of the TESS data, in a model where the individual transit times of planet b are included as free parameters. We employed a custom light curve extraction following the process described in \citet{Stalport2025a}, and confirmed visually that the linear ephemerides provides a good estimate of the transit timings until the last Sector 80. Therefore, to ease the analysis, we defined new light curves that contain only the transits of planet b and 0.3 day (i.e. 7.2 hours) of timespan before and after the transit centres as baseline. As a consequence, the two transits of planet c were masked as well, and we used a one-planet model focusing on planet b. For each transit timing, we employed wide normal priors around the linear ephemerides, with a standard deviation of 0.1 day. We did not include the period $P$ and time of inferior conjunction $T_0$, since they are computed a posteriori from the individual transit times. We set wide uniform priors on the other planet parameters ($R_b/R_{\star}$, $b_b$, $e_b$, $\omega_b$). We set a Gaussian prior on the stellar density from our stellar characterisation analysis, and on the limb darkening coefficients $q_1$, $q_2$ provided by the posterior of the linear ephemerides fit (Table \ref{Tab:gp}). We use a single set of TESS instrumental parameters for the entire dataset ($q_1$, $q_2$, offset, and jitter). To account for leftover red noise, we also include in the model a GP regression with Matern kernel, as implemented in \texttt{celerite} \citep{Foreman-Mackey2017}. We place wide uninformative priors on the two GP hyperparameters: the GP amplitude and timescale have log-uniform priors in [10$^{-4}$,10$^3$]$\%$ relative flux and [0.1,10$^3$] days, respectively. 

\begin{figure}
    \centering
    \includegraphics[width=\linewidth]{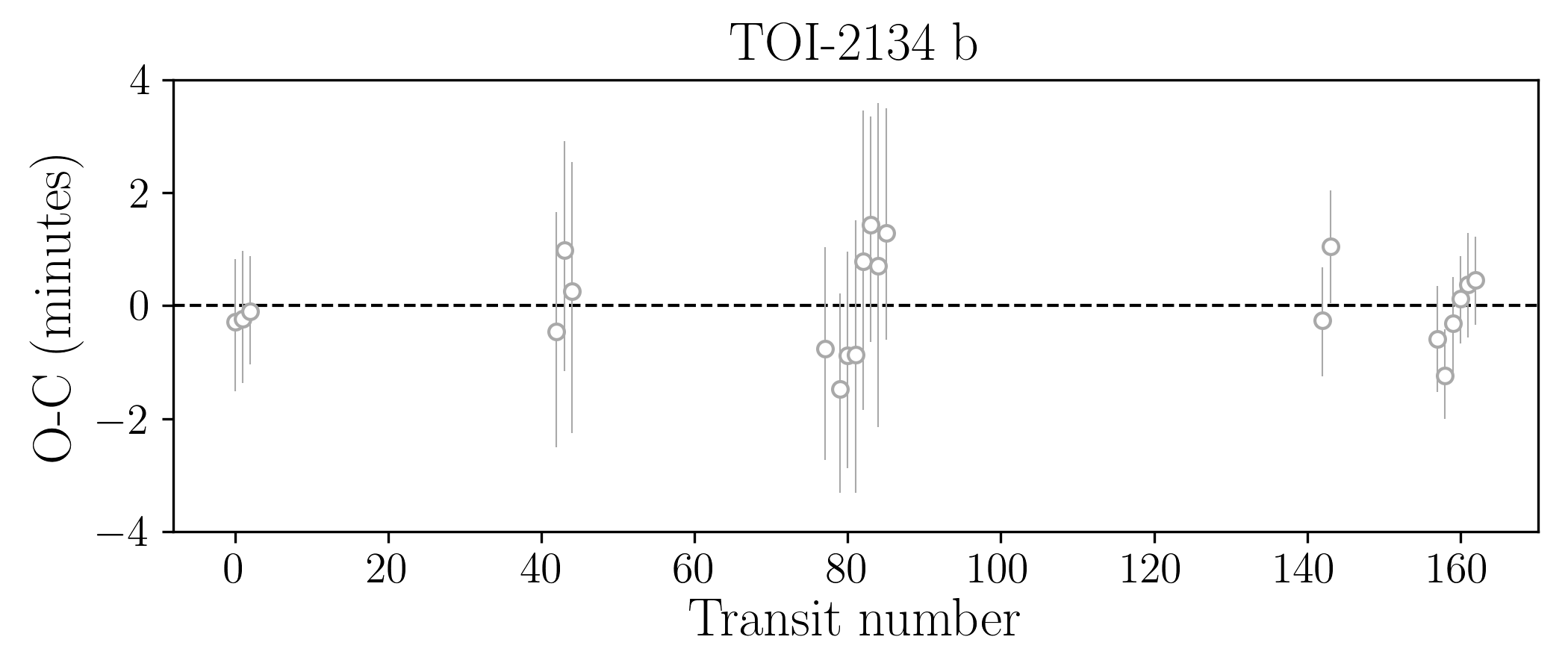}
    \caption{Observed - Calculated diagram of the transit times of planet b. The dataset includes 22 transits spread into 8 TESS sectors (26 to 80). Data are 120s cadence for sectors 26 up to 54, and 20s cadence for sectors 74, 79, and 80. This explains the lower error bars on the last transit times.}
    \label{fig:TTV}
\end{figure}

In total, 22 transits of planet b were fitted individually. We ran this photometric analysis with {\sc Juliet} \citep{Espinoza2019}. The parameter space was explored with the nested sampler {\sc dynesty} \citep{speagle_dynesty_2020}, with a number of live points given by $N^2$ (with $N$ the number of free parameters, 33 in total including 22 transit times). The result of our fit is presented in Fig. \ref{fig:TTV}. Unambiguously, no measurable TTVs were detected within the current dataset. We note a possible correlation in the O-C diagram. To establish its robustness, we repeated the analysis with an independent set of instrument-specific parameters per sector ($q_1$, $q_2$, offset, and jitter). In addition, we relaxed the priors on the stellar density and limb-darkening coefficients. To fasten the analysis, we split the dataset into 4 pairs of sectors and carry out independent fits in each pair, following the approach of \citet{Stalport2025a}. The resulting timings display a similar behaviour to Fig. \ref{fig:TTV}. The tentative correlation in the residual O-C diagram might therefore reflect a very low amplitude TTV signal, which remains currently out of reach. Forthcoming TESS sectors 118 and 119 (June - July 2027) will be helpful to investigate this aspect. 

We compared this result with dynamical simulations using {\sc TTVFast} \citep{Deck2014}. The absence of measurable TTVs on planet b is in agreement with an orbital eccentricity of planet c around 0.4, as estimated in Sect. \ref{Sec:transit_res}. However, the high-eccentricity model favoured in \citetalias{rescigno_hot_2023} clashes with this photometric analysis: for $e_c$ as high as 0.67, the dynamical simulations show significant TTVs with amplitudes up to 1 hour. This photometric analysis of the TESS data, as such, prioritizes the lower eccentricity model presented in \citetalias{rescigno_hot_2023}.

\section{Radial Velocity Analysis}
\label{Sec:rv}

We also undertook a series of analyses to better define the system characteristics using the radial-velocity data available. For these analyses we did not consider the EXPRES and PARAS-2 data: given their very short baseline and larger uncertainties, they would not significantly improve the analysis and they would only add more degrees of freedom. We describe the signals in the RVs with a series of models: Keplerians, polynomial trends, and two types of Gaussian processes. We also re-derived the radial velocities with two new extraction and stellar activity correction methods. All the analyses were compared against each other when possible.

\subsection{Defining the Number of Planets in the System}
\label{Sec:rv_number}
While the photometric analysis confirmed the presence of two orbiting bodies in the system, it cannot ascertain whether these signals can in fact be detected in the radial velocities, or whether other non-transiting planets could be present. Our first analysis thus attempted to describe all signals in the radial velocities with increasing numbers of Keplerians. Each Keplerian model depended on five parameters each: the orbital period $P$, the RV semi-amplitude $K$, the time of inferior conjunction $t_{\rm 0}$, and the eccentricity and argument of periastron (reparametrised as $S_k$ and $C_k$ as before). Both RV dataset are also described by a different constant offset, as well as a white-noise jitter term $\beta$. In all runs, the semi-amplitudes of any Keplerian were bound with a uniform prior spanning 0 and 50 \ms, the upper limit of which was defined to be comparable to the maximum scatter of the data. $S_k$ and $C_k$ were allowed to uniformly vary between -1 and 1, as per their definition.
All other parameters were allowed to explore the entire parameter space, unless explicitly stated. We used {\sc PyOrbit} with {\sc dynesty} nested sampling for all modelling in this Section.

We started with one Keplerian aimed at describing the signal with the leading power in both the RV time series at 95 days, tied to the two deep photometric transits. We imposed a uniform prior on $P$ spanning between 0 and 150 days. This fit yielded a $\log \hat{Z}_i$ of -892. This analysis recovered a Keplerian with a $95.74\pm0.19$ day orbital period, a semi-amplitude of $9.92\pm0.51$ \ms, and eccentricity $0.382\pm0.037$. The best-fit jitter parameters for both RV time series are $\sim$5\ms, most likely due to the presence of significant not modelled signals.

In the next step, we modelled the RVs with two Keplerians, as the transit analysis detected. For this run, we imposed informative Gaussian priors on both periods and times of transit, based on results of the photometric fit.
The $\log \hat{Z}_i$ of this two-Keplerians model was $-860$, yielding an overwhelming statistical preference against the one-Keplerian option ($\Delta \log B_0$ = 32) and confirming that the RVs were able to detect both signals present in the photometry.
The two recovered Keplerian signals had amplitudes of $3.03\pm0.38$ and $10.00\pm0.44$\ms\, for the inner and the outer planet respectively (8$\sigma$ and 23$\sigma$ detections). Masses were derived to be $8.1\pm1.0$ \mearth for TOI-2134~b and $55.1\pm2.7$ \mearth\, for TOI-2134~c. With this extended observing baseline of the higher-precision HARPS-N data, now covering twice the number of orbits of the outer planet as before, the phase space of TOI-2134~c was much more uniformly populated. This led to not only a drastically improved detection (from a mass significance of 5.4$\sigma$ in \citetalias{rescigno_hot_2023}, to one of 20$\sigma$ with this simplest model), but also to a unimodal distribution of the eccentricity of TOI-2134~c, centred at $0.347\pm0.022$. No model, in this or future Sections, prefers a different range of eccentricities. We were thus able to reject the high-eccentricity model proposed in the previous publication, which was slightly preferred by the original single transit in the photometry. These results highlight the importance of proper RV phase coverage for the accurate characterisation of planetary systems (more in Section \ref{sec:ecc}).

We then chose to assess the presence of any non-transiting bodies. We began by fitting the RVs with three Keplerians, once again using {\sc PyOrbit} with {\sc dynesty}. We allowed the period of this third Keplerian to explore the entire parameter space, and bound all other parameters as described for the previous fit. This model converged for a third Keplerian of period 5,842$_{-55}^{+15}$ days (or \mbox{16.0 $\pm$ 0.2 years}) and semi-amplitude of $8.65_{-0.4}^{+6.2}$ \ms. If confirmed a planet, it would be a $191_{-11}^{+124}$ \mearth\, (or \mbox{$0.60\pm0.4$ M$_{\rm J}$}) body orbiting at \mbox{$5.743\pm0.078$ AU}. This Keplerian modelled the long-term downward trend present in the data. Comparing this model to the two-Keplerian one led to a significant preference for the more complex three-Keplerian model \mbox{($\Delta \log B_0 = 40$)}.
We also tested a four-Keplerian model, which also converged to a solution. Besides the two known planets, the third Keplerian still attempted to explain the long-term trend,  with period $2248_{-81}^{+103}$ days and amplitude \mbox{$3.53_{-0.28}^{+0.22}$\ms}. The fourth Keplerian preferred a period of $19.25_{-0.59}^{+0.23}$ days, and semi-amplitude of \mbox{$0.332_{-0.086}^{+0.093}$\ms}, very high eccentricity of $0.79_{-0.23}^{+0.05}$. If a planet, its minimum mass would be $0.72_{-0.26}^{+0.19}$ \mearth. However, we could easily disfavour this model with respect to the three-Keplerian one \mbox{($\Delta \log B_0$ = 10)}. The four-Keplerian model is preferred to the two-Keplerian model, but the driving force of this preference was simply the better modelling of the long-term trend.

Overall, we found that the purely-Keplerian model that best described the signals present in the RVs was the three-planet model, where the final Keplerian mapped an unresolved long-term trend. However, the activity indicators also presented an unresolved long-term trend comparable to that in the radial velocities: we thus needed to better consider the effects of stellar variability on the RV datasets.

\subsection{An Unresolved Long-term Trend: Keplerian or Activity?}
\label{Sec:rv_poly}
When considering years-long trends, one must always be careful to differentiate between the signals generated by outer massive planets and the effects of stellar magnetic cycles. It can be tricky to prove or exclude the presence of long-term planets, especially in the cases where their period is comparable to the period of the magnetic cycle of their host star \citep{robertson_secretly_2013, pinamonti_gaps_2023}. In the case of TOI-2134, given the shared behaviour with the activity indicators, it was unlikely that this long signal was wholly of planetary origin. To further assess how this signal was most optimally modelled, we compared the three-Keplerians with a set of models comprising of two Keplerians and polynomial trends. On top of the two confirmed exoplanets, we fitted three polynomial models, allowing for a linear, a quadratic, and a cubic trend. The computed $\Delta \log \hat{Z}_i$ were, in order, $-809$, $-823$, and $-1534$. The models with linear and quadratic polynomials were strongly preferred to the simple two-Keplerian model ($\Delta \log B_0$ = 51 and 36). A linear trend was also significantly preferred to the three-Keplerian model ($\Delta \log B_0$ = 11), but the latter was marginally preferred against the quadratic trend ($\Delta \log B_0$ = 3). The cubic trend was strongly disfavoured. Between the polynomial models, the first order linear trend was statistically preferred ($\Delta \log B_0$ = 14 and 725).

We have then shown that when the long-term trend was described by simpler models, the two-Keplerian system structure returned to be preferred. This is partly expected for a signal not resolved in phase due to the lower number of degrees of freedom. The good match of the shape of the trend between RVs and activity indicators within the available baseline proved that at least part of this long-term behaviour must be stellar in nature, most likely the result of the star's magnetic cycle.
We thus investigated the best way to model the effects of stellar variability. Given the strong peak shown at the stellar rotation period in the HARPS-N RVs, we chose to describe the RVs as the sum of two Keplerians and a GP with a QP kernel to describe the stellar variability (as defined in Eq. 1 in Section \ref{Sec:st_act_gp}), implemented using the {\sc tinygp} package\footnote{Available at: \url{https://github.com/dfm/tinygp}} \citep{foreman-mackey_dfmtinygp_2024} in {\sc PyOrbit}.  We once again employed dynamic nested sampling as the parameter search algorithm. The Keplerian parameters were bound with the same priors described in Section \ref{Sec:rv_number}. We required the amplitude $A$ of the GP to be between 0 and 50\ms, and we bound the rotational period $P_{\rm rot}$ with a Gaussian prior centred on 48 days with a width of 2.5 days, from the activity indicator fit done in Section \ref{Sec:st_act_gp}.

This model yielded an $\Delta \log \hat{Z}_i = -765$, making it the preferred method to model the RVs ($\Delta \log B_0$ = 44 and 59 with the linear and the quadratic polynomial models respectively). The GP was simultaneously modelling the long-term trend and the RV variations modulated by the stellar rotation. This highlights the flexibility of GPs, but it is also a reminder of the dangers of this flexibility if not properly harnessed. $P_{\rm rot}$ was found to be \mbox{$46.73\pm0.52$ days}, comparable within 1$\sigma$ with the value extracted in \citetalias{rescigno_hot_2023}, and with the one extracted from the FWHM of the \mbox{HARPS-N} data. It was not, however, compatible with the period derived from the HARPS-N S-index within 1$\sigma$. In general, the stellar variability in the RVs was best modelled by a smaller rotation period than that extracted from fitting the activity indicators (although comparable within 2$\sigma$ based on the larger uncertainties of the activity fits). The two Keplerians were well-defined with periods of \mbox{$9.2291975\pm0.0000036$} and \mbox{$95.825851\pm0.000039$ days} (meaning that the posteriors were fully defined by the assigned priors from photometry), and semi-amplitudes of $3.37\pm0.24$ and $9.81\pm0.38$ \ms\, for the inner and the outer planet respectively. The eccentricity of TOI-2134~c was found to be $0.304\pm0.023$. Once again, the longer baseline of the observations allowed for more Gaussianly-shaped posterior distributions of $S_k$ and $C_k$, yielding a unimodal eccentricity distribution, and re-confirming the statistically-significant eccentricity of the outer planet.

\begin{figure*}
    \centering
    \includegraphics[width=16cm]{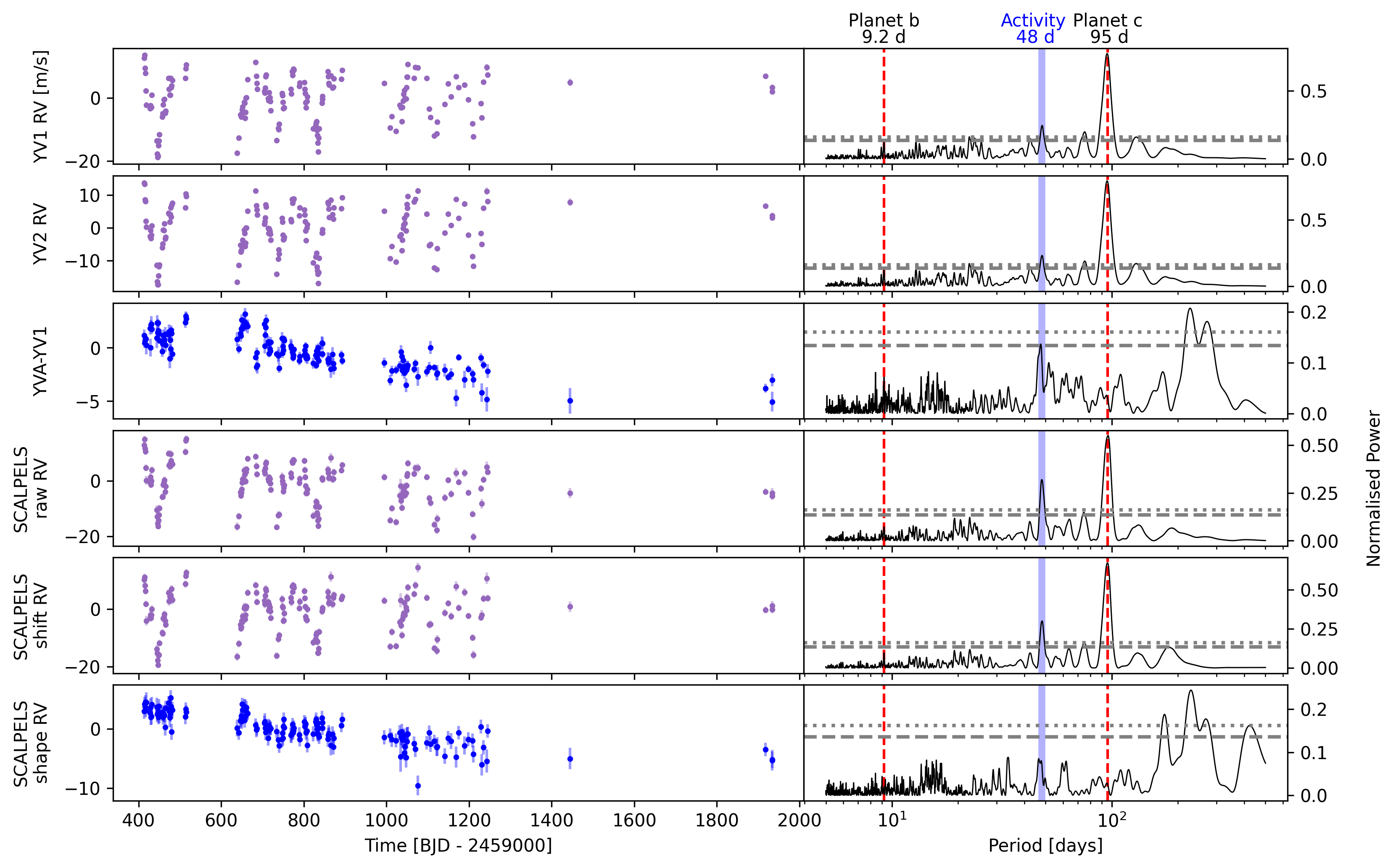}
    \caption{RVs derived from specially corrected HARPS-N spectra plotted against corrected Barycentric Julian Date. {\sc yarara}-derived RVs are plotted in the first three rows: YV1 is corrected mostly at the spectral level, while YV2 includes corrections in the time-domain. In the third row we plot in blue an approximation of the {\sc yarara}-computed stellar activity contribution, calculated as the difference between YVA (which includes planets and activity) and YV1 (which should include planets only).
    {\sc scalpels}-corrected RVs take over the final three rows, with in order the input raw RVs, the pure Doppler-shift RVs, and the shape RVs, the latter of which is plotted in blue to indicate that it represents the stellar activity contribution only.
    All RVs are expressed in \ms, with uncertainties plotted as errorbars (in cases too small to be visible). On the right of each time series, we plot their respective GLS periodograms, with the same conventions as for \mbox{Fig. \ref{fig:hn_rv}:} the horizontal dashed and dotted lines are the 1 and 0.1\% FAP levels, the vertical red dashed lines represent the orbital period of the known planets, and the blue solid band the expected stellar rotation period.}
    \label{fig:hn_rv_work}
\end{figure*}

\subsection{{\sc yarara} Analysis}
\label{Sec:rv-yarara}
To further assess the number of planets in the system, and further differentiate between Doppler-shifts caused by the presence of planets and distortions in the line profiles generated by stellar variability, we applied the {\sc yarara} correction to the spectra. {\sc yarara} \citep{cretignier_yarara_2021, cretignier_yarara_2023} is a post-processing method for high-resolution spectra that aims to address instrumental systematics, tellurics and stellar variability at the spectral level. The code requires as inputs continuum-normalised (done with {\sc rassine}, \citealt{cretignier_rassine_2020}) one-dimensional order-merged spectra. {\sc yarara} generates residual spectra time series starting from a master spectrum computed by co-adding all available data. Most flux variations are corrected at this level with a multilinear regression in either the stellar or the terrestrial frame, depending on the type of signal. In order to minimize the absorption of planetary signals, we shifted the spectra based on the preliminary Keplerian solution derived in Section \ref{Sec:rv_number}. The radial velocities of these corrected spectra were extracted using a line-by-line method \citep{dumusque_measuring_2018}. These RVs are called the {\sc yarara} V1 RVs (YV1). Since all spectral corrections are done separately, {\sc yarara} can also re-inject specific effects back in the spectra. It can thus produce another set of RVs, called {\sc yarara} Activity RVs (YVA), which include the effects of stellar variability.
A second layer of RV corrections are obtained by decorrelating the {\sc yarara} V1 RVs in a multilinear model using the shell time-series coefficients \citep{cretignier_stellar_2022} and Principal Component Analysis (PCA) \citep[for more information refer to ][]{cretignier_yarara_2023}. The RVs derived with this time-domain method are called {\sc yarara} V2 RVs (YV2).
{\sc yarara} requires very high signal-to-noise ratio spectra to be applied successfully, which made it only applicable to the HARPS-N high S/N spectra.
The HARPS-N YV1 and YV2 radial velocities are plotted at the top two rows of Fig. \ref{fig:hn_rv_work}, alongside their GLS periodograms. 
Both these time series still presented a peak in their periodograms at \mbox{$\sim$48 days}. This was primarily driven by the eccentricity of TOI-2134~c, rather then strong left-over stellar variability signals. However, this behaviour was once again symbolical of the difficulties of disentangling the effects of stellar activity and Keplerian signals when the rotation period of the star is close to the orbital period of the planet or to any of its harmonics. In an attempt to isolate the expected stellar activity component, the third row of Fig. \ref{fig:hn_rv_work} shows the subtraction between the YVA and the YV1 signals. As a reminder, YVA is expected to be fully corrected for instrumental effects, but to include stellar variability alongside planetary signals. On a simplistic level, the output of this subtraction should comprise of the activity component of the RV signal. The long-term trend is isolated to this "stellar-only" data, strengthening the hypothesis that it is a product of the stellar magnetic cycle.

As briefly described, YV1 and YV2 were both corrected for stellar activity, and thus they should be successfully modelled just by a simple sum of Keplerians. We thus repeated the same analysis undergone in Section \ref{Sec:rv_number}, modelling both {\sc yarara}-extracted HARPS-N RVs as the sum of one, two, three and four Keplerians. We bound the period and time of inferior conjunction of two known planets as before starting from information from photometry. All eccentricities and angles of periastron were reparametrised as $S_k$ and $C_k$ and only allowed to vary between -1 and 1, as per definition. All other parameters were bound by wide uniform priors. We perform parameter exploration using {\sc dynesty} implemented via {\sc PyOrbit}, as for all previous analyses.

Starting with the YV1 RVs, we could not find a significant preference between the two- and three-Keplerian models ($\Delta \log B_0$ = 0.5), but both were preferred to the one- ($\Delta \log B_0$ = 72 and 72.5) and the four-Keplerian models ($\Delta \log B_0$ = 3, and 3.5). All four models however agree within one sigma on the parameters of the two confirmed planets.
The semi-amplitude of TOI-2134~b was \mbox{$3.29\pm0.26$ \ms}. TOI-2134~c was modelled to have a semi-amplitude of \mbox{$10.43\pm0.32$ \ms} and eccentricity of $0.332\pm0.013$. The three-Keplerian model converged to a third signal with period $\sim$16 days and amplitude $\sim$1 \ms. The four-Keplerian model found the same third signal, but included a long period (>1000 days) signal with very small amplitude, hinting at possible residual power of the long-term trend or another long-period signal. For all models, the derived white noise amplitude was larger than expected in the case of perfect activity correction. The white noise term should include all variability not encoded in the model, which in stellar activity cases would include the red noise from convecting flows and photon noise. When modelling the data with two Keplerians $\beta$ was fit as $2.10\pm0.13$ \ms, and it was equal to $1.99\pm0.16$ \ms for the three-Keplerian model. Both values were significantly larger than the mean RV uncertainties, confirming that either {\sc yarara} was not able to successfully remove all non-planetary signals or a more complex analysis may be needed.

When fitting the YV2 RVs, the two-Keplerian model is once again the statistically preferred one, with $\Delta \log B_0$ = 66 against the one-Keplerian model, $\Delta \log B_0$ = 11 against the three-Keplerian model, and $\Delta \log B_0$ = 8 against the four-Keplerian model. It yielded semi-amplitudes for TOI-2134~b and c of \mbox{$3.23\pm0.24$ \ms} and \mbox{$10.51\pm0.29$ \ms} respectively. The eccentricity $e_{\rm c}$ was modelled to be \mbox{$0.330\pm0.012$}. The jitter term was still large, at \mbox{$1.93\pm0.12$ \ms}.

\subsection{{\sc tweaks} Analysis}
\label{Sec:rv_tweaks}
We also independently examined the HARPS-N RVs using {\sc tweaks} \citep{AnnaJohn2022, AnnaJohn2023}, a pipeline for planetary and stellar activity signals differentiation. 
In the wavelength domain, {\sc tweaks} uses {\sc scalpels} \citep{ACC2021} to separate RV variations arising from activity‑induced CCF line‑shape changes (shape signal) from true Doppler-shifts (shift signal) associated with planetary motion.
The time‑domain counterpart then applies the trans‑dimensional nested sampler {\sc kima} \citep{Faria2018} to the original RVs extracted from the CCFs, while simultaneously decorrelating them against the time‑domain coefficients of the leading principal components of the {\sc scalpels} shape signal.
This approach is necessary to guard against the risk of inadvertently suppressing genuine Doppler signatures, since the {\sc scalpels} shape vectors are not necessarily orthogonal to any planetary signals present in the data. By fitting original RV measurements using linear combinations of Keplerian models and the {\sc scalpels} shape vectors, {\sc tweaks} provides a rigorous framework for recovering planetary signals without attenuating them.

Our use of {\sc tweaks} is motivated by three key objectives: (1) to obtain independent mass measurements for the Keplerian signals, (2) to remove correlations with stellar activity–induced CCF line-shape variations, and (3) to marginalise over any additional low‑amplitude Keplerian signals using the nested sampler, thereby accounting for correlated noise contributions.

A set of basis vectors describing the dominant spectral‑line‑shape variations was obtained by performing PCA on the CCF auto‑correlation function with {\sc scalpels} \citep{ACC2021}. These vectors were subsequently implemented in {\sc kima} \citep{Faria2018} for stellar‑activity decorrelation, following the procedure outlined in \citet{AnnaJohn2022}. 

The separated {\sc scalpels} time series are shown in the bottom three rows of Fig. \ref{fig:hn_rv_work}, where the activity‑driven shape component (last row) clearly traces the long‑term stellar activity signal. To account for potential leakage of shift‑like activity signals through {\sc scalpels} separation, we modelled the remaining rotationally modulated RV variations with a simple one-dimensional Gaussian process regression. The GP was described by the same QP kernel form given in \mbox{Eq. \ref{eq:GP}.} 
For the hyperparameters, we imposed log‑uniform priors on the active‑region evolution timescale between 20 $<\lambda<$ 100\,days, and on the amplitude between \mbox{0\ms $<A<$ 4\ms.}
The stellar rotation period was assigned a uniform prior between \mbox{20 $<P_{\rm rot}<$ 60\,days}, based on the estimated rotational velocity (see Section \ref{Sec:st_act}) and previous analyses. To limit the harmonic complexity, we adopted a uniform prior of -1 $<$ ln($\sqrt{2}P_{\rm dec}$) $<$ 0. A white noise term $\beta$ was also included in the GP regression. It was bound with a uniform prior between \mbox{0.5 $<$ ln($\beta$) $<$ 50,} with the upper bound chosen to encompass the full range of RV observations. This term was incorporated following the same prescription as in Eq. \ref{eq:GP}. 

In short, {\sc tweaks} fitted the raw-RVs with a model containing a GP and up to five Keplerian components (number of which was treated as a free parameter), alongside simultaneous stellar activity decorrelation using the {\sc scalpels} basis vectors. 

A two‑Keplerian model was preferred by {\sc kima}’s analysis, and the stellar rotation period was constrained to \mbox{$P_{\rm rot} = 46.39_{-1.58}^{+1.84}$ days}. This rotation period agreed within 1$\sigma$ with the one derived in Section \ref{Sec:rv_number}, as well as with the period derived from activity indicators due to the larger uncertainties.
The joint posteriors showed clear detection of a Keplerian signal at orbital period $9.225 \pm 0.005$ days with a semi-amplitude of \mbox{$3.05\pm0.25$ \ms}. A second  Keplerian was detected with orbital period of $95.96 \pm 0.20$ days, and semi-amplitude of \mbox{$10.24\pm0.40$ \ms}. The outer planet's posteriors suggest it to be on an eccentric orbit with $e=0.33 \pm 0.04.$

\begin{figure}
    \centering
    \includegraphics[width=0.9\linewidth]{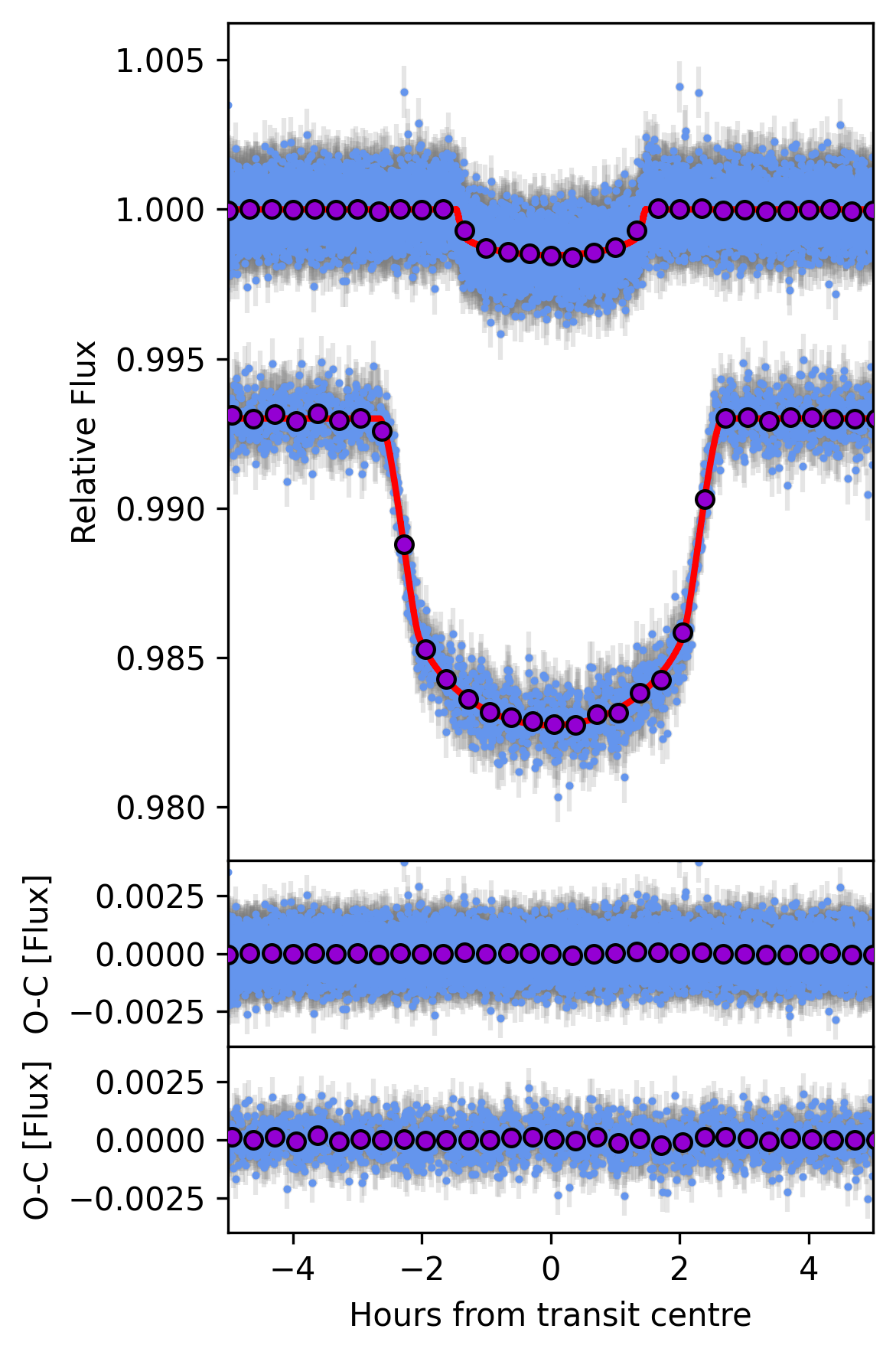}
    \caption{Transit fit results for planet b and planet c (respectively top and bottom of the first figure) plotted as relative flux against hours from the centre of transit. The TESS data is shown in light blue with uncertainties in grey, and the averaged values over a 15-minute window are plotted as larger dark purple dots. The best-fit transit models are shown in red. The transit of TOI-2134~c is artificially offset in relative flux by -0.007 for better readability. Underneath we also include the residuals between the TESS data and the predicted models for planet b (middle plot), and planet c (bottom plot).}
    \label{fig:tess_mod}
\end{figure}
\subsection{Multidimensional GP Analysis}
\label{Sec:rv_multigp}
Given the partial discrepancy between the $P_{\rm rot}$ measured by fitting the radial velocities and the one from the activity indicators, we also modelled the stellar variability with a multidimensional GP framework \citep{rajpaul_gaussian_2015, rajpaul_harps-n_2021}. We used the same quasi-periodic covariance form as described in the previous Sections. After some initial testing to define the most useful combination of activity indicators, we chose to simultaneously model the S-index, the bisector span of the CCF, and the RVs following the methods described in \cite{barragan_span_2022}:
\begin{equation}
\begin{aligned}
    \Delta {\rm RVs} &=  V_c G(t) + V_r \dot{G}(t) \\
    {\rm BIS} &= B_c G(t) + B_r \dot{G}(t) \\
    {\rm S-index} &= S_c G(t)
\end{aligned}
\label{eq:multiGP}
\end{equation}
in which $G(t)$ represents the Gaussian process, and $\dot{G}(t)$ its first derivative. $V$, $B$, and $S$ are amplitudes affecting the strength of the GP terms. They can be divided between convective ones (e.g., $V_c$), related to the GP itself, and rotational ones (e.g., $V_r$), related to its first derivative. We once again used nested sampling as a parameter searching algorithm, implemented in {\sc PyOrbit}. We bound the two Keplerians, and the hyperparameters of the covariance with the same priors as the previous Sections. $V_c$ was required to be positive, but all other amplitude hyperparameters were allowed to explore the entire parameter space. We fit both RV datasets simultaneously, including in the explored parameters an offset for each. The amplitudes of the GP components are also fit uniquely to each RV dataset, in order to account for the difference in spectral range, line selection and radial velocity extraction methods. 

With this new model, we extracted a rotation period of the star of $48.69_{-0.86}^{+0.97}$ days. This value did not always agree within 1$\sigma$ with the $P_{\rm rot}$ computed in previous analyses, but was in full agreement with the stellar rotation periods derived from the activity indicators only. It is possible that the previous stellar rotation estimates may have been biased by the lack of the first derivative term in the GP, which we found to be significant in this analysis.
Using this model, TOI-2134~b was computed to be in a \mbox{$9.2291975\pm0.0000036$ day} orbit, with a semi-amplitude of \mbox{$3.22\pm0.14$ \ms}. The period and semi-amplitude of the outer planet were derived to be \mbox{$95.825851\pm0.000039$ days} and \mbox{$10.40\pm0.21$ \ms} respectively, with an orbital eccentricity of $0.33\pm0.01$. This model could not be directly compared to all the previous models, as it used a different set of data, but its planetary results were overall in agreement, and the stellar variability was effectively mitigated. 
This more complex modelling technique led to a noticeable improvement in the planet detection relevance and in the measurement of the stellar rotation period.
Ultimately, we preferred the multidimensional GP model to the RV-only GP one, as it is the most physically-motivated solution and in the hope of best separating between the stellar activity signals modulated by rotation and the eccentricity of the outer planet.

\begin{figure}
    \centering
    \includegraphics[width=\linewidth]{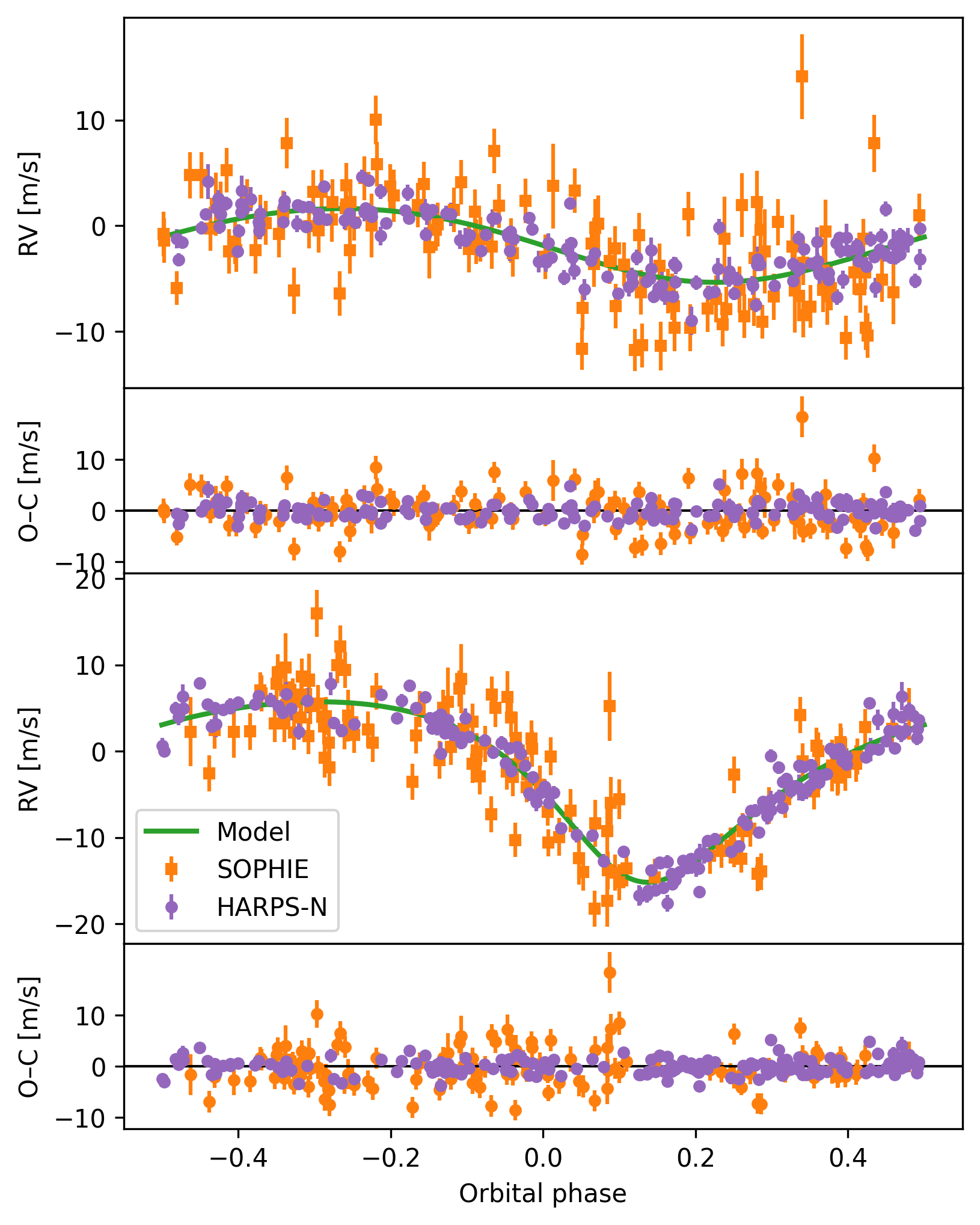}
    \caption{Phase-folded RVs (HARPS-N plotted as purple dots, and SOPHIE as orange squares) along the best-fit Keplerian models (shown as green lines). Uncertainties are plotted as errorbars, but may be too small to be visible. TOI-2134~b is on the top, and the eccentric planet c is on the bottom. The residuals of both phase-folded models are included underneath each plot.}
    \label{fig:kep}
\end{figure}

\section{Joint Photometric and Spectroscopic Analysis}
\label{Sec:joint}

Finally, we performed a last analysis jointly modelling the TESS photometry, both the DRS RV time series, and the activity indicators. We modelled the transit as described in Section \ref{Sec:transit}, and the RVs as described in Section \ref{Sec:rv_multigp}. We again used {\sc PyOrbit} and applied the same priors used in Sections \ref{Sec:transit} and \ref{Sec:rv_multigp}.

\begin{figure*}
    \centering
    \includegraphics{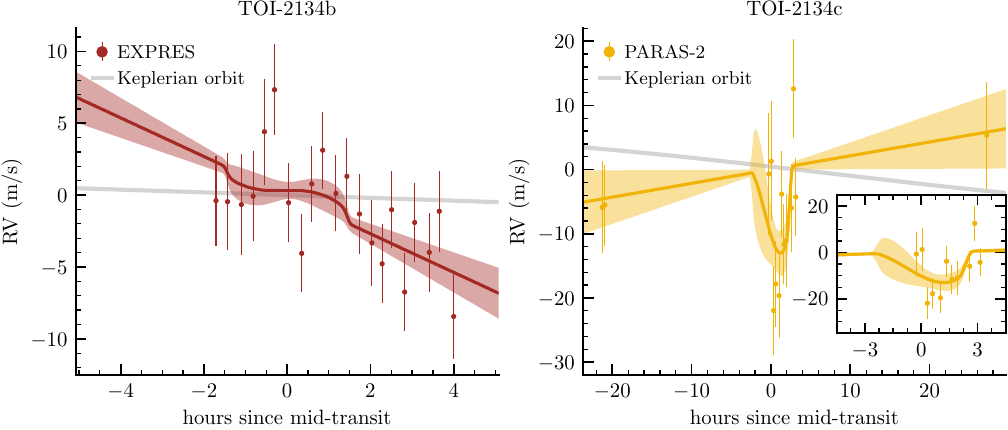}
    \caption{Rossiter-McLaughlin observations and fits to the data. The errorbars shown are the quadrature sum of the photon noise uncertainties and fitted white noise terms. The solid lines and shaded regions are the mean and $1\sigma$ standard deviations derived from the posterior of the model fits. The best-fit Keplerian models from the previous analyses are overplotted in gray. \textit{Left:} EXPRES observations of a transit of TOI-2134~b on local night 27 April 2023. The best model is plotted, but no significant detection could be made. \textit{Right:} PARAS-2 observations of a transit of TOI-2134~c on local night 7 April 2025. The inset axis is zoomed in on the transit night observations.}
    \label{fig:rm_fit}
\end{figure*}

We derived the radii ratios of the two planets to be \mbox{$0.03534\pm0.00020$} (186$\sigma$ detection) and \mbox{$0.09504\pm0.00061$} (153$\sigma$ detection)  respectively. The best fit transit models are shown in Fig. \ref{fig:tess_mod} as red lines, plotted against all the phase-folded data in blue, and the 15-minute averages in purple.
From the spectroscopy fit we derived a stellar rotational period of 48.78$_{-0.60}^{+0.65}$ days (in agreement with the one derived from the activity indicators). Planet b was found to have a semi-amplitude of 3.49$\pm0.18$ \ms (19$\sigma$ detection), while we computed the semi-amplitude of planet c to be \mbox{10.45$\pm0.24$ \ms} (44$\sigma$ detection). The eccentricity of the outer planet was found to be $0.31\pm0.01$, while the mini-Neptune orbits the star in a near-circular orbit with $e_{\rm b} = 0.067\pm0.022$. The time of inferior conjunction and periods for both TOI-2134~b and c are in perfect agreement with those derived from photometry only. All results are included in Table \ref{tab:planets}.
The best-fit models and the RV residuals are shown in the Appendix in Fig. \ref{fig:multigp}. The phase folded Keplerians and the spectroscopic data are plotted in Fig. \ref{fig:kep}. We highlight the added HARPS-N phase coverage with respect to \citetalias{rescigno_hot_2023}, for which we had no constraints on periastron passage, or the peak of the phase curve.

\section{Rossiter-McLaughlin Effect}
\label{Sec:rm}
We modelled the radial velocities from EXPRES and PARAS-2 during the transits of planet b and c, respectively, using the {\sc ARoME} software \citep{boue2013} with the {\sc python} wrapper {\sc PyARoME}\footnote{Available at: \url{https://github.com/andres-jordan/PyARoME}}. The software produces analytical RM effect models for CCFs-derived RVs. We adopted the orbital solution from the joint analysis in Table~\ref{tab:planets} using Gaussian priors on the transit times $t_0$, orbital periods $P$, orbital inclinations $i$, scaled semi-major axes $a/R_\star$, radius ratios $r/R_\star$, eccentricities $e$ and periastron angles $\omega$. The projected rotation $v\sin({i_\star})$ was shared between the two datasets and was left as a free parameter with a uniform prior in $[0, 10]\,\mathrm{km}\,\mathrm{s}^{-1}$. The projected obliquity of the star relative to the orbital planes of the two planets, $\lambda_b$ and $\lambda_c$ respectively, were further bound with uniform priors in $[-180^\circ, 180^\circ]$. Each dataset was also fitted with a slope $m$ and an additional white-noise term $\beta$, while the systemic velocity was calculated as the weighted average of out-of-transit data. We explored the parameter space around the maximum likelihood solution using a MCMC, specifically {\sc emcee}\footnote{Available at: \url{https://emcee.readthedocs.io/en/stable/}} \citep{foreman-mackey_emcee_2013}, to approximate the posterior distribution of our parameters. Before starting the analysis, we removed two PARAS-2 outliers from the transit night whose RVs differed by ${>}20\,\mathrm{m}\,\mathrm{s}^{-1}$ from the mean. We used 300 walkers over 100,000 steps, discarding the first 3,000 steps due to burn-in. We also thinned the chains by a factor of 225 due to autocorrelation. Convergence was confirmed by calculating the integrated autocorrelation time for each parameter for a number of steps up to 100,000. The best-fit median models with $1\sigma$ uncertainty regions are visualised in Fig. \ref{fig:rm_fit}.

For the transit of TOI-2134~b observed with EXPRES we did not detect any Rossiter-McLaughlin effect. The peak-to-peak amplitude of the ratio of the mean RM model to its uncertainty was only 1$\sigma$, and therefore compatible with no detection within the observed uncertainties. The solutions of $\lambda_b$ compatible with the data spanned the full prior range, with a preference for polar and retrograde solutions. The latter effect in particular can be produced by fitting RM models to random noise \citep{albrecht2011}. For reasonable values of $v\sin({i_\star})$, a priori we might have expected an amplitude of $1{-}2\,\mathrm{m}\,\mathrm{s}^{-1}$. Since the formal radial velocity uncertainties were in the same range, a detection might have been possible, however our analysis showed that given our model assumptions the uncertainties are underestimated by $2.5 \pm 0.6\,\mathrm{m}\,\mathrm{s}^{-1}$ due to scatter from unknown origin. This extra noise was most likely caused by variable observing conditions as the S/N changes by a factor 2 throughout the observing sequence.

For the transit of TOI-2134~c observed with PARAS-2 we found a potential detection, with $v\sin(i_\star) = 2.7 \pm 0.8\,\mathrm{km}\,\mathrm{s}^{-1}$, $\lambda_c = 59^\circ \pm 31^\circ$, at a detection significance of 4.7$\sigma$ (peak-to-peak amplitude of the mean model relative to the model posterior uncertainty). The white noise term for the quadratic model fit was found to be equal to $5 \pm 3\,\mathrm{m}\,\mathrm{s}^{-1}$.

The Rossiter-McLaughlin signal was superimposed on the signal of the Keplerian orbit of the planet, but in both cases the fitted slopes did not match the Keplerian model, pointing to a different origin. Active regions on the surface of the star can change the RV slope and even introduce curvature within the observing span of a transit \citep[e.g.][]{montet2020,palle2020,kunovac2025}. However, it was unlikely for the activity level for this star to be large enough to introduce these signals. Instead, the likely origin are the sub-standard observing conditions. While the FWHM of the PARAS-2 CCFs were relatively stable, the contrasts varied by more than a factor 2 between some exposures, ranging from ${\sim}15$ to $30\%$ relative line depths. Regardless of the cause of these line profile changes (e.g., clouds, seeing, variable fibre illumination, etc.) the S/N variation could introduce additional noise in the data that was not accounted for by the photon noise assumption of the individual exposures. 

The $v\sin({i_\star})$ measurement of $2.7 \pm 0.8\,\mathrm{km}\,\mathrm{s}^{-1}$ from the fit to the PARAS-2 data was somewhat at odds ($2.5\sigma$) with the estimated rotation period from the analysis of the other RV data. With $P_\mathrm{rot} = 48.3 \pm 1.5$ days, $R_\star = 0.709 \pm 0.017$\,R$_{\odot}$ and assuming a stellar inclination of $90^\circ$ the equatorial velocity would be $0.74 \pm 0.03\,\mathrm{km}\,\mathrm{s}^{-1}$. When we included this prior in the analysis, the best-fit RM signal amplitude was consistent with no detection. Previously reported upper limits on $v\sin({i_\star})$ derived from spectroscopic broadening are $2\,\mathrm{km}\,\mathrm{s}^{-1}$ which would be in $1.5\sigma$ agreement with the $v\sin({i_\star})$ the Rossiter-McLaughlin result. 

We conclude that we found a potential detection of a Rossiter-McLaughlin effect from the (partial) transit of TOI-2134~c at a modest significance of 4.7$\sigma$. If real, the data would be most compatible with $v\sin({i_\star}) = 2.8 \pm 0.8\,\mathrm{km}\,\mathrm{s}^{-1}$ and a stellar obliquity with respect to its orbit of $\lambda_c = 59^\circ \pm 31^\circ$. Given the model and the data, the probability that the orbit was prograde ($\lambda_c < 90^\circ$) was 84\%.

\begin{figure}
    \centering
    \includegraphics[width=0.9\linewidth]{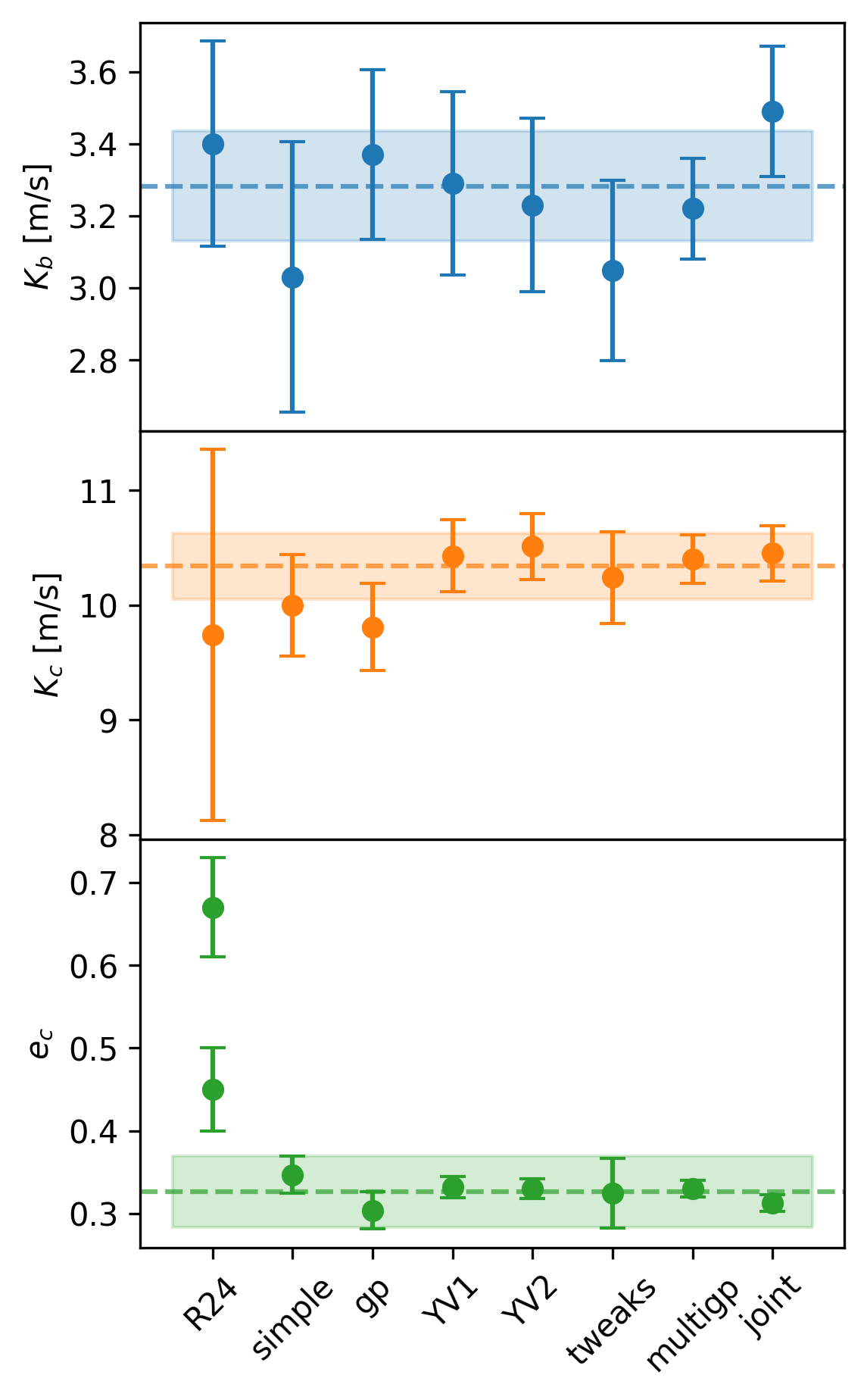}
    \caption{Comparison between the results of the previous publication (\citetalias{rescigno_hot_2023}) and the main model types tested in Sections \ref{Sec:rv} and \ref{Sec:joint}. We consider seven RV models in this comparison: a simple two-Keplerian model with no accounting for stellar variability beyond white noise, a two-Keplerian + GP model, the preferred two-Keplerian model on the activity-corrected YV1 and YV2 RVs, the {\sc tweaks} combined model, the two-Keplerian model with simultaneous fitting of activity indicators, and a joint TESS + multidimensional GP RV fit. From the top, we plot the semi-amplitude of planet b, the semi-amplitude of planet c, and its eccentricity. Each posterior is assumed to be Gaussian (confirmed by visual inspection) and its mean is plotted with its uncertainties as errorbars. The uncertainty-weighted mean value of each set is indicated by a dashed line, and the standard deviation by a shaded region. In the bottom plot, the eccentricity results of both the high-eccentricity and the medium eccentricity models of \citetalias{rescigno_hot_2023} are shown.}
    \label{fig:rv_comp}
\end{figure}
\begin{figure*}
    \centering
    \includegraphics[width=15cm]{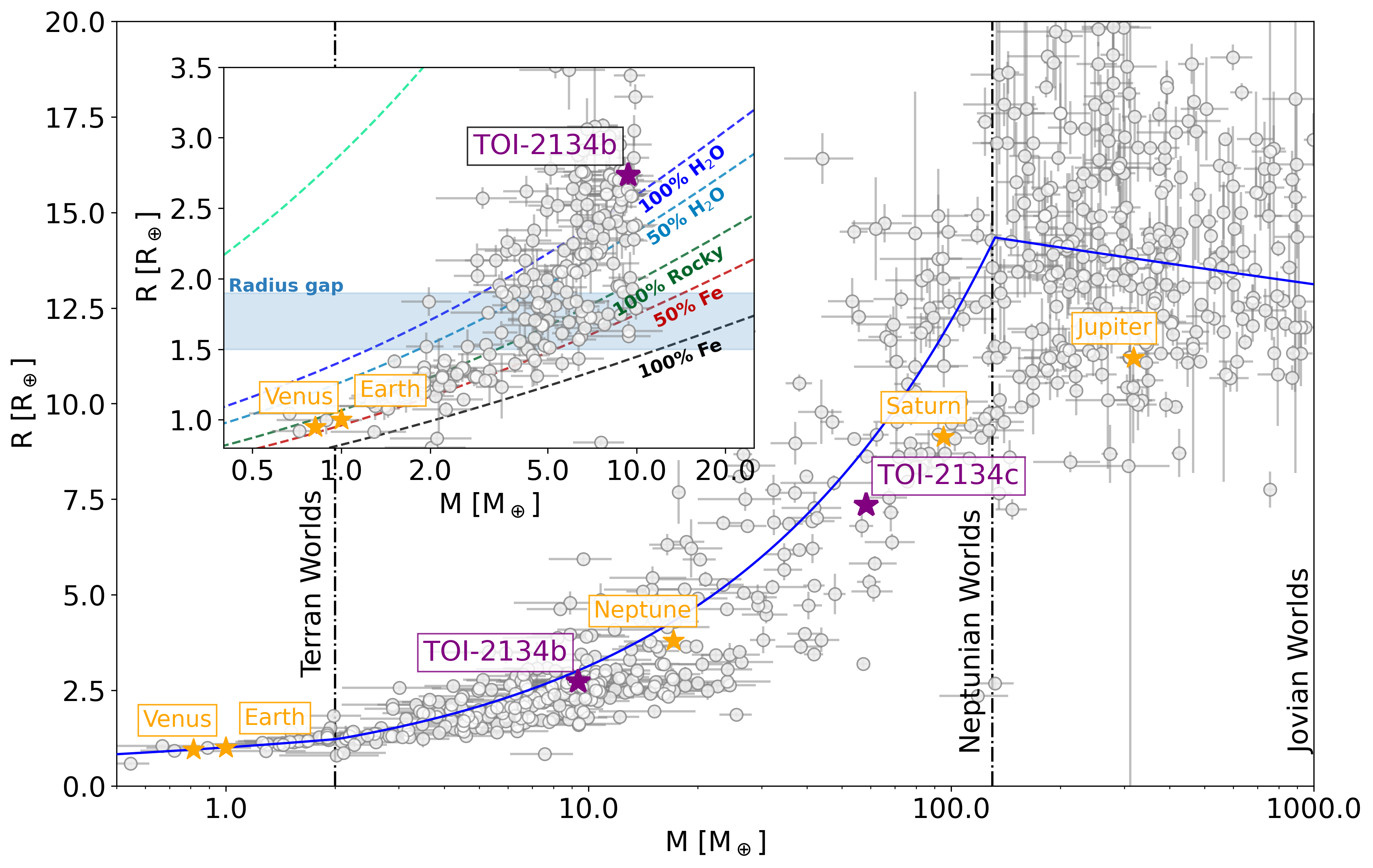}
    \caption{Mass-radius diagram with zoomed-in view for sub-Neptunian planets. The two TOI-2134 planets are plotted as purple stars, with errorbars to indicate their uncertainties (too small to be visible), and labelled by name in purple. Other confirmed exoplanets with mass detection better or equal to 5$\sigma$ are shown in gray. The data is taken from the EU exoplanet archives (\url{https://exoplanet.eu/}) on Jun 3 2025. The solid blue line in the outer plot shows the mass-radius relation developed by \protect\cite{chen_probabilistic_2017}. Vertical black dash-dotted lines separate the three exoplanetary regimes of Terran ($M_{\rm pl} < 2$ \mearth), Neptunian (2 \mearth $ < M_{\rm pl} < 0.4$ M$_{\rm J}$) and Jovian worlds ($M_{\rm pl} > 0.4$ M$_{\rm J}$). The zoomed-in plot includes composition lines taken from \protect\cite{zeng_massradius_2016}, colour-coded as labelled. The radius valley is also shown as a blue band. Solar system planets are plotted for scale as orange stars and labelled.}
    \label{fig:mr}
\end{figure*}
\begin{figure}
    \centering
    \includegraphics[width=8cm]{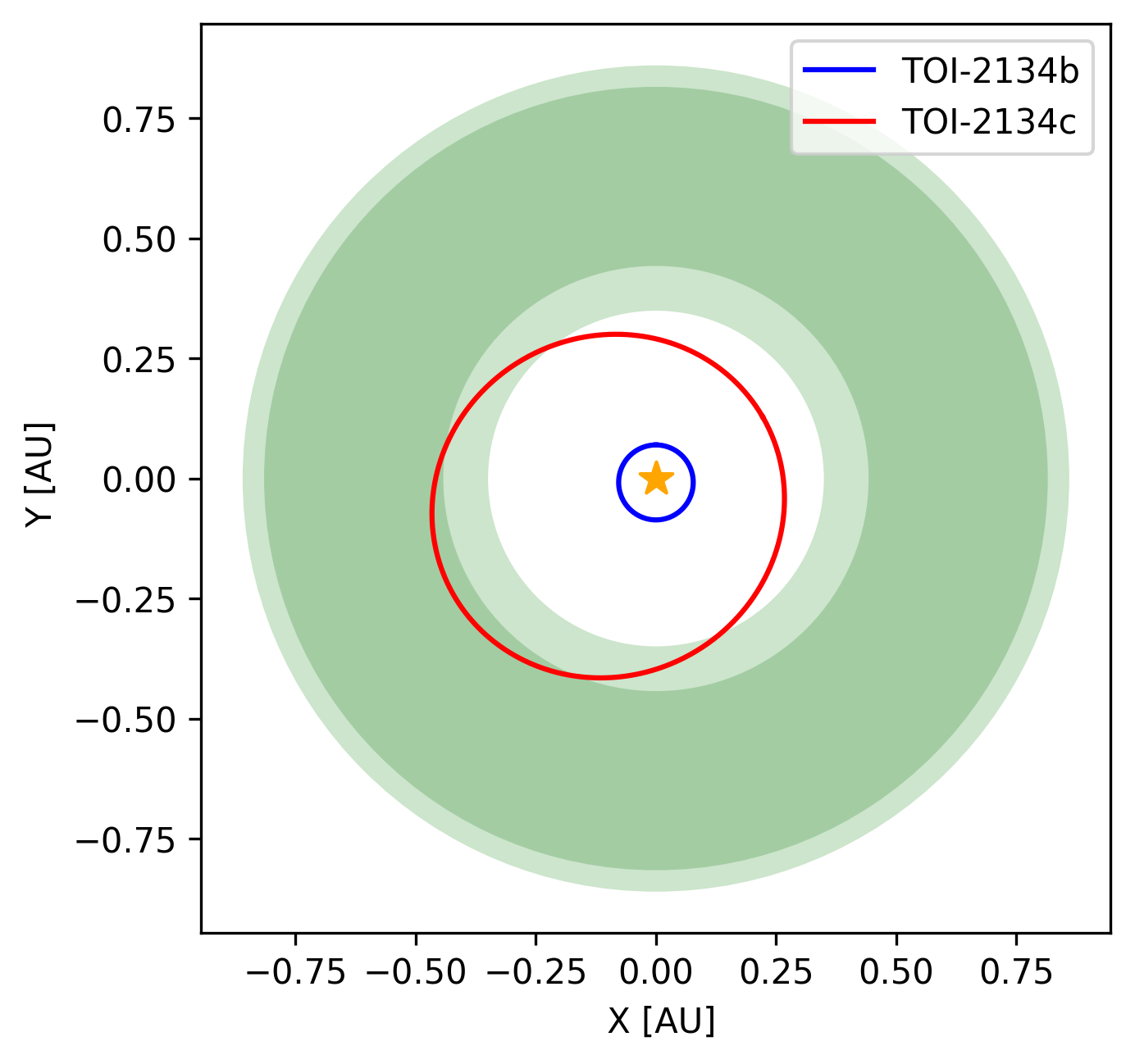}
    \caption{The TOI-2134 system structure. The star is placed at the centre as an orange star. The orbit of planet b is shown in blue, and the one of planet c in red. The empirical habitable zone is shown as the light green shaded region, while the narrow habitable zone is overplotted in darker green.}
    \label{fig:HZ}
\end{figure}

\input{Tables/results}

\section{Results and Discussion}
\label{Sec:disc}
We have undertaken extensive model testing to find the best way to model the available RVs and best define the system characteristics. We found that a two-Keplerian model with a GP described by a quasi-periodic kernel to model the stellar variability yielded the best $\Delta \log B$ when focusing on the RVs only. However, the slight disagreement between the extracted stellar rotation period by this model and the one derived using the activity indicators prompted us to undergo a multidimensional GP analysis, which modelled the S-index and the bisector span at the same time as the RVs. The results from all the different analyses were all in general agreement, as shown in \mbox{Fig. \ref{fig:rv_comp}}. All of the methods of this work agree within uncertainties on all parameters. In particular, the eccentricities of the outer planet all agreed within 1$\sigma$, as the last panel of the figure highlights.
Ultimately, we selected the results from the joint fit with {\sc batman} transit models for the TESS data, and a multidimensional GP approach with a QP kernel for the RVs as the final results. They are summarised in Table \ref{tab:planets}, including all other derived parameters. The best fit hyperparameters are included with all the remaining fitted parameters in Table \ref{tab:gp} in the Appendix.

The selection of the most complex model was driven by the need to properly account for the rotationally-modulated stellar signals, which have periodicity almost perfectly equal to half the orbital period of the outer planet, making deriving the eccentricity and the semi-amplitude of TOI-2134~c particularly tricky. The one-dimensional GP fit yielded the smallest RV residuals RMS (0.77 \ms, compared to the one derived from the multidimensional GP analysis of 1.51 \ms), but the quality of a fit should not be judged only based on RMS reduction in order to avoid overfitting. Simultaneously fitting the activity indicators and RVs is a common and physically-motivated method to minimise this possibility. In all cases, the derived $\beta$ value for the HARPS-N data were slightly larger than usually expected, with it reaching $0.9972^{+0.0021}_{-0.0047}$ \ms for the joint fit (more than 50$\sigma$ away from the mean data uncertainties), underlying the imperfect activity modelling. Part of this leftover "white" noise can be attributed to convective flows (granulation and supergranulation) and other non-rotationally modulated effects.

Nonetheless, we were able to confirm the presence of two planets in the system: TOI-2134~b and c. The longer baseline for both the HARPS-N and SOPHIE datasets allowed not only for a more precise determination of the orbital parameters of the outer planet, but also was able to break the multimodality of its eccentricity solutions. Both exoplanets are shown in the context of all confirmed exoplanets in  Fig. \ref{fig:mr}. We include a zoom in for the lower mass regime, showing the composition lines developed by \cite{zeng_massradius_2016}, which places TOI-2134~b above the 100$\%$ water line. TOI-2134~c resides in a less populated area of the mass-radius diagram, roughly following the mass-radius relation extracted by \cite{chen_probabilistic_2017}. 

The system is also presented on its plane of orbit in Fig. \ref{fig:HZ}. The star is placed in the middle, and the best-fit orbits for the two planets are depicted in blue and red. The plot also shows the empirical habitable zone as a light green shaded region (an optimistic estimate based on observations of Mars and Venus), and the narrow habitable zone in darker green (a more pessimistic range defined by one-dimensional climate models) \citep{kopparapu_habitable_2014}. Due to its eccentricity, TOI-2134~c spends about half of its orbit within the empirical habitable zone of its system, which has implications of potentially habitable exomoons in the system.

As a summary, planet b was derived to be orbiting the star with a \mbox{$9.229198\pm0.000003$} day period, an inclination of $89.69\pm0.24$$^{\circ}$, and a semi-major axis of \mbox{$0.07803\pm0.00095$ AU}. We computed its radius to be \mbox{$2.735\pm0.068$ \rearth} and its mass to be \mbox{$9.37\pm0.54$ \mearth}. TOI-2134~b is thus confirmed to be a hot sub-Neptune, with equilibrium temperature of \mbox{$660\pm41$ K} for an incident flux of $32\pm8$ F$_{\rm inc,\odot}$ (assuming albedo = 0 and perfect circulation). The derived parameters are in 1$\sigma$ agreement with the results from \citetalias{rescigno_hot_2023}, but we more than doubled the significance of the radius detection to the point of being limited by the precision on the stellar radius, and also partially improved the mass determination.

TOI-2134~c was confirmed to be a sub-Saturn of mass \mbox{$58.3\pm1.9$ \mearth} and radius \mbox{$7.35\pm0.18$ \rearth}. With the new data, we were able to improve the mass precision by more than 4 times, and the radius precision by more than twice. The planet was found to orbit TOI-2134 in a \mbox{$95.852840\pm0.000041$ days} orbit at the semi-major axis of \mbox{$0.3715\pm0.0045$ AU}, with a similar inclination as planet b of \mbox{$89.688\pm0.014^{\circ}$}. Most crucially the eccentricity was found to be $0.31\pm0.01$ (an in-depth discussion regarding the new preferred eccentricity for planet c is included in Section \ref{sec:ecc}). As shown in \mbox{Fig. \ref{fig:HZ},} planet c spends a large fraction of its orbit in the habitable zone of the system, with the average $T_{\rm eq}$ of $303\pm19$ K for a mean incident flux of $1.4\pm0.4$ F$_{\rm inc,\oplus}$. At distance of closest approach it orbits at $0.255\pm0.004$ AU from the star, receiving a flux of $2.95\pm0.74$ F$_{\rm inc,\odot}$ and with $T_{\rm eq} = 364.7\pm23$ K. At the opposite edge of the orbit, at a distance of $0.487\pm0.006$ AU, it receives a flux of $0.81\pm0.20$ F$_{\rm inc,\odot}$ and and has $T_{\rm eq} = 264.0\pm17$ K. The eccentricity of this planet thus yields a flux variation of 2.14 F$_{\rm inc,\odot}$ over the orbit, or an equilibrium temperature change of $\sim$100 K. This excursion could promote atmospheric circularisation and mixing. At transit, we would expect a $T_{\rm eq} \approx 340$ K, assuming no albedo or thermal inertia. If either of those factors would come into play, as it is likely, this extreme temperature variation would be suppressed.

We note that while the majority of orbital parameters for TOI-2134~c agree within 1$\sigma$ with the results presented in \citetalias{rescigno_hot_2023} (radius included), the masses do not and instead only overlap within 2$\sigma$ due to the large uncertainty on the original measurement and the bimodality of its posteriors.

The assessment of the stellar obliquity was less straightforward. We analysed radial velocities taken during transit by EXPRES for TOI-2134~b and with PARAS-2 for TOI-2134~c. While no detection could be made for the inner planet, we computed a $59\pm31^{\circ}$ obliquity for the outer sub-Saturn.
From a disc-migration perspective we would a priori assume a prograde and aligned system if no mechanism was responsible for significantly tilting the disk itself \citep[e.g.,][]{Batygin2012}. However, we have found evidence of multi-transiting systems that are both coplanar and retrograde \citep[e.g.,][]{Hjorth2021}, and of systems with high mutual inclinations \citep[e.g.,][]{Dalal2019,Bourrier2021}, suggesting that this assumption may not always be true.  At this stage, since we could not detect the spin-orbit angle of planet b, we could not confirm whether the two planets are coplanar. 
With a very simple geometric model based on the orbital parameters of this system, we found that only 2.81$\%$ of all transiting models were also coplanar within $5^\circ$. However, this simple model does not include any information tied to planet formation and orbital stability. For this reason, an assumption of coplanarity is still reasonable for this system as the simplest explanation.
Our results indicated that TOI-2134~c was likely in a prograde orbit, though with a wide range of allowed spin-orbit angles. Further assessment of the likely formation and evolution pathways for the TOI-2134 system are addressed in Section \ref{sec:migration}.

\subsection{The Eccentricity of TOI-2134~c}
\label{sec:ecc}
\begin{figure}
    \centering
    \includegraphics[width=8cm]{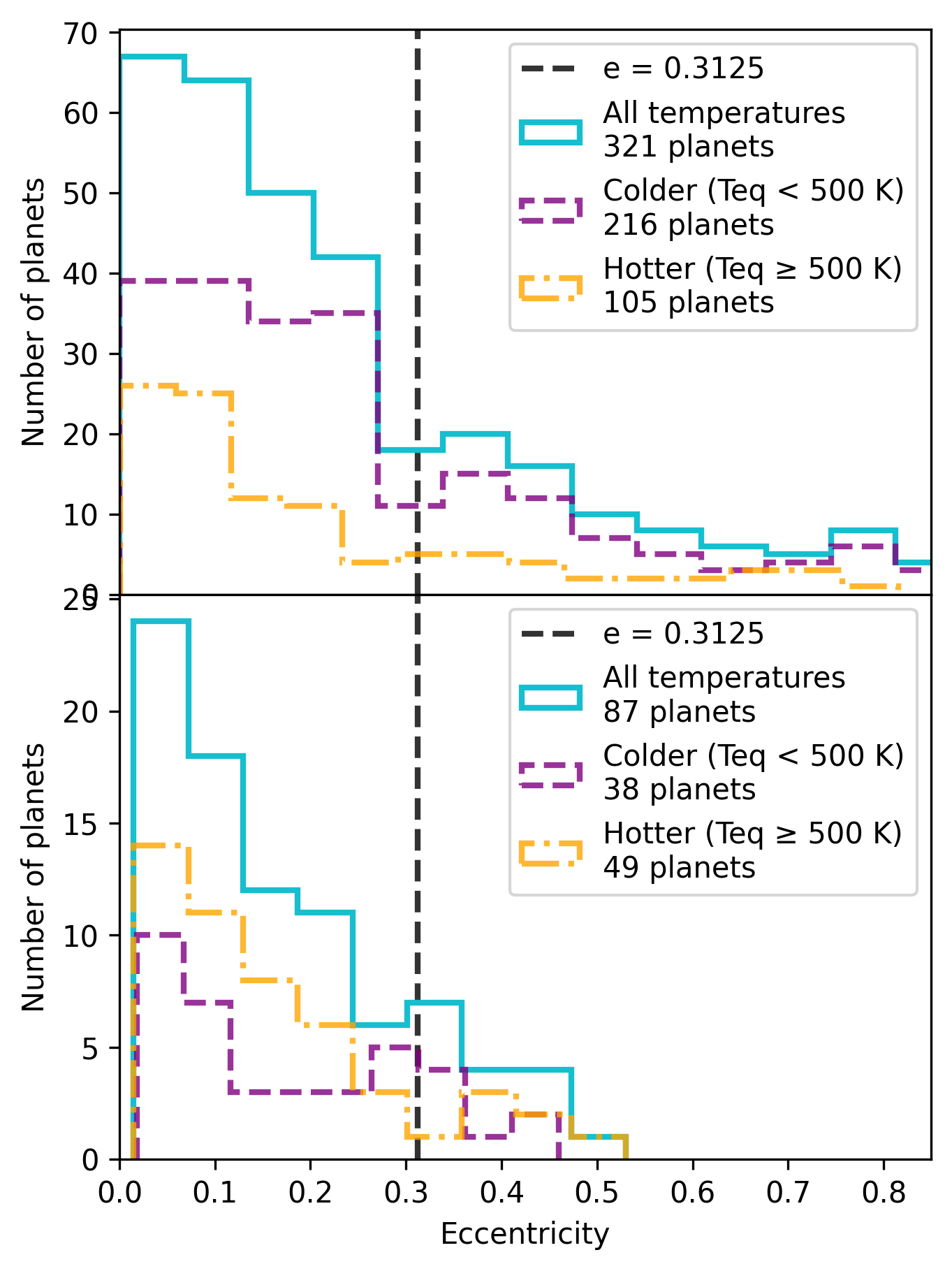}
    \caption{Eccentricity distribution histograms of all confirmed planets with 3$\sigma$ mass detection, known eccentricity, stellar luminosity and orbital semi-major axis, as stored by NASA's exoplanet archive. The derived eccentricity of TOI-2134~c is shown as a vertical dashed line. \textit{Top:} distribution including all temeperate ($P_{\rm pl} > 10$ days) gas giants ($M_{\rm pl}>10$ \mearth). The entire sample is shown by the cyan solid histogram. The sample is further split into warmer ($T_{\rm eq} > 500$ K) and colder planets, respectively in orange dash-dotted line and purple dashed line. \textit{Bottom:} Similar distributions for only sub-Saturn planets (10 \mearth$<M_{\rm pl}<100$ \mearth).}
    \label{fig:ecc_hist}
\end{figure}
The eccentricity of the outer planet was and has remained a source of questions for this system. Analyses done on a subset of the currently available RV data in \citetalias{rescigno_hot_2023} could not hone in on a single solution, as the posterior distribution for the eccentricity of planet c was inherently multimodal. 
The two signals at $P_{\rm c}$ and $P_{\rm rot} \sim P_{\rm c}/2$ generated a beating pattern hard to properly model, especially with the low temporal baseline in comparison with the orbit length.
This is a known behaviour, as two resonant near sinusoidal signals can often masquerade as one eccentric Keplerian \citep{Anglada2010} and vice-versa.
With only the RVs available in \citetalias{rescigno_hot_2023} no preferred model could be found, and all eccentricity options were dynamically viable for the system. \citetalias{rescigno_hot_2023} ultimately preferred the highest eccentricity solution ($e \sim 0.6$) after analysing the shape of \mbox{TOI-2134~c's} monotransit in TESS. 

The analyses described in Section \ref{Sec:rv} of this work were instead able to better constrain the orbital parameters of the system and yielded a well-behaved posterior distribution preferring the intermediate eccentricity of $0.31\pm0.01$ (compatible within $3\sigma$ with the middle eccentricity presented in \citetalias{rescigno_hot_2023}). This improvement was possible due to two main factors: firstly the second transit of planet c with higher cadence allowed the photometric analysis to better define the shape and period of the transit. This was able to better constrain the likely eccentricity space allowed by the duration of the transit, and lost its preference for the high-eccentricities. The reasons for this change can be numerous, but are beyond the scope of this work. Secondly, after the first publication we undertook a dedicated campaign with HARPS-N to better fill the phase coverage of the outer planet. In \citetalias{rescigno_hot_2023}, the centre of the Keplerian around phase 0 was dominated by the higher-scatter SOPHIE data, which drove the possibility of the high-eccentricity case. With one more observing season, we were able to more uniformly populate the orbital phase of TOI-2134~c with HARPS-N data (see the purple dots in the lower half of Fig. \ref{fig:kep}). This better coverage allowed us to more strongly bind the shape of the Keplerian around orbital phase 0.1, and led to the statistically significant preference for a lower eccentricity. These results also agree with the conclusions of previous studies, finding a tendency to overestimate the eccentricity value when fitting Keplerian models to RV time series with missing or low-precision data near periastron \citep{Shen2008,Zakamska2011}.
All in all, this work highlighted the importance of even phase coverage for the accurate and precise determination of planetary eccentricity \citep[see also][]{Burt2018}, especially when coupled with flexible modelling tools such as Gaussian processes. This is doubly relevant for longer period planets, for which we cannot assume circular orbits.

This analysis proved that a more moderate eccentricity is preferred for the outer planet. Nevertheless, TOI-2134~c remains a planet with significant eccentricity. This fact bears a series of implications for the completeness and the formation of the system that we will further address in Sections \ref{sec:detection_limits} and \ref{sec:migration}. The eccentricity of the sub-Saturn follows an identified trend in temperate and cold gas giants. To put planet c in context, the upper panel of Fig. \ref{fig:ecc_hist} shows the eccentricity distribution of all temperate (orbital period > 10 days) gas giants (minimum mass > 10 \mearth) in blue. Exoplanet data is taken from NASA's exoplanet archive\footnote{Available at: \url{https://exoplanetarchive.ipac.caltech.edu}, and accessed on 10/01/2026}. We removed all exoplanets with a controversial flag, and we required mass (or minimum mass) and eccentricity of the planet to be known, as well as stellar luminosity and planetary semi-major axis (needed for computing the equilibrium temperature of the planets). In total this selection yielded 321 planets. The eccentricity of TOI-2134~c is indicated by a vertical dashed line. The sample selected is further subdivided into hotter or colder planets, with the cut-off set at $T_{\rm eq} = 500$ K \citep{Alqasim2025}. We choose equilibrium temperature as the separating parameter as it simultaneously encodes orbital period and stellar luminosity.
The number of planets left in each sub-sample is indicated in the legend. Several works have noted the existence of a higher-eccentricity overdensity in temperate gas giants \citep[e.g.,][]{Gilbert2025}, often centred around $e=0.4$, and visible in Fig. \ref{fig:ecc_hist}, with long tails reaching very high eccentricities. This distribution is equally driven by both hotter and colder planets, showing similar sub-distribution when normalised. \cite{Alqasim2025} analysed the eccentricity distribution of transiting gas planets and found dependence on stellar metallicity (which increases the probability of higher eccentricities, see also \cite{Morgan2026}), planet radius, and planet multiplicity. By including planet radius in the analysis they also found two separate planet populations, one composed of larger, Jupiter-sized gas giants, and another of mini-giants, more similar to the size of our target TOI-2134~c. To further highlight this sub-population, the bottom panel of Fig. \ref{fig:ecc_hist} includes only mini-giants, defined as gas giant planets with mass $< 100$ \mearth. Via visual inspection, we still find a significant overdensity at higher eccentricity for this subset, although centred around 0.3 rather than 0.4. We however do not see the same well-populated tail for very high eccentricities. Also differently from the more general case, the eccentricity overdensity is here mainly driven by colder mini-giants. It is postulated that although high-eccentricity in mini-giants can be excited by the same mechanisms as for the Jupiter-sized cases, scattering efficiency decreases with mass \citep{Carrera2019}, removing the long tail and lowering the overdensity centre. 
Overall, TOI-2134~c further supported this overdensity and its eccentricity fits neatly in the population distribution as another cold mini-giant with the significant eccentricity of $\sim0.3$.

\subsection{Planet Detection Limits}
Section \ref{Sec:rv} concluded that the radial velocities for this system were best described by the sum of two Keplerians (the two confirmed planets b and c) and stellar activity with a significant magnetic cycle component. Nevertheless, additional planets could have evaded detection. To assess the completeness of this system, we computed detection limits with the publicly available code {\sc ardent}\footnote{Available at: \url{https://github.com/manustalport/ardent}} \citep{Stalport2025b}. {\sc ardent} derives two kinds of detection limits: the RV detection limits based on planet injection-recovery tests, and the dynamical detection limits that include the orbital stability constraint. 
This additional information impacts the detection limits by further rejecting areas of the period-mass space where injected planets are dynamically not viable. To evaluate orbital stability, {\sc ardent} combines analytic criteria with short N-body simulations coupled with a chaos indicator. This process fastens the stability estimation compared to classic N-body simulations.

\label{sec:detection_limits}
\begin{figure}
    \centering
    \includegraphics[width=8cm]{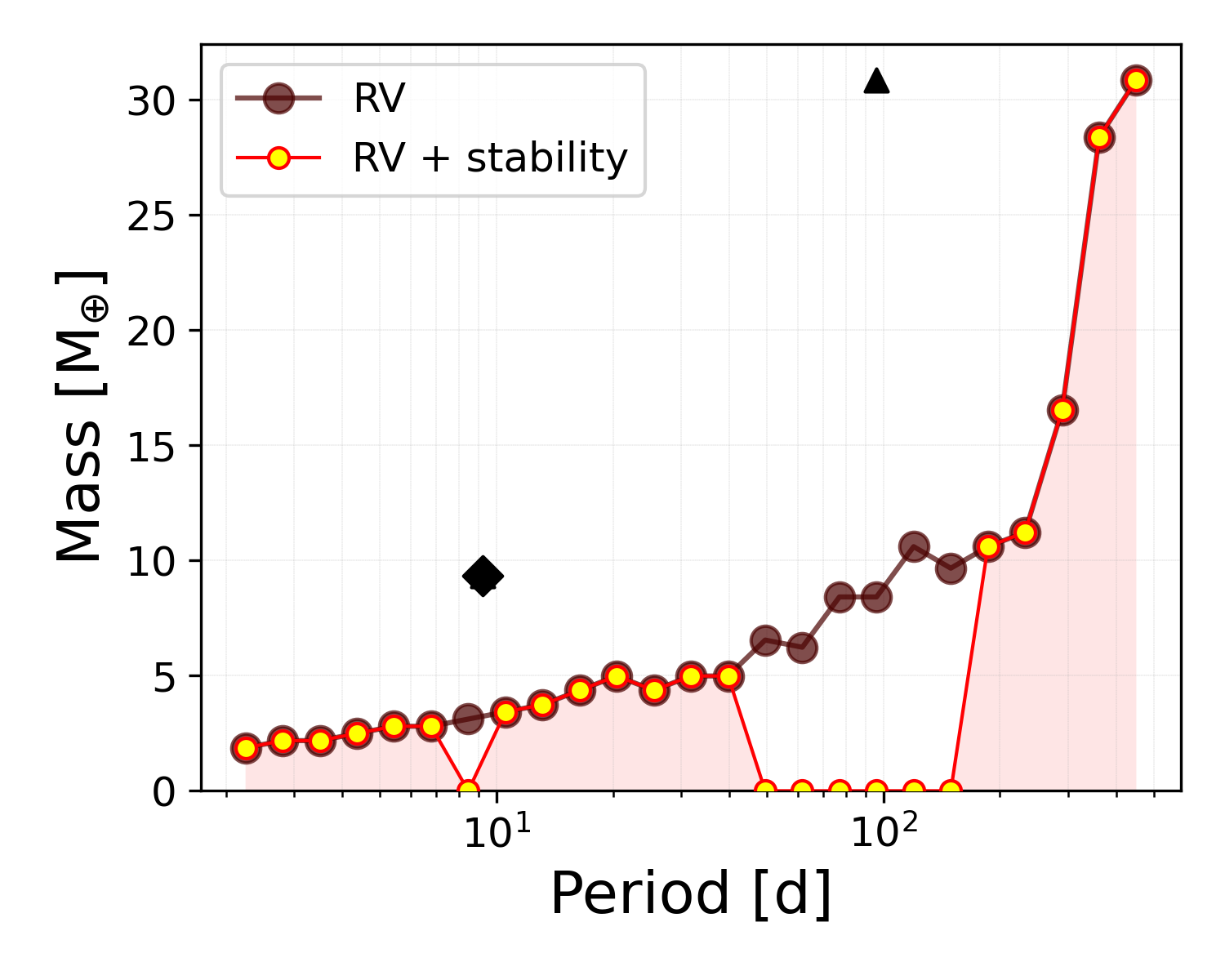}
    \caption{Detection limits for the TOI-2134 system. TOI-2134~b and c are plotted as black symbols (the triangle for TOI-2134~c indicates its mass lies above the plotted range). The detection limits based only on the information in the residual radial velocities are shown as brown dots. Yellow dots represent the detection limits for the system once information from both the residual RVs and the stability constraints are included. The red shaded regions highlight the parameter space for which a possible planetary body can elude detection given the current dataset.}
    \label{fig:detection_limit}
\end{figure}

We computed detection limits for planets orbiting with periods between 2 and 500 days. Using the residual RVs from the best-fit results of Section \ref{Sec:joint}, we sampled the $P$-$K$ space with $10^4$ points and, for each point, performed 10 injection-recovery tests after assigning different orbital phases to the injected planets (evenly spread in $[0,2\pi]$). From these tests, we derived the RV 95$\%$ detection limits, as shown in Fig.\ref{fig:detection_limit} (brown dots).
We computed the dynamical detection limits based on analytical stability criteria and numerical integrations over $10^4$ years. The result is illustrated in Fig. \ref{fig:detection_limit} with red-yellow dots. We also include Fig. \ref{fig:limits2} in the appendix as a different representation of the detection limits, mapped instead on the orbital plane of the system. We repeated the computation with integration times of $10^5$ years and obtained identical results, demonstrating that $10^4$ years is long enough to reveal short-term instabilities in the system. 

Fig. \ref{fig:detection_limit} shows that any planetary body orbiting inner to planet b with mass larger than 3 \mearth\, would have been detected in the radial velocities. Similarly for planets orbiting between b and c, we can exclude the presence of any body with mass larger than 5 \mearth. Any retained planet inner to the sub-Saturn would also have large probability to transit. However, while the photometry shows no hint of further transiting planets and is well fit with the preferred 2-planet model, the small size of these planets could still evade detection.

Most importantly, the stability analysis allows us to confirm that no stable orbit for any planet could be found between periods of $\sim$50 and 185 days due to the eccentricity of TOI-2134~c. Any massive gas giant outer companion would have been fully detectable out to periods of 400 days.

We however want to caution that this analysis relies on the assumption that the stellar variability of this target is very well described by the best-fit model. The residuals produced by the multidimensional GP analysis were well-centred around 0 \ms\, and present no statistically relevant deviation from a Gaussian distribution to indicate the possibility of a significant left-over trend (as shown in the leftmost panel in the second row of Fig. \ref{fig:multigp}). As mentioned in Section \ref{Sec:rv_poly}, in this work all the long-term trend was assigned to a magnetic cycle signal, which is shared between RVs and activity indicators. This choice is further supported by the shared phase of the short ascending section and longer descending parts of the long-term trend between RVs and S-index and BIS (Section \ref{Sec:rv_multigp}), and by the fact that both the wavelength-domain and the CCF-domain stellar mitigation methods also independently assigned most (or all) of this trend to their non-coherent activity components (Sections \ref{Sec:rv-yarara} and \ref{Sec:rv_tweaks}). These analyses all in all proved that the 5-year period possible planetary signal found in Section \ref{Sec:rv_number} cannot exist for the fitted semi-amplitude, but it must at least in part be stellar activity. However, we also could not ultimately assert that all this trend was only activity. GPs are excellent at absorbing unresolved signals, even when further bound by simultaneous fitting of activity indicators. In fact, the amplitudes of the latent GPs were simply fitted by the optimisation routine, and were not bound by strong physical priors. No literature could be found focusing on assessing the likelihood of long-period planetary signal absorption by long-term trends in a GP.
Moreover, a temporary agreement in phase between a long-term signal in the RVs and the activity indicators can still hide the presence of a planet, as it would do for our own Solar system with the similar amplitudes and periods of the solar magnetic cycle and the signal Jupiter imprints.
With the current dataset no further assessment of this possible degeneracy could be done. In order to fully describe the system and detect any possible outermost body, further observations are required. In particular, a few spectra every couple of months may help identify a possible phase separation, and a longer baseline could better refine the parameter space allowed to any cold giant companion.

\begin{figure}
    \centering
    \includegraphics[width=\linewidth]{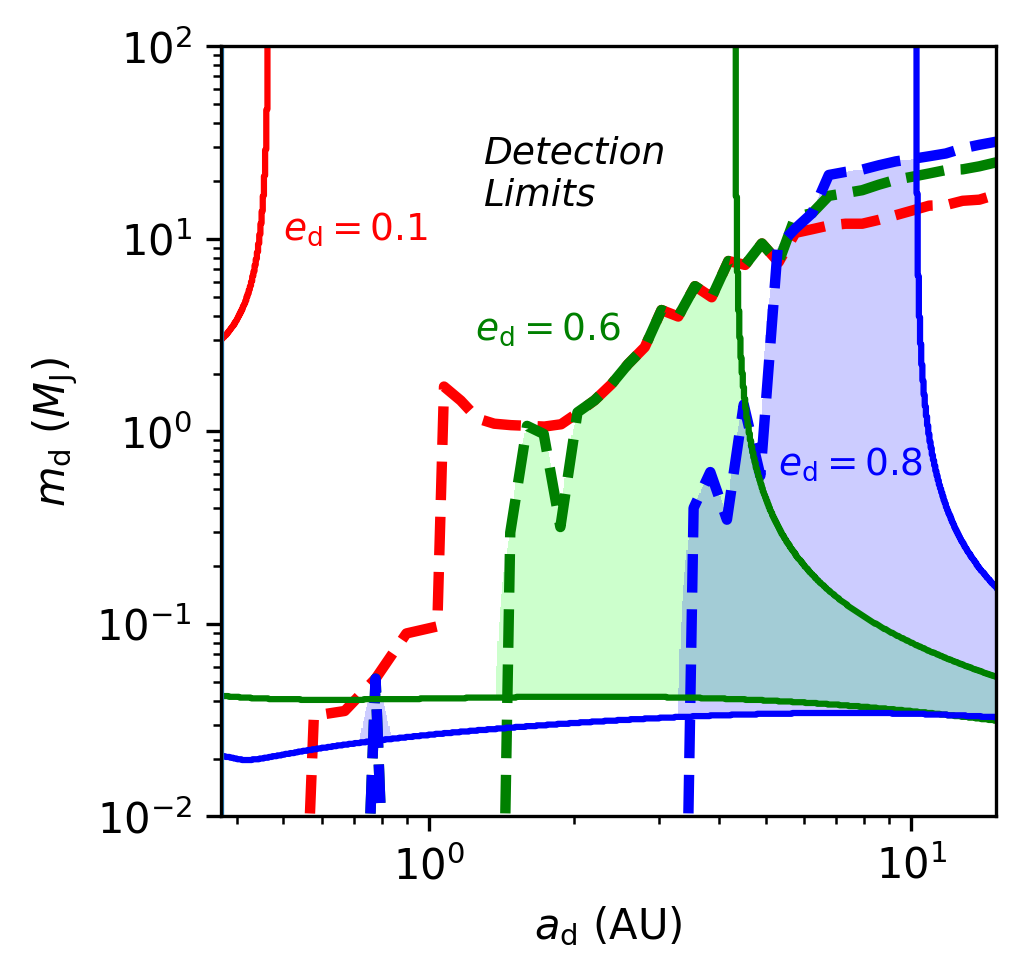}
    \caption{Parameter space of potential TOI-2134 ds which could feasibly excite TOI-2134 c's eccentricity to its present-day value via coplanar high-eccentricity migration. Solid lines delineate theoretical dynamical boundaries, while dashed lines delineate boundaries derived from RVs, stability constraints, and \textit{Gaia} DR3 detection limits. Red, green, and blue are associated with perturbers with $e_\mathrm{d} = 0.1, 0.6, 0.8$ respectively. Shaded regions denote areas of parameter space where perturbers could have both evaded detection and sufficiently excited TOI 2134 c's eccentricity. We see that there is no overlap for the red $e_d = 0.1$ scenario -- hence, the companion must be endowed with significant eccentricity in this scenario.}
    \label{fig:chem}
\end{figure}

Finally, we also utilized the Renormalized UnitWeight Error (RUWE) statistics from the \textit{Gaia} DR3 archive \citep{gaia_collaboration_vizier_2020} to identify further detection limits for cold giant companions. To this end, we used the approach described in \citet{Sozzetti2023}. In brief, we performed a Monte Carlo experiment fitting a single-star model to synthetic \textit{Gaia} DR3 astrometric data generated based on the values of the observing times, along-scan parallax factors, and scan angles obtained from the GOST tool\footnote{Available at: \url{https://gaia.esac.esa.int/gost/}}, the nominal \textit{Gaia} DR3 astrometric parameters for TOI-2134, the effects of orbital motion produced by companions over a range of masses and semi-major axis, random draws from uniform distributions of the other orbital elements, and prescriptions for transit-level \textit{Gaia} measurement errors appropriate for a star with the G mag of TOI-2134. For each mass-separation pair, the fraction of systems with RUWE $> 1.02$ was recorded \citep{Penoyre2022}. Through this analysis, we were able to exclude the presence of $M > 20$ M$_{\rm J}$ companions for all orbital periods, as well as $M > 2.5$ M$_{\rm J}$ planet with semi-major axes $< 3$ AU. The 99\% confidence level boundaries are included in Fig. \ref{fig:chem}. 

\subsection{Planet formation and migration}
\label{sec:migration}
The architecture of the TOI-2134 system is perplexing, and places strong constraints on its possible dynamical histories. Planets are naively expected to form in perfectly circular orbits due to strong eccentricity damping from the protoplanetary disk \citep{cresswell2007evolution}. Although modest eccentricities can, in principle, be excited through planet–disk interactions \citep{duffell2015eccentric}, the eccentricity of TOI-2134 c is likely too large to be produced by this mechanism. Mechanisms such as planet-planet scattering \citep[e.g.][]{chatterjee2008dynamical, juric2008eccentricity, ford2008eccentric, Carrera2019, lu2025hatp11}, and secular interactions such as von Ziepel-Lidov-Kozai oscillations \citep{vonzeipel1910zlk,lidov1962evolution,kozai1962secular, wu2003migration, naoz2016eccentric} can more readily generate higher eccentricities. However, the fact that both TOI-2134~b and TOI-2134~c transit their host star rules out most of these dynamical mechanisms, as in addition to eccentricity they tend to impart large mutual inclinations and break the coplanarity of the system.

One method of generating significant eccentricity without inclination is to invoke secular interactions in hierarchical systems, an effect coined by \cite{petrovich2015coplanar} as coplanar high-eccentricity migration. In this scenario, an exterior coplanar companion is required. Using Equations 1 and 3 from \cite{petrovich2015coplanar}, we computed the parameter space of putative planet ds which could reasonably excite \mbox{TOI-2134~c's} eccentricity from $0.1$ -- the nominal reasonable maximum expected from planet-disk interactions \citep[e.g.][]{duffell2015eccentric} -- to its present day value, shown as solid lines in Fig.  \ref{fig:chem}. Overplotted are also constraints on the parameter space from RV detection sensitivity\footnote{At such large orbital distances, the RV detection limits are ruled by the absence of an observable trend (that is not explained by stellar activity).} and orbital stability, which are discussed in depth in Section \ref{sec:detection_limits} and extended here to large semi-major axes. We carried out a battery of planet injection-recovery tests to find the mass limits below which no RV trend could be detected. We combined these detection limits at large orbital distances ($a_\mathrm{d}> 1$ AU) with classic Keplerian detection limits at small orbital distances ($a_\mathrm{d}< 1$ AU), the latter of which are described in Section \ref{sec:detection_limits}. We investigated orbital stability below this limit curve, and excluded the unstable regions of the parameter space. We also included detection limits derived from GR3 data. The dashed curves plotted in Fig. \ref{fig:chem} are the result of these combined RV, GR3, and stability constraints.

 The combined constraints rule out a low eccentricity for the potential planet d ($e_\mathrm{d} \sim 0.1$), but show that a massive eccentric companion is indeed capable of generating TOI-2134~c's eccentricity, which we verified with \textit{N}-body simulations. 

Of course, the necessity that the perturbing companion also be eccentric raises its own questions. One possibility is that the companion may be a brown dwarf\footnote{While our RV and stability constraints exclude perturbers more massive than the commonly cited $13 M_\mathrm{J}$ boundary between giant planets and brown dwarfs, this threshold should not be interpreted as a sharp physical limit \citep{spiegel2011deuterium} and our constraints do not rule out bodies with \mbox{$m \sim10M_\mathrm{J}.$}}, whose mass would be conducive to eccentricity \textit{growth} from planet-disk interactions \citep{goldreich2003eccentricity} which would maintain the coplanarity of the system. Indeed, there is population-level evidence of elevated eccentricities in brown dwarfs \citep{bowler2020population} to support this hypothesis, though we note that the eccentricities necessary in our scenario are higher than the limits suggested by numerical work \citep{papaloizou2001eccentricity}. A detailed analysis of the formation and dynamical history of the system is warranted, but shall be deferred to future work. Fortuitously much of the feasible parameter space will be readily detectable by the upcoming data releases of the \textit{Gaia} mission \citep{lammers2026gaia}, though the specific conclusion of a single eccentric giant planet detection must be treated with caution in astrometric data \citep{yahalomi2026astrometric}.

\begin{figure*}
    \centering
    \includegraphics[width=\linewidth]{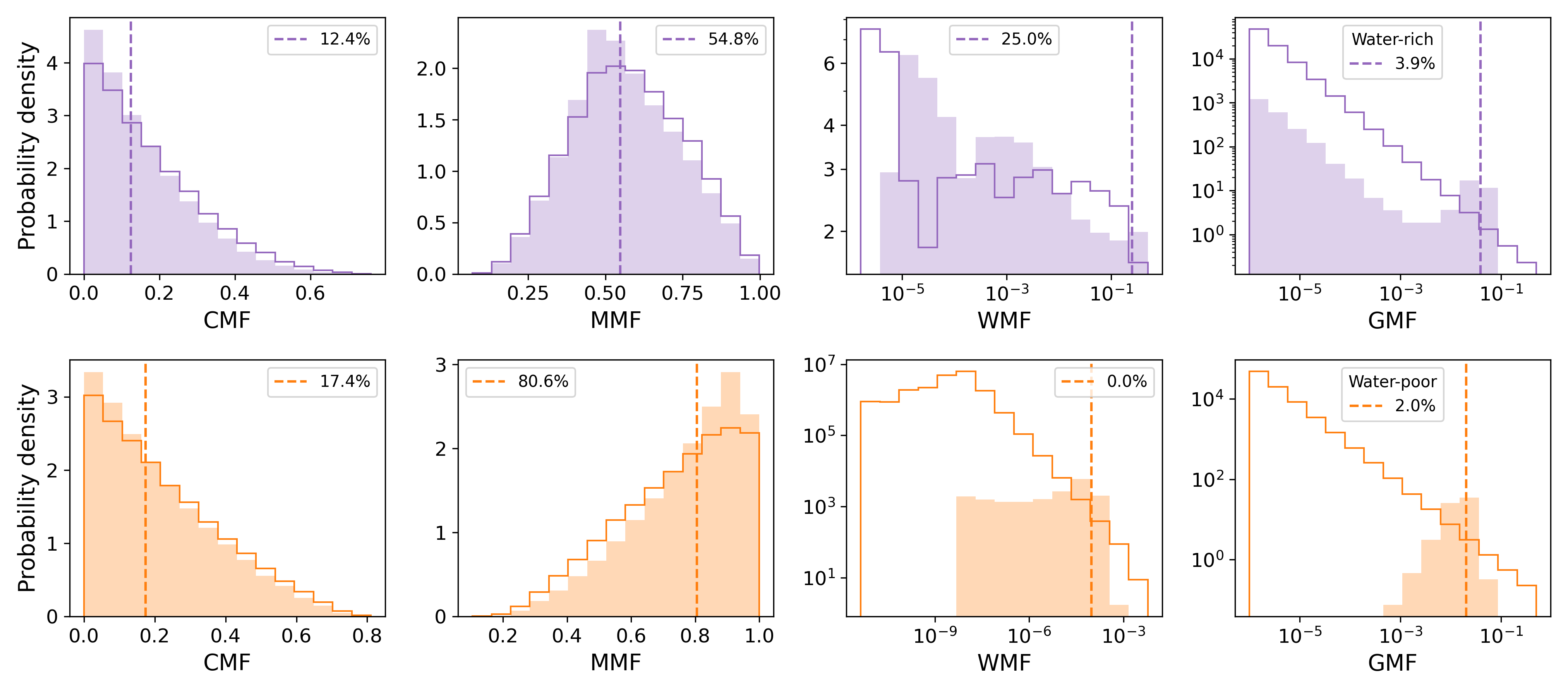}
    \caption{Modelled interior composition of TOI-2134~b from {\sc plaNETic}. Results for the water-rich and water-poor scenarios are shown in the top and bottom rows, respectively. Prior (step histograms) and posterior (filled histograms) distributions are shown for each component. The median posterior value (dashed vertical line) is provided in each legend.}
    \label{fig:b_interior}
\end{figure*}
\subsection{Interior composition analysis}
\label{sec:interiors}

The calculated mean bulk density of TOI-2134~b is \mbox{2.52$\pm$0.47 \,g\,cm$^{-3}$}, placing it between a terrestrial density (5.5\,g\,cm$^{-3}$) and the lower density of a Neptune-like planet (1.64\,g\,cm$^{-3}$).
This intermediate density suggested that the planet likely comprises an interior with an iron-rich core, a silicate mantle, and a significant volatile envelope.
To investigate the plausible range of compositions, we modelled the interior of \mbox{TOI-2134~b} using {\sc plaNETic} \citep{Egger2024}, a Bayesian framework based on the forward model of {\sc biceps} \citep{Haldemann2024}.

The interior of the planet is modelled using a three-layer structure, consisting of a core of iron (Fe) and sulphur (S), a rocky mantle containing magnesium (Mg), oxygenated silicon (Si), and Fe, and a uniformly mixed envelope of hydrogen (H), helium (He), and water.
This approach subsequently allows for the derivation of mass fractions for each key planetary component: the core (CMF), the mantle (MMF), and an envelope of water (WMF) and gas (GMF).
The ratios of Mg, Si, and Fe determine the mantle composition.
We employed a free prior on these ratios, modelling them independently of the composition of the host star.

We modelled the planet twice to account for two possible distinct formation pathways (accretion of water primarily as solid ice versus accretion in gaseous form) since it is unknown whether TOI-2134~b formed outside or inside the system's H$_2$O iceline.
We refer to the solutions as water-rich and water-poor, respectively.
These results are presented in Fig.~\ref{fig:b_interior}, with the priors and posteriors depicted by the step histograms and filled histograms, respectively, and the vertical dashed lines denoting the median values.
The CMF, MMF, and WMF were sampled uniformly on a simplex, which accounts for the non-uniform appearance of the priors when visualised in this space (see \citet{Egger2024} for a more detailed explanation).

As expected for a planet with the density of \mbox{TOI-2134~b}, both model types predicted a non-zero gaseous envelope (\mbox{$\mathrm{GMF}=0.04\pm0.01$} for the water-rich pathway and \mbox{$\mathrm{GMF}=0.02\pm0.01$} for the water-poor pathway).
Additionally, the water-rich formation scenario converged to a solution with a substantial water mass fraction ($\mathrm{WMF}=0.25\pm0.18$), indicating that the planet could have a significant hydrosphere.
However, different interior compositions can result in the same observed planetary mass and radius, especially for sub-Neptunes, meaning additional information about the nature of the planet is needed to resolve the degeneracy: atmospheric observations offer a viable means of breaking it.

\begin{figure}
    \centering
    \includegraphics[width=0.9\linewidth]{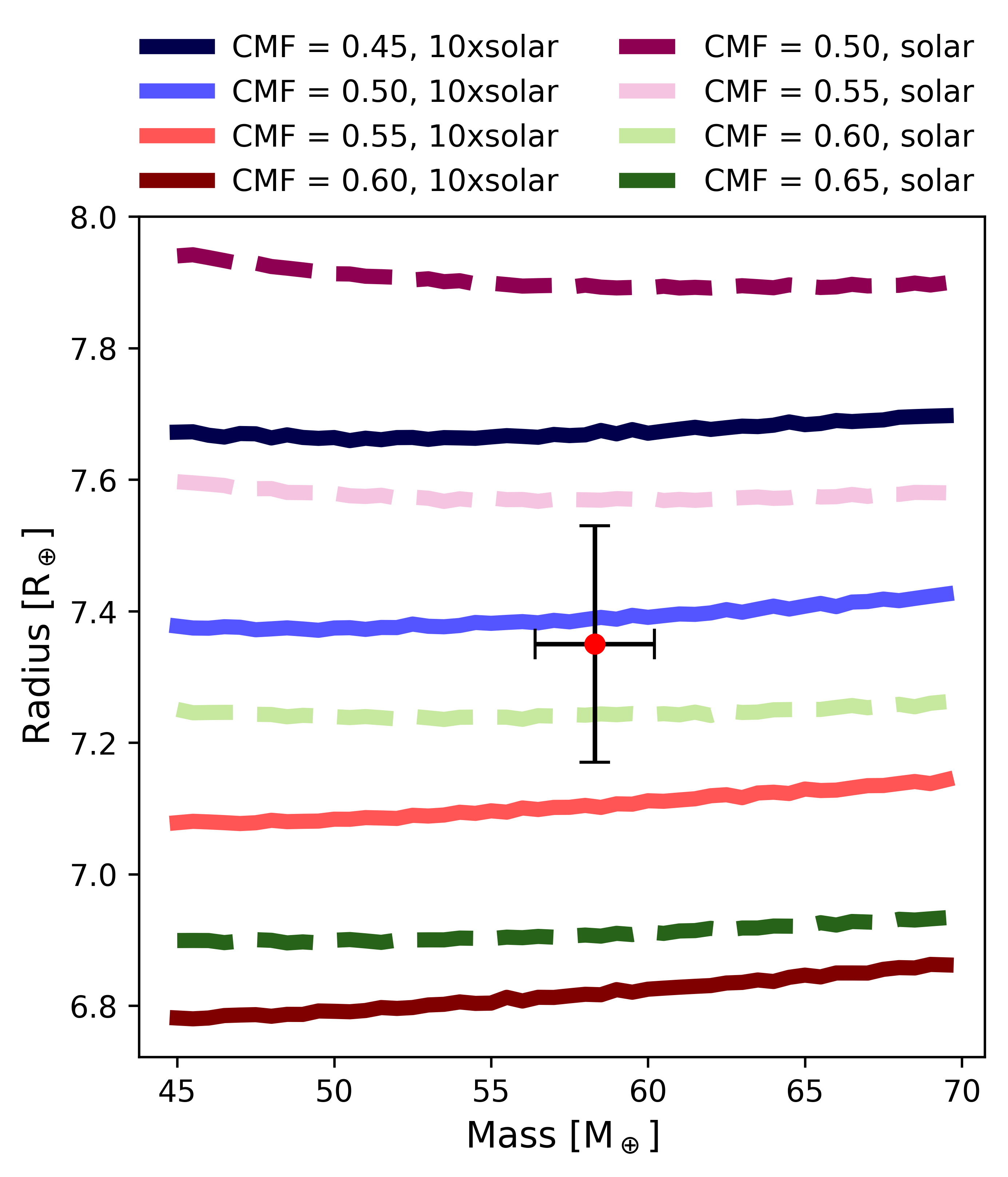}
    \caption{Mass-radius diagram for varying core mass fraction (CMF) for the sub-Saturn TOI-2134~c. Solid lines represent an assumed 10$\times$ solar metallicity, while dashed lines assume 1$\times$ solar metallicity. The planet is plotted as a red dot with black error bars. For all tracks we assumed an internal temperature of 100~K, coherent with the age of the system, and solar C/O ratio of 0.55.}
    \label{fig:gastli}
\end{figure}

The bulk density of TOI-2134~c was computed to be \mbox{$0.81\pm0.15$\,g cm$^{-3}$}. This places planet c between the densities of the two puffiest planets in the Solar system, Saturn and Uranus (with mean bulk densities of $0.6871$ and $1.270$\,\mbox{g cm$^{-3}$} respectively). To investigate the interior composition of this planet we used GAS gianT modeL for Interiors \citep[{\sc gastli}: ][]{acuna2021,acuna2024}\footnote{Available at: \url{https://gastli.readthedocs.io/en/latest}}, a Python package designed to model the interior structure of volatile-rich exoplanets. {\sc gastli} describes the planet with a two-layer structure, composed of a core and an envelope. The core is assumed to be composed of a 1:1 water and rock mixture due to the high pressures, and the envelope of a mixture of H/He and water. In order to assess the most likely composition of TOI-2134~c, we computed the mass-radius curves for a selection of CMFs and metallicities, as shown in Fig. \ref{fig:gastli}. For these calculations we assumed an internal temperature of 100~K, consistent with the age of the system, and the solar C/O ratio of 0.55. Using a grid of $\Delta$CMF = 0.05 for the core mass fraction, we considered three possible metallicities: 1$\times$ solar, 10$\times$ solar, and 100$\times$ solar. In Fig. \ref{fig:gastli} the first two are represented by dashed and solid lines respectively. The radius of TOI-2134~c is larger than its simulated radius for a composition of 100$\times$ solar envelope and CMF = 0 (meaning no core). This means that its density is too low to present an atmospheric metallicity greater than 100$\times$ solar, which is consistent with the mass-metallicity trend at a population level \citep{Wakeford17,Welbanks19}. Via this analysis, TOI-2134~c was found to be coherent with a CMF of 0.60$\pm$0.05 for a log(Fe/H) = 0, or with a CMF of 0.50$\pm$0.05 for a log(Fe/H) = 1. The low-metallicity model yielded an envelope mass fraction (EMF) of 0.4 with an envelope water mass fraction of circa $Z_{\rm env}$ = 0.013 (and thus a He/H envelope mass fraction of 0.987). The total metal mass was computed to be circa 35 \mearth, or 60\% of the total mass. In the 10$\times$ solar metallicity case, we found a CMF = 0.50 and thus \mbox{EMF = 0.50}. This atmospheric metallicity yielded a water mass fraction in the envelope of $Z_{\rm env}$ = 0.12 (0.88 He/H mass fraction), and a total metal mass of 33 \mearth, or 57\% of the total mass.

The interior composition of TOI-2134~c is degenerate between the two parameters, CMF and atmospheric metallicity, and further atmospheric characterisation would be needed here as well to break this degeneracy (we further investigate this avenue in Section \ref{sec:atmo}). We however note that large age uncertainties can still prevent atmospheric metallicity measurements from solving the CMF-$Z_{\rm env}$ degeneracy \citep{acuna2024, MH23}. A better age measurement may in fact be necessary for further analysis.

\begin{figure}
    \centering
    \includegraphics[width=\linewidth]{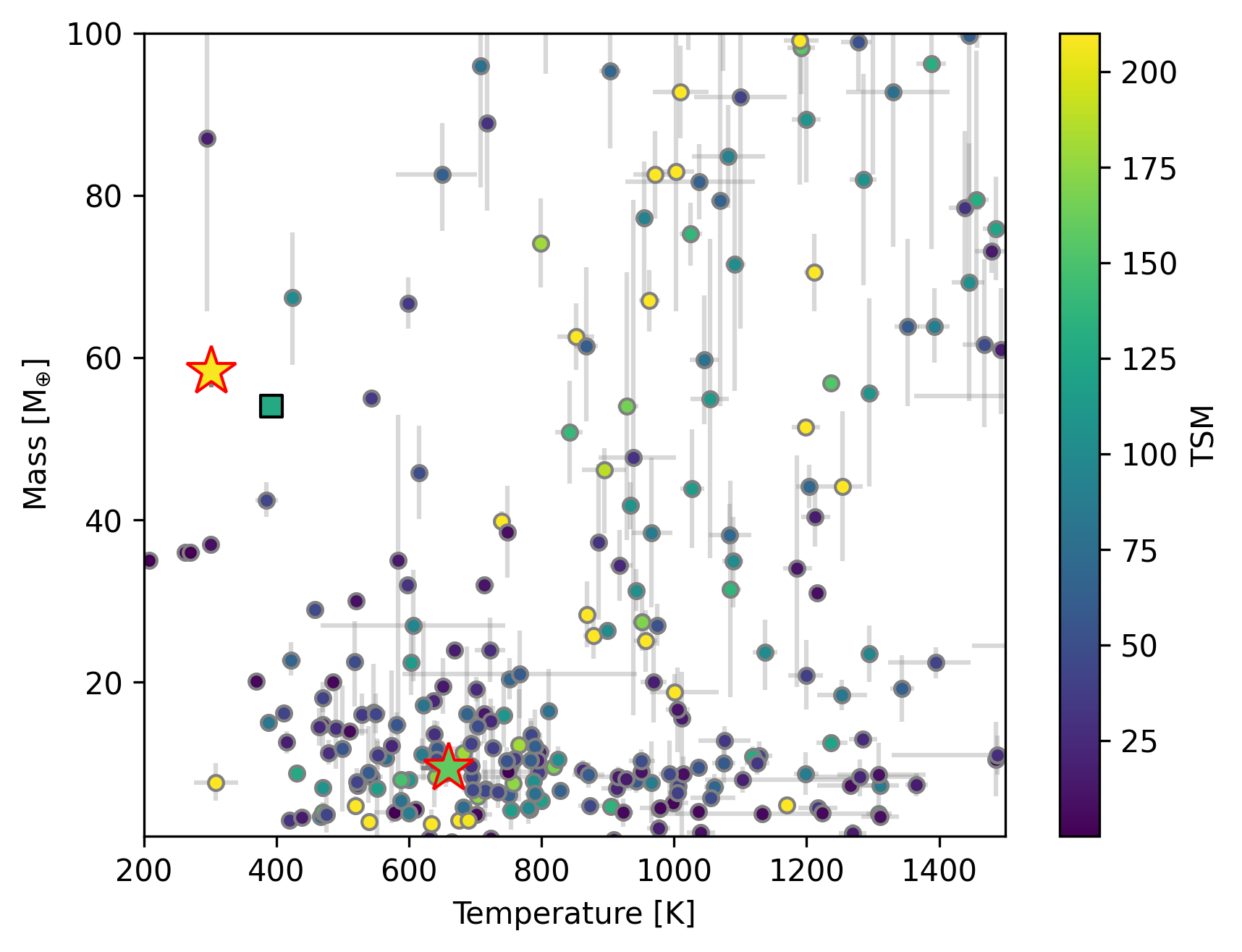}
    \caption{Mass-equilibrium temperature diagram for all confirmed exoplanets with better than 3$\sigma$ mass detection. The TOI-2134 planets are plotted as stars with red borders. Uncertainties are included but not visible. All other confirmed planets orbiting FGK-type stars are plotted as circles with gray errorbars indicating the quoted uncertainties. Data is colour coded based on computed planetary transmission spectroscopy metric (TSM). TOI-199~b is also plotted as a square.}
    \label{fig:M_T}
\end{figure}

\begin{figure*}
    \centering
    \includegraphics[width=18cm]{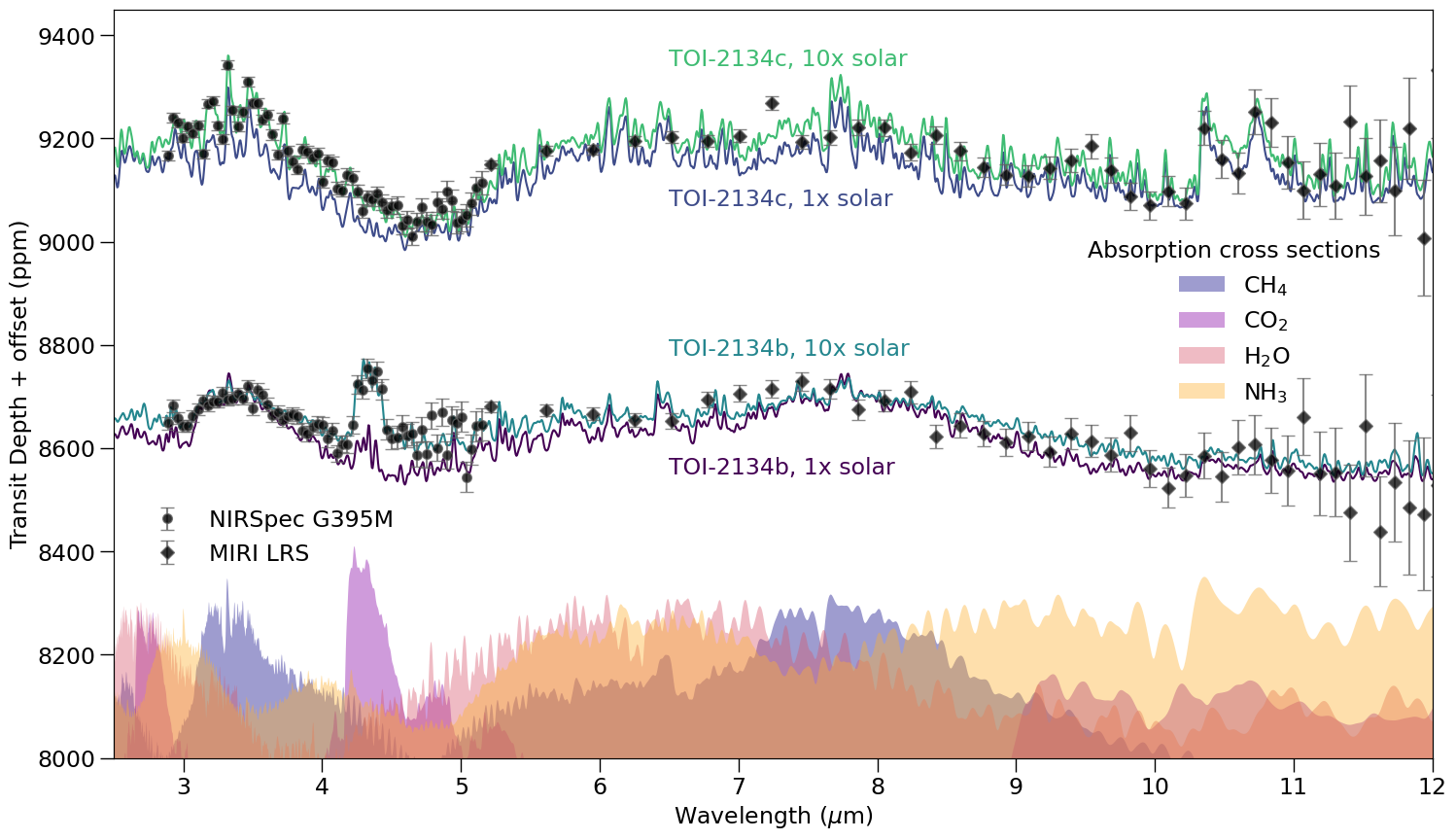}
    \caption{Model atmospheric spectra for planets b (bottom) and c (top), with either a 1$\times$ solar (darker lines) or 10$\times$ solar (lighter lines) atmospheric metallicity. Shaded regions show the relative absorption cross sections of CH$_4$, CO$_2$, H$_2$O and NH$_3$ for reference. Circle/diamond markers and error bars show simulated NIRSpec G395M/MIRI LRS observations assuming a single transit per instrument.}
    \label{fig:atm_sim}
\end{figure*}

\subsection{Atmospheric simulations}
\label{sec:atmo}

We investigated the potential of planets b and c for atmospheric characterisation by modelling their near- to mid-infrared transmission spectra. Fig. \ref{fig:M_T} shows the Transmission Spectroscopy Metric (TSM: \citealt{Kempton2018}) of known exoplanets in the $\sim$10-100 \mearth\, mass range with equilibrium temperature between 200 and 1500~K (assuming albedo = 0 and perfect circulation). Both TOI-2134~b and c have notable TSM values (159 and 208, respectively), and planet c in particular stands out as a uniquely favourable target in the <400~K regime. In fact, both planets have been observed before. TOI-2134~b was followed by Keck's high resolution NIRSPEC spectrograph, detecting escaping helium \citep{Zhang2023}. Planet c has been observed by the Hubble Space Telescope WFC3 G141 (GO 17920), and will also be followed-up to look for escaping hydrogen.

We modelled representative atmospheric spectra for these planets using the {\sc Genesis} atmospheric modelling code \citep{Gandhi2017,Piette2020_MN} coupled with the {\sc FastChem Cond} equilibrium chemistry solver \citep{Kitzmann2024_fastchemcond}. The temperature profiles for each planet were assumed to be isotherms at their zero-albedo, orbit-averaged equilibrium temperatures (660~K and 302~K for planets b and c, respectively, included in Table \ref{tab:planets}). We include opacity due to H$_2$O \citep{Rothman2010}, CH$_4$ \citep{Yurchenko2013, Yurchenko2014}, C$_2$H$_2$ \citep{Rothman2013, Gordon2017}, CO$_2$ \citep{Rothman2010}, CO \citep{Rothman2010}, HCN \citep{Harris2006}, NH$_3$ \citep{Yurchenko2011}, N$_2$ \citep{Barklem2016, Western2018}, O$_2$ \citep{Gordon2017}, and H$_2$S \citep{Azzam2016,Chubb2018}, and collision-induced absorption (CIA) due to H$_2$- H$_2$ and H$_2$-He \citep{Richard2012}.

Fig. \ref{fig:atm_sim} shows the model transmission spectra for planets b and c. We show nominal atmospheric compositions corresponding to 1$\times$ and 10$\times$ solar elemental abundances, following the results of the previous Sections, where the molecular abundances are calculated for each scenario according to thermochemical equilibrium using {\sc FastChem Cond}. Fig. \ref{fig:atm_sim} also shows the relative cross sections of CH$_4$, CO$_2$, H$_2$O and NH$_3$ to illustrate their contributions to these spectra. We additionally show simulated JWST observations with NIRSpec~G395M and MIRI~LRS, calculated using \texttt{PandExo} \citep{Batalha2017} and assuming one transit for each instrument.

For both planets b and c, the transmission spectrum is dominated by H$_2$O and CH$_4$, given their relatively low temperatures. For the cooler planet c, NH$_3$ opacity also becomes important, with a notable feature visible at $\sim$10-11~$\mu$m and smaller features at 2$\mu$m and 1.5$\mu$m. In the case of a higher atmospheric metallicity (e.g., 10$\times$ solar elemental abundances, as shown in Fig. \ref{fig:atm_sim}), CO$_2$ features are also visible, especially at $\sim$4.5~$\mu$m. The simulated JWST observations show that these spectral features could feasibly be constrained with one or more transits per instrument.

TOI-2134~c is therefore a prime candidate for probing nitrogen chemistry in exoplanet atmospheres. To date, this has been a challenging task, as NH$_3$ only becomes the dominant nitrogen carrier in H$_2$-rich atmospheres below $\sim$500~K, above which undetectable N$_2$ is instead the dominant carrier \citep[e.g.][]{Moses2013}. Exoplanets in this temperature range can be more challenging to characterise due to their longer orbital periods and smaller atmospheric scale heights. Nevertheless, NH$_3$ was recently detected in the atmosphere of the transiting warm Neptune WASP-107~b \citep{Welbanks2024} and in directly-imaged planets such as GJ~504~b \citep{Malin2025} and HR8799~b \citep{Xuan2026}. Non-detections have been reported in smaller, sub-Neptune planets (e.g. K2-18~b and TOI-270~d, \citealt{Madhusudhan2023,Benneke2024,Holmberg2024}), which may be a result of atmosphere-interior interactions. TOI-2134~c also resides in the temperature range in which H$_2$O can condense \citep{Lodders2002}, allowing for the possible probing of Earth-like water clouds for the first time in an exoplanet.

The potential of planet c for atmospheric characterisation with JWST is particularly compelling given recent results for the similar, temperate sub-Saturn TOI-199~b \citep{Bello-Arufe2025_TOI-199b}, highlighted by a square marker in Fig. \ref{fig:M_T}. With a TSM lower than TOI-2134~c (126 compared to 208), TOI-199~b showed strong evidence of CH$_4$ absorption from its atmosphere based on a single JWST/NIRSpec G395M transit. TOI-199 b's atmospheric chemistry will be further refined with two upcoming transits with NIRSpec/G395M (GO 7188, PI: Acu\~na, Acu\~na-Aguirre et al. in prep). While the presence of NH$_3$ and/or HCN was unconstrained by the current data, both TOI-199~b and TOI-2134~c are promising targets to uncover nitrogen chemistry in the temperate Saturn to sub-Saturn mass range.   

\section{Summary and Conclusions}
\label{Sec:concl}

In this work, we presented the re-analysis of the orbital structure of the TOI-2134 system comprising of an inner mini-Neptune and a temperate sub-Saturn. Following a previous publication (\citetalias{rescigno_hot_2023}) and in order to better define the eccentricity of the outer planet, we obtained 98 further spectral observations and analysed three new sectors of TESS data. We investigated the number of planets in the system via a rigorous analysis of 160 HARPS-N RVs and 120 SOPHIE RVs. The statistically preferred model described the data as the sum of two Keplerians and a GP with a quasi-periodic kernel. The data included an unresolved long-term trend that was found to be mainly derived from the magnetic activity cycle of the star. However, since the extracted stellar rotation period sightly disagreed with the one extracted from a GP analysis of the stellar activity indicators, we also tested a multidimensional GP framework, which simultaneously modelled the RVs from both spectrographs, and the S-index and the bisector span. The resulting planetary parameters agreed within 1$\sigma$ and the stellar rotation period was better defined. We also tested the preferred models on two other reductions of the HARPS-N spectra: {\sc yarara} and {\sc tweaks}. In all cases the agreed best results only included two Keplerians mapping the two observed sets of transits. Final orbital characteristics were taken from a joint analysis of the photometry, described by {\sc batman} transit models, and of the RVs, modelled with a multidimensional GP. 

In order to study the spin-orbit alignment of the system, we also acquired radial velocities for both planets during transit. We observed the 28 April 2023 transit of the inner planet with EXPRES, acquiring 21 observations, and the 7 April 2025 transit of the outer one with PARAS-2, obtaining 13 spectra over 3 nights. We analysed the data in order to measure the Rossiter-McLaughlin effect of either planets, but some particularly bad weather on both nights made the modelling tricky. We were not able to detect any RM signal in the EXPRES data for TOI-2134~b, and we only have a detection in the PARAS-2 data for TOI-2134~c, dependent on the chosen baseline model.

As results of these analyses we reached the following conclusions:
\begin{itemize}
    \item We characterise the mini-Neptune TOI-2134~b to be in a \mbox{$9.229198\pm0.000003$} days orbit. We measure its mass and radius as \mbox{$9.37\pm0.54$ \mearth} and \mbox{$2.735\pm0.068$ \rearth}. The newly-derived values all agree within 1$\sigma$ with those of \citetalias{rescigno_hot_2023}, but the new data significantly improved their precision and significance.
    We could not measure TOI-2134~b's orbital obliquity due to poor data quality. Its bulk density places it between terrestrial and Neptune-like worlds, and modelling its interior composition with {\sc plaNETic} predicted a non-zero gaseous envelope for both water-rich and water-poor formation pathways (GMF = $0.04\pm0.1$ and $0.02\pm0.1$ respectively). For the water-rich case, the analysis also preferred a substantial water mass fraction of $0.25\pm0.18$, pointing to a significant hydrosphere.

    \item Planet c is confirmed to be a sub-Saturn in a \mbox{$95.852840\pm0.000041$ days} orbit, with mass equal to \mbox{$58.3\pm1.9$ \mearth}, and radius \mbox{$7.35\pm0.18$ \rearth}. The much improved completeness over the orbital phase space for this planet allowed to break the eccentricity posterior multimodality, and resulted in well-converged Gaussian-like posteriors centred on the lower eccentricity value of $0.31\pm0.01$, instead of the split results presented in \citetalias{rescigno_hot_2023}. We underwent extensive model testing to avoid incorrectly deflating or inflating the planet's semi-amplitude by incorrect signal attribution between this outer Keplerian and the stellar rotation period found to be at its first harmonic. This eccentricity places TOI-2134~c squarely in the overdensity in eccentricity space typical of gas giants with $P>10$ days. A detection of a RM effect could be found with a significance of 4.7$\sigma$ for a $59\pm31^\circ$ obliquity, although the result is model-dependent. Overall TOI-2134~c has a 84$\%$ probability to be in a prograde orbit. The bulk density of the planet lies between those of Saturn and Uranus. We modelled the planet interior composition using {\sc gastli}, and found the planet to be consistent with core mass fractions of $0.60\pm0.05$ and $0.50\pm0.05$ for 1$\times$solar and 10$\times$solar metallicities respectively. The total metal mass was computed to be between 57 and 60$\%$ of the total mass. Although higher metallicities of 100$\times$solar were also considered, no compatible models could be found.

    \item Via detection limits analysis, we could exclude the presence of any undetected body with mass larger than 3 \mearth\, orbiting inner to planet b, and any with mass larger than 5 \mearth\, orbiting between the two planets. In fact the stability analysis found that no stable orbits could be found for any planet with orbital periods between $\sim$50 and 185 days due to the significant eccentricity of TOI-2134~c. Any massive companion ($M>25$ \mearth) with period under 400 days would also have been detected and can thus be excluded. \textit{Gaia} data further restricted the parameter space available for undetected bodies: any planet with $M > 20$ M$_{\rm J}$ would have been detected, as well as any with $M > 2.5$ M$_{\rm J}$ orbiting at all semi-major axes $< 3$ AU.

    \item The architecture of the system is peculiar, with a high-eccentricity outer gas giant and a surviving inner planet. The eccentricity of TOI-2134~c is likely too large to be excited via planet-disk interactions, and too small for extreme high-eccentricity migration. Other mechanisms such as planet-planet scattering are also unlikely due to the co-planarity of both planets. The most likely method given the current data was found to be coplanar high-eccentricity migration, for which a third coplanar companion would be required. While an unresolved long-term trend is present in the radial velocities, the analyses preferably assigned it to a magnetic cycle effect, and no third body could be found.

    \item Finally, we investigate the potential of both planets for atmospheric characterisation. TOI-2134~b has a TSM of 159, and is a very interesting candidate in the 600-700 K equilibrium temperature regime. Previous observations confirmed escaping helium. Its simulated transmission spectrum is dominated by H$_2$O and CH$_4$, and planetary metallicity is shown to have a significant effect around 4.3$\mu m$. TOI-2134~c is a uniquely favourable target within the $T_{\rm eq}<400$K regime. It has the largest TSM of 208 and occupies a significantly underpopulated mass-temperature parameter space. At these temperatures the opacity of NH$_3$ becomes important, with a notable feature around 10.5$\mu m$, making planet c a prime candidate for probing nitrogen chemistry. This is particularly compelling given the recent results for the sub-Saturn TOI-199 b, which with a TSM of only 126 was found to show evidence of CH$_4$ absorption. Finally planet c is also a promising target for the detection of water clouds for the first time in an exoplanet.

\end{itemize}

\section*{Acknowledgements}

FR and AM acknowledge a UK Science and Technology Facilities Council (STFC) small grant ST/Y002334/1.

The HARPS-N project was funded by the Prodex Program of the Swiss Space Office (SSO), the Harvard University Origin of Life Initiative (HUOLI), the Scottish Universities Physics Alliance (SUPA), the University of Geneva, the Smithsonian Astrophysical Observatory (SAO), the Italian National Astrophysical Institute (INAF), University of St. Andrews, Queen’s University Belfast, and University of Edinburgh.

Funding for the TESS mission is provided by NASA's Science Mission Directorate.KAC acknowledges support from the TESS mission via subaward s3449 from MIT.

This research has made use of the Exoplanet Follow-up Observation Program website, which is operated by the California Institute of Technology, under contract with the National Aeronautics and Space Administration under the Exoplanet Exploration Program.

This paper includes data collected by the TESS mission that are publicly available from the Mikulski Archive for Space Telescopes (MAST).

This publication makes use of The Data $\&$ Analysis Center for Exoplanets (DACE), which is a facility based at the University of Geneva (CH) dedicated to extrasolar planets data visualisation, exchange and analysis. DACE is a platform of the Swiss National Centre of Competence in Research (NCCR) PlanetS, federating the Swiss expertise in Exoplanet research. The DACE platform is available at \url{https://dace.unige.ch}.

We are grateful to PRL-DOS (Department of Space, Government of India) as well as the Director, PRL, for their generous support. Their support has been instrumental in funding the PARAS-2 spectrograph. We express our gratitude to all the Mount Abu Observatory staff and the PARAS-2 instrument team for their invaluable support throughout the observations.

These results made use of data provided by the EXPRES team using the EXtreme PREcision Spectrograph at the Lowell Discovery telescope, Lowell Observatory. Lowell is a private, non-profit institution dedicated to astrophysical research and public appreciation of astronomy and operates the LDT in partnership with Boston University, the University of Maryland, the University of Toledo, Northern Arizona University and Yale University. Lowell Observatory sits at the base of mountains sacred to tribes throughout the region. We honor their past, present, and future generations, who have lived here for millennia and will forever call this place home.

This work made use of {\sc tpfplotter} by J. Lillo-Box (publicly available in \url{www.github.com/jlillo/tpfplotter}), which also made use of the python packages {\sc astropy}, {\sc lightkurve}, {\sc matplotlib} and {\sc numpy}.

AB acknowledges PhD studentship funding from the Science and Technology Facilities Council (STFC), UK.

MC acknowledges funding from the European Research Council under the European Union’s Horizon 2020 research and innovation programme (grant agreement no. 865624, GPRV).

DAT acknowledges the support of the Science and Technology Facilities Council (STFC).

ACC acknowledges support from STFC consolidated grant number ST/V000861/1, UKRI/ERC Synergy Grant EP/Z000181/1 (REVEAL) and UKSA grant number ST/X002217/1.

AP acknowledges funding from a UK Science and Technology Facilities Council (STFC) Small Award, grant number UKRI/ST/B001171/1.

This work was funded by the European Union (ERC, FIERCE, 101052347). Views and opinions expressed are however those of the author(s) only and do not necessarily reflect those of the European Union or the European Research Council. Neither the European Union nor the granting authority can be held responsible for them. This work was also supported by FCT - Funda\c c\~ ao para a Ci\^ encia e a Tecnologia through national funds by these grants: UIDB/04434/2020 DOI: 10.54499/UIDB/04434/2020, UIDP/04434/2020 DOI: 10.54499/UIDP/04434/2020, PTDC/FIS-AST/4862/2020, UID/04434/2025.

JM acknowledges support from the Italian Ministero dell’Universit\'a e della Ricerca and the European Union – Next Generation EU through project PRIN 2022 PM4JLH “Know your little neighbours: characterizing low-mass stars and planets in the Solar neighbourhood”.

VK acknowledges funding from the Royal Society through a Newton International Fellowship with grant number NIF/R1/232229.

MS acknowledges financial support from the Belgian Federal Science Policy Office (BELSPO) in the framework of the PRODEX Programme of the European Space Agency (ESA) under contract number C4000140754.

TGW acknowledges support from the University of Warwick and UKSA.

MLM is supported by individual research time under NASA contracts NAS5-26555 and NAS5-03127 to the Associated Universities for Research in Astronomy for the operation of the Hubble Space Telescope and James Webb Space Telescope Science Operations Centers at STScI.

CZ acknowledge support from STFC consolidated grant number ST/V000861/1 and UKRI/ERC Synergy Grant EP/Z000181/1 (REVEAL)

FPE and CLO would like to acknowledge the Swiss National Science Foundation (SNSF) for supporting research with HARPS-N through the SNSF grants nr. 140649, 152721, 166227, 184618 and 215190. The HARPS-N Instrument Project was partially funded through the Swiss ESA-PRODEX Programme.

AAJ and AM acknowledge funding from a UKRI Future Leader Fellowship, grant number MR/X033244/1.

GM acknowledges support by the Space It Up project funded by the Italian Space Agency, ASI, and the Ministry of University and Research, MUR, under contract n. 2024-5-E.0 - CUP n. I53D24000060005.

The SOPHIE program benefits from CNES and Région Ile-de-France (IDF-DIM-ORIGINES-2024-2-03) supports, as well as staff at Observatoire de Haute-Provence.

\section*{Data Availability}

The observational data presented in this publication are openly available. The TESS data are available at: \url{https://mast.stsci.edu/portal/Mashup/Clients/Mast/Portal.html}. The HARPS-N and SOPHIE data, alongside their mentioned activity proxies, and the EXPRES and PARAS-2 data are included as Supporting Information.



\bibliographystyle{mnras}
\bibliography{Bib/references} 




\appendix
\section{Some extra figures and tables}

\begin{figure*}
    \centering
    \includegraphics[width=16cm]{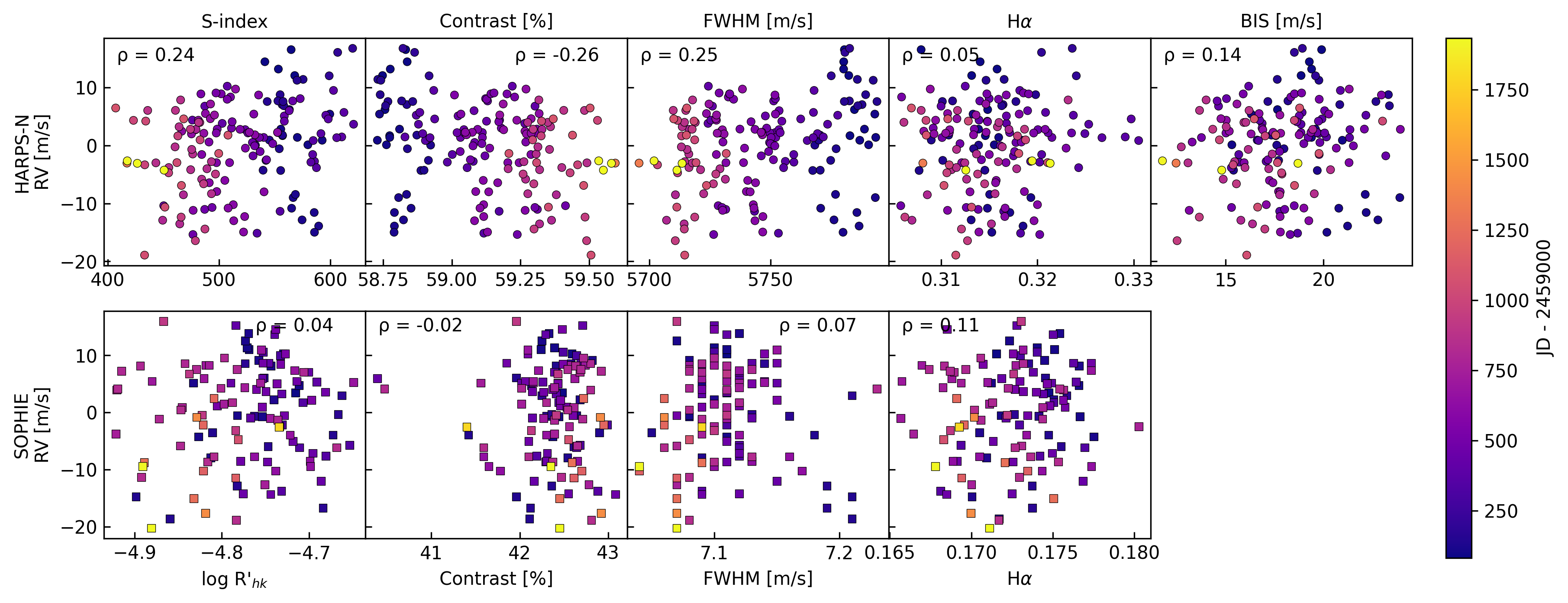}
    \caption{Correlation plots of the RVs, shown on the y-axes, against their activity indicators, plotted on the x-axes. \textit{Top row}: HARPS-N RVs against bisector span, contrast and FWHM of the CCF, S-index and H$\alpha$ index. \textit{Bottom row}: SOPHIE RVs against $\log R'_{\rm hk}$, contrast, FWHM and H$\alpha$. All data points are colour-coded by the date of the observation, as shown in the colour bar. For each activity indicator, the Spearman correlation coefficient is also included.}
    \label{fig:corr}
\end{figure*}

In this Section we include all supplementary materials.

Fig. \ref{fig:corr} depicts the correlation between the HARPS-N and the SOPHIE radial velocities with their available activity indicators, as described in Section \ref{Sec:st_act}. The Spearman correlation coefficients are included as text in each plot.

\begin{figure*}
    \centering
    \includegraphics[width=18cm]{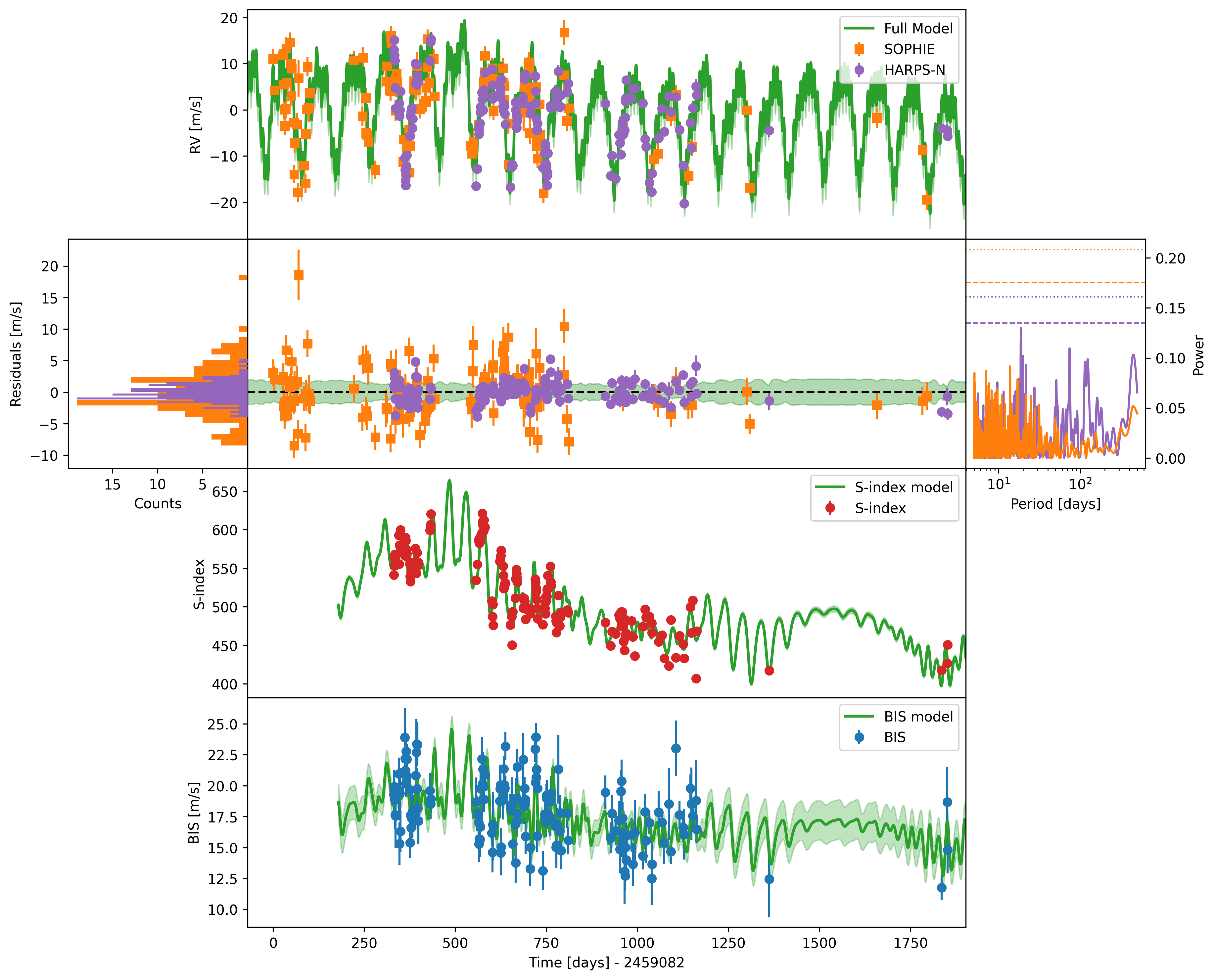}
    \caption{Results of the multidimensional GP analysis undertaken in Section \ref{Sec:joint}. On the first row the RVs from HARPS-N and SOPHIE are plotted with uncertainties as purple dots and orange squares respectively. The two RV time series are aligned vertically with the derived best offsets. The combined GP and Keplerians model is shown as a green line, with the shaded green areas representing the model uncertainties. On the second row, the residuals between the RVs and the full model are shown. The colour scheme is the same as the plot above, with the zero value indicated by a black dashed line. The green shaded region around the zero-point represents the uncertainties on the full RV model. Histograms of the residuals are shown on the left. GLS periodograms are shown on the right, with the respective 1 and 0.1$\%$ FAPs indicated in the respective colours as dashed and dotted lines. On the third and fourth row, we show the two activity indicators used for the multidimensional GP analysis: the S-index in red, and the BIS in blue. As above, the GP model and its uncertainties are shown in green. We note that we only use activity proxies from the HARPS-N data.}
    \label{fig:multigp}
\end{figure*}
In Fig. \ref{fig:multigp} we present the full multidimensional GP RV model from which final results are derived in Section \ref{Sec:joint}. The SOPHIE and HARPS-N RVs are depicted as orange squared and purple dots respectively, while the model is shown by the green line, with uncertainties highlighted by shaded regions. The residuals of the RVs are shown in the second plot, with related distributions and periodograms. We also include the derived models for the S-index and the BIS.

\input{Tables/gp}
In Table \ref{tab:gp} we list all the fitted parameters for the joint photometry and multidimensional GP RV fit not already included in Table \ref{tab:planets}. These include the jitter values, $\beta$, for both RV datasets, the activity indicators, and for the TESS photometry, the offsets for the radial velocities and the activity indicators, the GP amplitudes of the convective and the rotational terms for the RVs and the activity indicators (as described in Eq. \ref{eq:multiGP}). The TESS limb darkening coefficients, $q_1$ and $q_2$ are also present, together with the hyperparamters of the QP kernel: the stellar rotational period, $P_{\rm rot}$, its evolution timescale, $P_{\rm dec}$, and its harmonic complexity, $\lambda$.

\begin{figure}
    \centering
    \includegraphics[width=\linewidth]{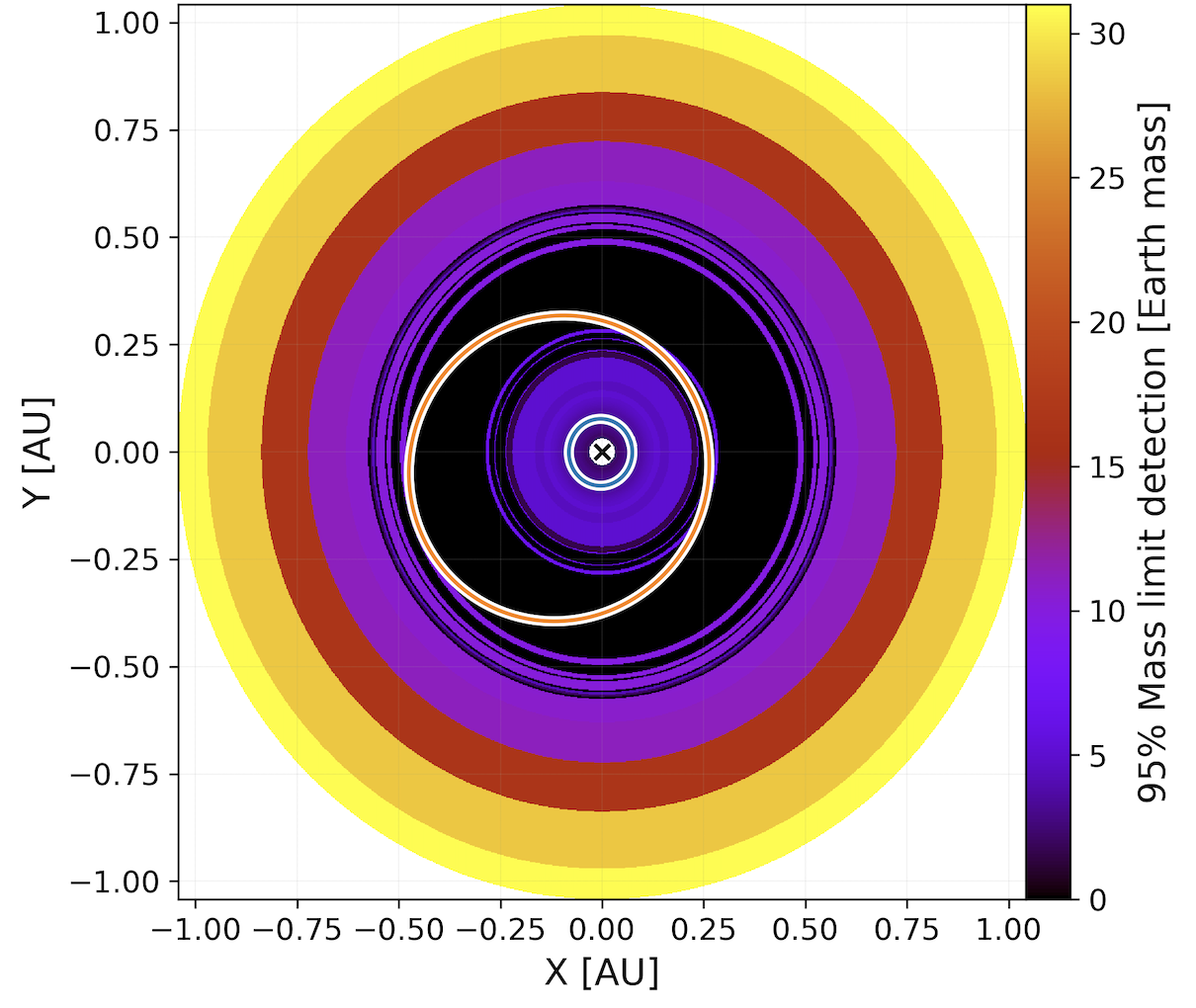}
    \caption{Alternative visualisation of the computed detection limits for the TOI-2134 system mapped on the orbital plane. The star is represented by an X, while the two confirmed planets are plotted as blue and orange obits with white borders. The rest of the orbital space is colour coded based on the mass of the detection limit. Black areas represent spaces for which no stable orbit could be found. For this plot, we used a higher resolution in the period sampling (100 additional period bins between 0.5$P$ and 2$P$ of each planet). It reveals islands of stability associated with the 2:1, 3:2, 4:3, and 5:3 motion mean resonances.}
    \label{fig:limits2}
\end{figure}

Finally, Fig. \ref{fig:limits2} is an alternative representation of the results depicted in Fig. \ref{fig:detection_limit} in Section \ref{sec:detection_limits}.


\bsp	
\label{lastpage}
\end{document}

%% file: Tables/stellar_characteristics.tex
\begin{table}
\centering

\caption{Stellar parameters of TOI-2134.
\label{bigtable}\label{tab:star}}

\begin{tabular}{c@{\hskip6pt}c@{\hskip7pt}c}

\rule{0pt}{0ex} \\
\hline
\hline
\rule{0pt}{0ex} \\
Parameter & Value & Source \\
\rule{0pt}{0ex} \\
\hline
\rule{0pt}{0ex} \\
	RA [h:m:s]      & 18:07:44.52 &  \cite{gaia_collaboration_vizier_2020} \\
	DEC [d:m:s]     & +39:04:22.54 & \cite{gaia_collaboration_vizier_2020}  \\
	Spectral type   & K5V  & \cite{stephenson_dwarf_1986} \\
	$m_V$ [mag]   & 8.933$\pm$0.003 & TESS Project\textsuperscript{*} \\
	Parallax [mas]  & 44.1087$\pm$  0.0144 & \cite{gaia_collaboration_vizier_2020} \\
	Distance [pc]   & 22.655$\pm$0.007 & \citetalias{rescigno_hot_2023}\\
    Proper motion RA [mas/yr] & 54.536$\pm$0.016 & \cite{gaia_collaboration_vizier_2020}\\
    Proper motion Dec [mas/yr] & -283.051$\pm$0.016 & \cite{gaia_collaboration_vizier_2020}\\
    $L_{\star}$ [$L_{\odot}$] & 0.192$\pm$0.008 & \citetalias{rescigno_hot_2023}\\
    $F_{\rm bol}$ [erg cm$^2$ s$^{-1}$] & 1.198$\pm$0.048 & \citetalias{rescigno_hot_2023}\\
	$T_{\rm eff}$ [K]  & 4580$\pm$50  & \citetalias{rescigno_hot_2023} \\
	log($g$) [cm s$^{-1}$]& 4.8$\pm$0.3 & \citetalias{rescigno_hot_2023} \\
	${\rm [Fe/H]}$  & 0.12$\pm$0.02 & \citetalias{rescigno_hot_2023} \\
	Mass $[M_{\odot}]$ & 0.744$\pm$0.027 & \citetalias{rescigno_hot_2023} \\
	Radius $[R_{\odot}]$ & 0.709$\pm$0.017 & \citetalias{rescigno_hot_2023} \\
	Density [$\rho_{\odot}$] & 2.09$\pm$0.10 & \citetalias{rescigno_hot_2023} \\
    Age [Gyr] & $3.8^{+5.5}_{-2.7}$ & \citetalias{rescigno_hot_2023} \\ [2pt]
	$v\sin(i_\star)$ [km~s$^{-1}$] & $< 2$\kms & this work \\ [2pt]
	$<\log{R'_{\rm HK}}>$ & -4.87$\pm0.41$ & this work \\
	P$_{\rm rot}$ [days] & $48.94_{-0.81}^{+0.88}$ & this work \\

\rule{0pt}{0ex} \\
\hline
\rule{0pt}{0ex} \\

\end{tabular}
\textsuperscript{*}See ExoFOP: \url{https://exofop.ipac.caltech.edu/tess/target.php?id=75878355}

\end{table}

%% file: Tables/results.tex
\begin{table}
\centering
\caption{System parameters for the TOI-2134 system. The transit and radial-velocity parameters are computed in Section \ref{Sec:joint}. Derived parameters are addressed in Section \ref{Sec:disc} and its subsections alongside the necessary assumptions. \label{tab:planets}}
\begin{tabular}{ccc}
\hline
\hline
\rule{0pt}{0ex} \vspace{-0.2cm} \\
Parameter & Value\\
\rule{0pt}{0ex} \vspace{-0.2cm} \\
\hline
\rule{0pt}{0ex} \vspace{-0.2cm} \\
\multicolumn{2}{c}{\textbf{TOI-2134 b}}\\
\multicolumn{2}{c}{\emph{Transit and Radial-Velocity Parameters}}\\
\rule{0pt}{0ex} \vspace{-0.3cm} \\
    Orbital period $P_{\rm b}$~[days] & $9.229198\pm0.000003$\\
    Time of transit $t_{\rm 0,b}$~[BJD] & $2459010.68937\pm0.00038$\\
    Radius ratio $(R_{\rm b}/R_\star)$ & $0.0353\pm0.00020$ \\
    Transit impact parameter $b_{\rm b}$ & $0.13\pm$0.11  \\
    RV Amplitude $K_{\rm b}$ [\ms] & $3.49\pm0.18$\\
    Eccentricity $e_{\rm b}$ & $0.067\pm0.022$ \\
    Argument of periastron $\omega_{\rm p,b}$ [deg] & $163\pm18$\\
\rule{0pt}{0ex} \vspace{-0.2cm} \\
\hline
\rule{0pt}{0ex} \vspace{-0.2cm} \\
\multicolumn{2}{c}{\emph{Derived Parameters}}\\
\rule{0pt}{0ex} \vspace{-0.3cm} \\
    Orbital inclination $i_{\rm b}$~[deg] & $89.69\pm0.24$ \\
    Radius $R_{\rm b}$ [\rearth] & $2.735\pm0.068$ \\
    Mass $M_{\rm b}$ [\mearth] & $9.37\pm0.54$\\
    Density $\rho_{\rm b}$~[g cm$^{-3}$] &  $2.52\pm0.47$ \\
    Density $\rho_{\rm b}$ [$\rho_{\oplus}$] &  $0.456\pm0.085$ \\
    Semi-major axis $a_{\rm b}$ [AU] &  $0.07803\pm0.00095$\\
    Incident Flux $F_{\rm inc,b}$~[$F_{\rm inc,\oplus}$] &  $32\pm8$\\
    Equilibrium temperature $T_{\rm eq,b}$~[K] &  $660\pm41$ \\
    Transmission Spectroscopy Metric & 159 \\
\rule{0pt}{0ex} \vspace{-0.2cm} \\
\hline
\rule{0pt}{0ex} \vspace{-0.2cm} \\
\multicolumn{2}{c}{\textbf{TOI-2134 c}}\\
\multicolumn{2}{c}{\emph{Transit and Radial-Velocity Parameters}}\\
\rule{0pt}{0ex} \vspace{-0.3cm} \\
    Orbital period $P_{\rm c}$~[days] & $95.852840\pm0.000041$\\
    Time of transit $t_{\rm 0,c}$~[BJD] & $2459718.96933\pm0.00027$ \\
    Radius ratio $(R_{\rm c}/R_\star)$ & $0.09504\pm0.00061$ \\
    Transit impact parameter $b_{\rm c}$ & $0.511\pm0.023$ \\ 
    RV Amplitude $K_{\rm c}$ [\ms] & $10.45\pm0.24$ \\ 
    Eccentricity $e_{\rm c}$ & $0.3125\pm0.01$\\
    Argument of periastron $\omega_{\rm p,c}$ [deg] & $160\pm2.8$\\
\rule{0pt}{0ex} \vspace{-0.2cm} \\
\hline
\rule{0pt}{0ex} \vspace{-0.2cm} \\
\multicolumn{2}{c}{\emph{Derived Parameters}}\\
\rule{0pt}{0ex} \vspace{-0.3cm} \\
    Orbital inclination $i_c$~[deg] & $89.688\pm0.014$ \\ 
    Radius $R_{\rm c}$~[\rearth] & $7.35\pm0.18$\\
    Mass $M_{\rm c}$~[\mearth] & $58.3\pm1.9$ \\
    Density $\rho_{\rm c}$~[g cm$^{-3}$] &  $0.81\pm0.15$ \\
    Density $\rho_{\rm c}$ [$\rho_{\oplus}$] &  $0.146\pm0.026$\\
    Semi-major axis $a_{\rm c}$ [AU] &  $0.3715\pm0.0045$ \\
    Incident Flux $F_{\rm inc,c}$~[$F_{\rm inc,\oplus}$] &  $1.4\pm0.4$ \\
    Equilibrium temperature $T_{\rm eq,c}$~[K] & $302\pm19$ \\
    Transmission Spectroscopy Metric & 208 \\
\rule{0pt}{0ex} \vspace{-0.2cm} \\
\hline
\end{tabular}	
\end{table}

%% file: Tables/gp.tex
\begin{table}
\centering
\label{Tab:gp}

\caption{Results for the joint fit performed in Section \ref{Sec:joint}. This table includes all non-planetary parameters.
\label{tab:gp}}

\begin{tabular}{l@{\hskip6pt}c@{\hskip7pt}}

\rule{0pt}{0ex} \\
\hline
\hline
\rule{0pt}{0ex} \\
Parameter & Value \\
\rule{0pt}{0ex} \\
\hline
\rule{0pt}{0ex} \\
    SOPHIE RV $\beta$ [\ms] & $3.51\pm0.37$ \\
	SOPHIE RV offset [\ms] & $-3.50\pm0.81$ \\
	SOPHIE $V_{c}$ & $3.65\pm0.76$ \\
	SOPHIE $V_{r}$ & $20.0\pm7.8$ \\ 
	HARPS-N RV $\beta$ [\ms] & $0.9972_{-0.0047}^{+0.0021}$ \\ [2pt]
	HARPS-N RV offset [\ms] & $-1.66_{-0.71}^{+0.77}$ \\ [2pt]
	HARPS-N $V_{c}$ & $3.70_{-0.49}^{+0.61}$ \\ [2pt]
	HARPS-N $V_{r}$ & $19.4_{-3.8}^{+4.5}$ \\ [2pt]
    HARPS-N BIS $\beta$ [\ms] & $0.82\pm0.22$ \\ 
	HARPS-N BIS offset [\ms] & $-17.25\pm0.39$ \\
	HARPS-N $B_{c}$ & $1.78_{-0.27}^{+0.33}$ \\ [2pt]
	HARPS-N $B_{r}$ & $-17.5_{-4.0}^{+3.3}$ \\ [2pt]
    HARPS-N S-index $\beta$ & $14.5\pm1.1$ \\
	HARPS-N S-index offset & $499\pm12$ \\
	HARPS-N $S_{c}$ & $57.0_{-7.6}^{+9.0}$ \\ [2pt]
    TESS $\beta$ [ppm] & $0.0000087_{-0.0000024}^{+0.0000038}$ \\ [2pt]
    $q_{1}$ & $0.537\pm0.071$ \\ 
    $q_{2}$ & $0.06\pm0.12$ \\
    $P_{\rm rot}$ [days] & $48.78_{-0.60}^{+0.65}$ \\ [2pt]
    $P_{\rm dec}$ [days] & $74.3_{-7.9}^{+8.3}$ \\ [2pt]
    $\lambda$ & $0.624_{-0.59}^{+0.69}$ \\

\rule{0pt}{0ex} \\
\hline
\rule{0pt}{0ex} \\

\end{tabular}
\end{table}